\documentclass[useAMS,usenatbib]{mn2e}
\usepackage[utf8]{inputenc}
\usepackage{fixltx2e}
\usepackage{color}
\usepackage[usenames,dvipsnames,svgnames,table]{xcolor}
\usepackage{graphicx}
\usepackage{amsmath}
\usepackage{placeins}
\usepackage{morefloats}
\usepackage{amssymb}

\title[Spin period change and the magnetic fields of neutron stars in Be X-ray binaries in the SMC]{Spin period change and the magnetic fields of neutron stars in Be X-ray binaries 
in the Small Magellanic Cloud}
\author[H. Klus et al.]{H. Klus$^{1}$\thanks{E-mail: hvk1g11@soton.ac.uk (HK)}, W.C.G. Ho$^{2}$, M.J. Coe$^{1}$, R.H.D. Corbet$^{3}$ and L.J. Townsend$^{1}$\\
$^{1}$Physics and Astronomy, University of Southampton, Southampton SO17 1BJ, UK\\
$^{2}$Mathematical Sciences, University of Southampton, Southampton SO17 1BJ, UK\\
$^{3}$University of Maryland Baltimore County, X-ray Astrophysics Laboratory, Mail Code 662, NASA Goddard Space Flight Center,\\Greenbelt, MD 20771, USA}

\begin{document}

\date{November 2013}

\pagerange{\pageref{firstpage}--\pageref{lastpage}} \pubyear{2013}

\maketitle

\label{firstpage}

\begin{abstract}
We report on the long-term average spin period, rate of change of spin period and X-ray luminosity during outbursts for 42 Be X-ray binary systems in the Small Magellanic 
Cloud. We also collect and calculate parameters of each system and use these data to determine that all systems contain a neutron star which is accreting via a disc, 
rather than a wind, and that if these neutron stars are near 
spin equilibrium, then over half of them, including all with spin periods over about 100 s, have magnetic fields over the quantum critical level of 
4.4$\times$10\textsuperscript{13} G. If these neutron stars are not close to spin equilibrium, then their magnetic fields are inferred to be much lower, of the order of 
10\textsuperscript{6}-10\textsuperscript{10} G, 
comparable to the fields of neutron stars in low-mass X-ray binaries. Both results are unexpected and have implications for the rate of magnetic field decay and the 
isolated neutron star population. 
\end{abstract}


\section{Introduction}
X-ray binaries contain a compact star - a white dwarf, neutron star or black hole - and a mass donor companion. 
They are generally divided into two groups depending on the mass of their companion. Low-mass X-ray binaries (LMXB) 
contain a companion comparable in mass to the Sun or less, whilst high-mass X-ray binaries (HMXB) contain a companion star over 10 times the mass of the Sun \citep{b26}. 
This is either a supergiant star – in the case of supergiant X-ray binaries (SGXB) – or an OBe star - in the case of Be X-ray binaries (BeXB). OBe stars are 
fast-rotating O- or B-type stars that show Balmer lines in emission, indicating the presence of a circumstellar disc. The compact star 
in all confirmed BeXB is a neutron star \citep{b1}, although there are several white dwarf candidates \citep{Haberl1, deOliv, Sturm}. BeXB typically have eccentric orbits and at periastron 
the neutron star briefly passes through the edge of the OBe star's circumstellar disc where it can accrete matter, 
either via a Keplerian accretion disc or via a wind, causing X-ray outbursts. An accretion disc can only form if the net angular momentum per unit mass of accreted matter 
is large enough (see Section 3). Once it is known how the neutron star in each system accretes, then an appropriate theory of accretion can be used to determine its 
magnetic field based only on the long-term average spin period and X-ray luminosity (for disc accretion) or the 
long-term average spin period, X-ray luminosity, orbital period and relative velocity of accreted matter (for wind accretion). This assumes that the neutron stars in each system are 
near spin equilibrium 
with a rate of change of spin period near zero.
Since we actually measure the rate of change of spin period (or determine upper limits in a few systems), we do not need to assume spin equilibrium, and we can obtain more 
rigorous results.

We use archival Rossi X-ray Timing Explorer (RXTE) data taken with the Proportional Counter Array to determine 
the long-term average spin period, rate of change of spin period and X-ray luminosity during outbursts for 42 BeXB in the Small Magellanic Cloud (SMC). 
We then determine the most likely magnetic field of the neutron star in each of these systems. This is the first time that the rate 
of change of spin period and the long-term average X-ray luminosity has been accurately measured for so many systems. 
The Magellanic Clouds provide astronomers with a valuable resource for studying
BeXB because, not only do they provide whole galactic populations, but they 
are close enough for relatively faint optical sources to be resolved from the ground and they are at well-known distances and are relatively 
un-obscured by interstellar dust unlike most BeXB in the Milky Way.

An outline of the paper is as follows: our observations are discussed in Section 2. An evaluation of disc versus wind accretion is considered in Section 3. We briefly describe 
different models used to determine the magnetic field of the neutron star in Section 4, present our results in Section 5 and discuss possible consequences in Section 6.

\section{Observations}
The observations used in this paper come from the study of the SMC carried out using RXTE over the period 1997-2012. 
The SMC was observed once or twice a week and the activity of the neutron stars determined from timing analysis. See Laycock et al. (2005) and 
Galache et al. (2008) for detailed reports on this work; note that we report here on observations which extend the published record by several further years. 
As discussed in Laycock et al. and Galache et al., the quality of any single observation depends upon the significance of the detected period 
combined with the collimator response to the source. We remove any period detections with a significance less than 99\%, a 
collimator response less than 0.2 (with the exception of SXP15.3 which reached a high enough X-ray luminosity to compensate for the low collimator value) 
and data sets with less than five detections. This leaves 42 systems with the number of detections between 5 and 88 (see Table 1). Fig. 1 shows the location of our sources.

The average count rate is converted to X-ray luminosity using
\begin{equation}
  L=0.4\times10^{37}\times3\times{CR},
\end{equation}
where we assume a distance of 60 kpc to the SMC and an average pulsed fraction of 33\% \citep{Coe}. L is the X-ray luminosity in erg s\textsuperscript{-1} and CR is the 
RXTE count rate in counts/PCU/second.
We then calculate a weighted $\dot{P}$ by fitting the time evolution of the spin period using MPFITEXPR\footnote{www.physics.wisc.edu/$\sim$craigm/idl/down/mpfitexpr.pro}. 
[See the Appendix B for plots of P and L against Modified Julian Date (MJD) for all sources].

All H$\alpha$ measurements were obtained as part of the Southampton SMC X-ray binary pulsar (SXP) optical monitoring campaign that has been running for several years. 
The data were collected primarily at the South African Astronomical Observatory 1.9m telescope in South Africa and also at the ESO New Technology Telescope in Chile. 
The instrumental set-ups and the data reduction in both cases are the same as those described in Coe et al. (2012). 

The orbital periods are mostly taken from Bird et al. (2012). We determine the relative velocity of accreted matter from 
the eccentricity of the system - which is known in six cases \citep{4,11,12} and otherwise assumed to be 0.3$\pm$0.2, the H$\alpha$ equivalent width and 
the total mass of the system. The mass of the neutron star is assumed to be 1.4 M\textsubscript{$\odot$} and we determine the mass of the 
OBe star from spectral type and luminosity class, mostly taken from McBride et al. (2008).

The values obtained for L, P, $\dot{P}$ and the H$\alpha$ equivalent width are shown in Table 1. The orbital period and eccentricity of each system are given in Table 2, 
as well as the spectral type, luminosity class, V band magnitude and the mass and radius of the OBe star in each system.

\begin{figure*}
 \centering
  \includegraphics[height=100mm,angle=0]{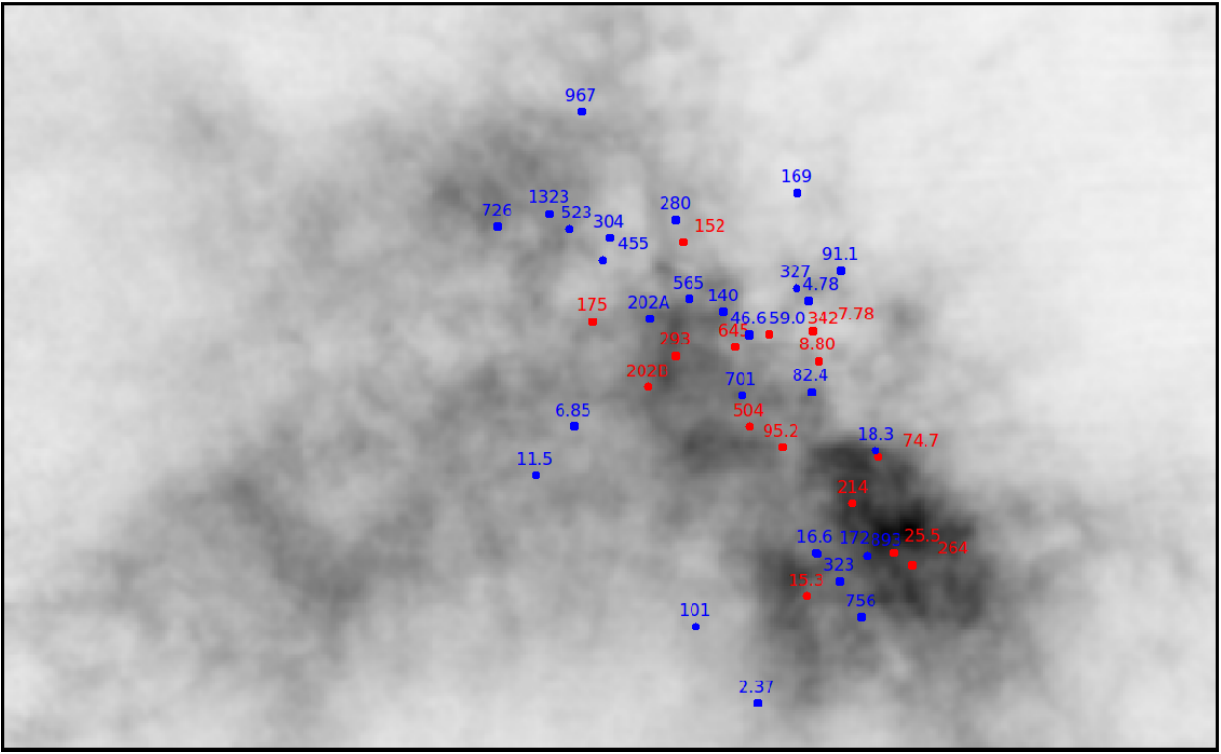}
  \caption{Image of the SMC from Stanimirović et al. (1999), taken by combining Parkes telescope observations of neutral hydrogen with an 
Australia Telescope Compact Array (ATCA) aperture synthesis mosaic, both in the radio spectrum. 
Neutron stars that are spinning up ($\dot{P}<0$; blue circles) and down ($\dot{P}>0$; red circles) are shown. Numbers indicate the spin period of each SMC X-ray binary pulsar (SXP).}
\end{figure*}

\begin{table*}
 \centering
\begin{tabular}{@{}ccccccc@{}}
  \hline
BeXB&No. of&Range of data&Average P&Average L&Average $\dot{P}$&EW H$\alpha$\\
&detections&(yr)&(s)&(10\textsuperscript{37}erg s\textsuperscript{-1})&(s yr\textsuperscript{-1})&(Angstrom)\\
\\
  \hline
SXP2.37	&	24	&	11.22	&	2.37230	$\pm$	0.00001	&	3.70	$\pm$	0.06	&	-0.0036827	$\pm$	0.0000003	&	-7.9	$\pm$	0.6	\\
SXP4.78	&	9	&	12.70	&	4.78015	$\pm$	0.00004	&	0.71	$\pm$	0.03	&	-0.00085	$\pm$	0.00001	&	-43.7	$\pm$	1.1	\\
SXP6.85	&	61	&	8.03	&	6.85206	$\pm$	0.00006	&	1.03	$\pm$	0.02	&	-0.00022	$\pm$	0.00001	&	-3.8			\\
SXP7.78	&	29	&	12.30	&	7.7836	$\pm$	0.0001	&	0.37	$\pm$	0.01	&	0.00262	$\pm$	0.00003	&	-14.3	$\pm$	2.3	\\
SXP8.80	&	46	&	11.23	&	8.89961	$\pm$	0.00009	&	1.58	$\pm$	0.02	&	0.001224	$\pm$	0.000007	&	-5.1	$\pm$	0.4	\\
SXP11.5	&	18	&	0.15	&	11.4806	$\pm$	0.0007	&	1.43	$\pm$	0.04	&	-0.047	$\pm$	0.006	&				\\
SXP15.3	&	10	&	11.13	&	15.2538	$\pm$	0.0009	&	0.66	$\pm$	0.03	&	0.0070	$\pm$	0.0001	&	-25.1	$\pm$	1.5	\\
SXP16.6	&	12	&	5.46	&	16.555	$\pm$	0.001	&	0.23	$\pm$	0.01	&	-0.0131	$\pm$	0.0005	&				\\
SXP18.3	&	74	&	7.39	&	18.3751	$\pm$	0.0003	&	0.67	$\pm$	0.01	&	-0.00118	$\pm$	0.00006	&				\\
SXP25.5	&	35	&	10.56	&	25.5456	$\pm$	0.0007	&	0.36	$\pm$	0.01	&	0.0003	$\pm$	0.0003	&				\\
SXP46.6	&	76	&	13.25	&	46.508	$\pm$	0.003	&	0.45	$\pm$	0.01	&	-0.0155	$\pm$	0.0002	&	-21.9	$\pm$	0.7	\\
SXP59.0	&	88	&	13.10	&	58.859	$\pm$	0.005	&	0.84	$\pm$	0.02	&	-0.0206	$\pm$	0.0005	&	-23.4	$\pm$	1.4	\\
SXP74.7	&	28	&	12.31	&	74.647	$\pm$	0.008	&	1.10	$\pm$	0.03	&	0.0300	$\pm$	0.0004	&	-18.3	$\pm$	2.3	\\
SXP82.4	&	21	&	12.24	&	82.46	$\pm$	0.02	&	0.50	$\pm$	0.02	&	-0.022	$\pm$	0.002	&	-25.9	$\pm$	1.1	\\
SXP91.1	&	59	&	13.48	&	88.38	$\pm$	0.01	&	0.83	$\pm$	0.01	&	-0.4417	$\pm$	0.0006	&	-26.7	$\pm$	2.6	\\
SXP95.2	&	10	&	11.01	&	95.21	$\pm$	0.04	&	0.63	$\pm$	0.05	&	0.027	$\pm$	0.005	&				\\
SXP101	&	5	&	13.32	&	101.77	$\pm$	0.04	&	0.33	$\pm$	0.04	&	-0.05	$\pm$	0.01	&	-7.8			\\
SXP140	&	5	&	6.67	&	140.4	$\pm$	0.7	&	0.4	$\pm$	0.1	&	-0.16	$\pm$	0.10	&	-47.3	$\pm$	3.1	\\
SXP152	&	23	&	11.94	&	151.68	$\pm$	0.06	&	0.39	$\pm$	0.02	&	0.02	$\pm$	0.01	&	-17.3	$\pm$	1.7	\\
SXP169	&	35	&	11.97	&	167.0	$\pm$	0.1	&	0.69	$\pm$	0.02	&	-0.238	$\pm$	0.006	&	-29.2	$\pm$	2.6	\\
SXP172	&	42	&	10.39	&	171.86	$\pm$	0.05	&	0.39	$\pm$	0.02	&	-0.123	$\pm$	0.006	&	-15.0	$\pm$	1.3	\\
SXP175	&	11	&	8.50	&	175.0	$\pm$	0.1	&	0.50	$\pm$	0.05	&	0.15	$\pm$	0.01	&				\\
SXP202A	&	16	&	13.28	&	201.5	$\pm$	0.1	&	0.50	$\pm$	0.03	&	-0.13	$\pm$	0.01	&	-18.1			\\
SXP202B	&	5	&	13.24	&	202.3	$\pm$	0.4	&	0.27	$\pm$	0.05	&	0.21	$\pm$	0.04	&				\\
SXP214	&	16	&	13.26	&	213.7	$\pm$	0.1	&	0.29	$\pm$	0.03	&	0.12	$\pm$	0.02	&				\\
SXP264	&	6	&	10.13	&	262.6	$\pm$	0.4	&	0.21	$\pm$	0.03	&	0.06	$\pm$	0.08	&	-30.1	$\pm$	1.7	\\
SXP280	&	6	&	8.24	&	280.0	$\pm$	0.3	&	0.29	$\pm$	0.05	&	-0.37	$\pm$	0.06	&	-42.0	$\pm$	3.1	\\
SXP293	&	12	&	11.08	&	293.9	$\pm$	0.3	&	0.28	$\pm$	0.02	&	0.03	$\pm$	0.05	&				\\
SXP304	&	7	&	6.09	&	304.1	$\pm$	0.4	&	0.68	$\pm$	0.09	&	-0.5	$\pm$	0.2	&	-70.4	$\pm$	6.2	\\
SXP323	&	20	&	9.42	&	318.7	$\pm$	0.2	&	0.55	$\pm$	0.03	&	-0.95	$\pm$	0.02	&	-30.9	$\pm$	1.1	\\
SXP327	&	5	&	1.76	&	327.5	$\pm$	0.5	&	0.17	$\pm$	0.02	&	-0.8	$\pm$	0.8	&				\\
SXP342	&	20	&	10.29	&	341.0	$\pm$	0.4	&	0.43	$\pm$	0.03	&	0.96	$\pm$	0.06	&				\\
SXP455	&	7	&	12.05	&	452.3	$\pm$	1.3	&	0.7	$\pm$	0.1	&	-0.2	$\pm$	0.3	&	-15.1	$\pm$	2.0	\\
SXP504	&	31	&	13.29	&	502.0	$\pm$	0.6	&	0.35	$\pm$	0.02	&	0.34	$\pm$	0.05	&	-52.9	$\pm$	3.9	\\
SXP565	&	8	&	7.48	&	564.1	$\pm$	1.2	&	0.19	$\pm$	0.04	&	-0.9	$\pm$	0.4	&	-37.4	$\pm$	2.9	\\
SXP645	&	13	&	11.74	&	644.6	$\pm$	2.2	&	0.25	$\pm$	0.03	&	0.3	$\pm$	0.3	&				\\
SXP701	&	27	&	11.71	&	695.8	$\pm$	1.3	&	0.27	$\pm$	0.02	&	-0.0	$\pm$	0.3	&	-37.1	$\pm$	3.5	\\
SXP726	&	7	&	4.20	&	726.3	$\pm$	4.1	&	0.48	$\pm$	0.08	&	-1	$\pm$	1	&				\\
SXP756	&	29	&	11.20	&	754.6	$\pm$	0.8	&	0.63	$\pm$	0.02	&	-0.01	$\pm$	0.08	&	-27.0	$\pm$	3.6	\\
SXP893	&	29	&	10.44	&	890.8	$\pm$	1.5	&	0.24	$\pm$	0.02	&	-1.9	$\pm$	0.3	&				\\
SXP967	&	7	&	2.85	&	962.9	$\pm$	4.0	&	0.71	$\pm$	0.09	&	-1	$\pm$	3	&	-12.3			\\
SXP1323	&	26	&	4.89	&	1323.7	$\pm$	2.6	&	0.97	$\pm$	0.04	&	-6.2	$\pm$	0.7	&	-17.1	$\pm$	1.5	\\
  \hline
\end{tabular}
\caption{Long-term average spin period P, X-ray luminosity during outbursts L, rate of change of spin period $\dot{P}$, and H$\alpha$ 
equivalent width -EW H$\alpha$, for all the systems in our data set satisfying the criteria described in Section 2. P, L and $\dot{P}$ are obtained 
from RXTE data and H$\alpha$ measurements were 
obtained as part of the Southampton 
SXP optical monitoring campaign. Detections are defined as having a significance $>$99\% and a collimator response of $>$0.2 (except in the case of SXP15.3 as discussed in Section 2).}
\end{table*}

\begin{table*}
 \centering
\begin{tabular}{@{}ccccccc@{}}
  \hline
BeXB&P\textsubscript{orb}&Eccentricity&Spectral type&V\textsubscript{mag}&M\textsubscript{OB}&R\textsubscript{OB}\\
&(d)&&\& luminosity class&&(M\textsubscript{$\odot$})&(R\textsubscript{$\odot$})\\
\\
  \hline
SXP2.37	&	18.62	$\pm$	0.02 [1]	&	0.07	$\pm$	0.02 [4]	&	O9.5 III-V [13]	&	16.38	$\pm$	0.02 [13]	&	19.9	$\pm$	4.3	&	11.7	$\pm$	5.3	\\
SXP4.78	&	23.9	$\pm$	0.06 [3]	&				&	B0-B1 V [10]	&	15.8 [10]			&	14.3	$\pm$	3.5	&	6.6	$\pm$	1.4	\\
SXP6.85	&	21.9	$\pm$	0.1 [4]	&	0.26	$\pm$	0.03 [4]	&	O9.5-B0 IV-V [13]	&	14.59	$\pm$	0.02 [13]	&	16.7	$\pm$	2.9	&	7.2	$\pm$	1.2	\\
SXP7.78	&	44.93	$\pm$	0.01 [2]	&				&	B1-B1.5 IV-V [13]	&	14.91	$\pm$	0.02 [13]	&	12.2	$\pm$	2.6	&	6.0	$\pm$	1.2	\\
SXP8.80	&	28.47	$\pm$	0.04 [7]	&	0.41	$\pm$	0.04 [4]	&	O9.5-B0 IV-V [13]	&	14.87	$\pm$	0.12 [13]	&	16.7	$\pm$	2.9	&	7.2	$\pm$	1.2	\\
SXP11.5	&	36.3	$\pm$	0.4 [6]	&	0.28	$\pm$	0.03 [11]	&	O9.5-B0 IV-V [11]	&	14.8 [13]			&	16.7	$\pm$	2.9	&	7.2	$\pm$	1.2	\\
SXP15.3	&	74.32	$\pm$	0.03 [2]	&				&	O9.5-B0 III-V [13]	&	14.67	$\pm$	0.04 [13]	&	18.9	$\pm$	5.1	&	11.3	$\pm$	5.4	\\
SXP16.6	&	33.72	$\pm$	0.05 [7]	&				&		&	 		 	&				&				\\
SXP18.3	&	17.79	$\pm$	0.03 [2]	&	0.43	$\pm$	0.03 [12]	&	B1-B3 V [4]	&	15.6 [10]			&	10.7	$\pm$	4.2	&	5.6	$\pm$	1.6	\\
SXP25.5	&	22.53	$\pm$	0.01 [2]	&				&	 	&	15.2 [10]			&				&				\\
SXP46.6	&	137.4	$\pm$	0.2 [2]	&				&	O9.5-B1 IV-V [13]	&	14.72	$\pm$	0.03 [13]	&	15.2	$\pm$	4.3	&	6.8	$\pm$	1.6	\\
SXP59.0	&	122.1	$\pm$	0.38 [7]	&				&	O9 V [13]	&	15.28	$\pm$	0.01 [13]	&	19.5	$\pm$	2.0	&	7.8	$\pm$	1.0	\\
SXP74.7	&	33.387	$\pm$	0.006 [2]	&	0.40	$\pm$	0.23 [4]	&	B3 V [13]	&	16.92	$\pm$	0.06 [13]	&	8.5	$\pm$	2.0	&	4.9	$\pm$	1.0	\\
SXP82.4	&	362.3	$\pm$	4.1 [7]	&				&	B1-B3 III-V [13]	&	15.02	$\pm$	0.02 [13]	&	12.1	$\pm$	5.6	&	8.7	$\pm$	4.7	\\
SXP91.1	&	88.37	$\pm$	0.03 [2]	&				&	B0.5 III-V [13]	&	15.06	$\pm$	0.06 [13]	&	16.1	$\pm$	3.9	&	10.3	$\pm$	4.8	\\
SXP95.2	&	280	$\pm$	8 [8]	&				&	 	&				&				&				\\
SXP101	&	21.949	$\pm$	0.003 [2]	&				&		&	15.67	$\pm$	0.15 [13]	&				&				\\
SXP140	&	197	$\pm$	5 [5]	&				&	B1 V [13]	&	15.88	$\pm$	0.03 [13]	&	12.9	$\pm$	2.0	&	6.2	$\pm$	1.0	\\
SXP152	&				&				&	B1-B2.5 III-V [13]	&	15.69	$\pm$	0.03 [13]	&	12.6	$\pm$	5.2	&	8.9	$\pm$	4.7	\\
SXP169	&	68.37	$\pm$	0.07 [2]	&				&	B0-B1 III-V [13]	&	15.53	$\pm$	0.02 [13]	&	16.2	$\pm$	5.4	&	10.4	$\pm$	5.1	\\
SXP172	&	68.78	$\pm$	0.08 [2]	&				&	O9.5-B0 V [13]	&	14.45	$\pm$	0.02 [13]	&	16.7	$\pm$	2.9	&	7.2	$\pm$	1.2	\\
SXP175	&	87.2	$\pm$	0.2 [9]	&				&	B0-B0.5 IIIe [9]	&	14.6 [9]			&	19.0	$\pm$	3.0	&	14.5	$\pm$	1.5	\\
SXP202A	&	71.98	$\pm$	5 [10]	&				&	B0-B1 V [13]	&	14.83	$\pm$	0.02 [13]	&	14.3	$\pm$	3.5	&	6.6	$\pm$	1.4	\\
SXP202B	&	224.6	$\pm$	0.3 [2]	&				&	B0-5 III [10]	&	15.6 [10]			&	13.5	$\pm$	8.5	&	11.5	$\pm$	4.5	\\
SXP214	&	4.5832	$\pm$	0.0004 [2]	&				&	B2-B3 III [14]	&	 			&	11.9	$\pm$	3.2	&	11.0	$\pm$	1.7	\\
SXP264	&	49.12	$\pm$	0.03 [2]	&				&	B1-B1.5 V [13]	&	15.85	$\pm$	0.01 [13]	&	13.7	$\pm$	4.1	&	6.4	$\pm$	1.6	\\
SXP280	&	127.62	$\pm$	0.25 [2]	&				&	B0-B2 III-V [13]	&	15.65	$\pm$	0.03 [13]	&	14.9	$\pm$	6.4	&	9.8	$\pm$	5.3	\\
SXP293	&	59.726	$\pm$	0.006 [2]	&				&	B2-B3 V [10]	&	14.9 [10]			&	9.5	$\pm$	3.0	&	5.2	$\pm$	1.3	\\
SXP304	&	520	$\pm$	12 [5]	&				&	B0-B2 III-V [13]	&	15.72	$\pm$	0.01 [13]	&	14.9	$\pm$	6.4	&	9.8	$\pm$	5.3	\\
SXP323	&	116.6	$\pm$	0.6 [7]	&				&	B0-B0.5 V [13]	&	15.44	$\pm$	0.04 [13]	&	15.0	$\pm$	2.8	&	6.8	$\pm$	1.2	\\
SXP327	&	45.93	$\pm$	0.01 [2]	&				&	 	&	16.3 [10]			&				&				\\
SXP342	&				&				&		&				&				&				\\
SXP455	&	74.56	$\pm$	0.05 [2]	&				&	B0.5-B2 IV-V [13]	&	15.49	$\pm$	0.02 [13]	&	12.4	$\pm$	3.9	&	6.1	$\pm$	1.5	\\
SXP504	&	270.1	$\pm$	0.5 [2]	&				&	B1 III-V [13]	&	14.99	$\pm$	0.01 [13]	&	14.5	$\pm$	3.7	&	9.7	$\pm$	4.5	\\
SXP565	&	152.4	$\pm$	0.3 [2]	&				&	B0-B2 IV-V [13]	&	15.97	$\pm$	0.02 [13]	&	13.1	$\pm$	4.7	&	6.3	$\pm$	1.7	\\
SXP645	&				&				&	B0-B0.5 III–V [10]	&	14.6 [10]			&	17.0	$\pm$	4.8	&	10.7	$\pm$	5.1	\\
SXP701	&	412	$\pm$	5 [10]	&				&	O9.5 V [13]	&	15.87	$\pm$	0.05 [13]	&	17.5	$\pm$	2.0	&	7.4	$\pm$	1.0	\\
SXP726	&				&				&	B0.5-B3 III-V [10]	&	15.6 [10]			&	12.9	$\pm$	6.4	&	9.0	$\pm$	5.0	\\
SXP756	&	393.6	$\pm$	1.2 [2]	&				&	O9.5-B0.5 III-V [13]	&	14.98	$\pm$	0.02 [13]	&	18.0	$\pm$	5.7	&	11.0	$\pm$	5.4	\\
SXP893	&	3.7434	$\pm$	0.0005 [2]	&				&	 	&	16.3 [10]			&				&				\\
SXP967	&	101.4	$\pm$	0.2 [2]	&				&	B0-B0.5 III-V [10]	&	14.6 [10]			&	17.0	$\pm$	4.8	&	10.7	$\pm$	5.1	\\
SXP1323	&	26.174	$\pm$	0.002 [2]	&				&	B0 III-V [13]	&	14.65	$\pm$	0.02 [13]	&	17.9	$\pm$	4.1	&	11.0	$\pm$	5.0	\\
  \hline
\end{tabular}
\caption{Orbital period P\textsubscript{orb}, and eccentricity e, of each system in our data set. Also shown are the spectral type, luminosity class, V band magnitude and 
the mass and radius of the OBe star in each system. References are given in brackets and are as follows; 1 \citep{1}, 2 \citep{2}, 3 \citep{3}, 
4 \citep{4}, 5 \citep{5}, 6 \citep{6}, 7 \citep{7}, 8 \citep{8}, 9 \citep{9}, 10 \citep{10}, 11 \citep{11}, 12 \citep{12}, 13 \citep{13}, 14 \citep{14}.}
\end{table*}

\section{Disc or Wind Accretion}
From the parameters of each system,
we determine whether each neutron star is accreting via a disc or wind. An accretion disc will form if the net angular momentum per unit mass of accreted matter J, 
is too large for it to accrete spherically or quasi-spherically. This occurs at the circularization radius R\textsubscript{circ}, where
\begin{equation}
R\textsubscript{circ}=\frac{J^2}{GM}.
\end{equation}
Here G is the gravitational constant and M is the mass of the neutron star. 
If the neutron star and its magnetosphere are fully engulfed in the OBe star's circumstellar disc then
\begin{equation}
J=-\frac{1}{4}(n_{\rho}+1/2)V_{rel}\frac{R_B^2}{a};
\end{equation}
if the OBe star's circumstellar disc is truncated so that only approximately half the neutron star's magnetosphere is exposed to accreting material at a time 
(as is illustrated in Fig. 2 and discussed in Reig et al. (1997), Negueruela \& Okazaki (2001), Okazaki \& Negueruela (2001) and Okazaki et al. (2002)) then
\begin{equation}
J_{t}=-V_{rel}R_B\left[\frac{2}{3\pi}+\frac{1}{8}(n_{\rho}+1/2)\frac{R_B}{a}\right]
\end{equation}
[see Appendix A for the derivation of equations (3) and (4)]. Here V\textsubscript{rel} is the relative velocity of accreted matter, n\textsubscript{$\rho$} 
depends on the density gradient and is taken to be $2.5\pm0.5$ and a is the semimajor axis of the system, which we determine using
\begin{equation}
a=\left[\frac{P_{orb}^{2}G(M+M_{OB})}{4\pi^{2}}\right]^{1/3},
\end{equation}
where P\textsubscript{orb} is the orbital period of the neutron star and M\textsubscript{OB} is the mass of the OBe star. R\textsubscript{B} is the Bondi radius given by
\begin{equation}
R_{B}=\frac{2GM}{V_{rel}^{2}}.
\end{equation}
In order for matter to be accreted it must first penetrate the neutron star's magnetosphere. The radius of the neutron star's magnetosphere is approximately equal to the 
Alfv\'{e}n radius R\textsubscript{A} - which occurs where the magnetic pressure 
of the neutron star is balanced by the ram pressure of infalling matter - and is given by
\begin{equation}
R_{A}=\left(\frac{\mu^{4}}{2GM\dot{M}^{2}}\right)^{1/7}.
\end{equation}
Here, $\mu$ is the magnetic moment of the neutron star ($\sim$BR\textsubscript{NS}\textsuperscript{3}), where R\textsubscript{NS} is the radius of the 
neutron star, assumed throughout to be 10 km. We assume a magnetic field in the range 10\textsuperscript{7}-10\textsuperscript{15} G, corresponding to 
$\mu\approx$10\textsuperscript{25}-10\textsuperscript{33} G cm\textsuperscript{3}. $\dot{M}$ (=LR\textsubscript{NS}/GM) is the mass accretion rate. Thus, disc accretion occurs if 
R\textsubscript{circ}$>$R\textsubscript{A}.
This inequality can be rearranged to find the maximum relative velocity of accreted matter  
for which disc accretion can take place V\textsubscript{Crel}. 
We determine V\textsubscript{Crel} for each system using equations (2)-(7).

\begin{figure}
\centering
\includegraphics[scale=0.55,angle=0]{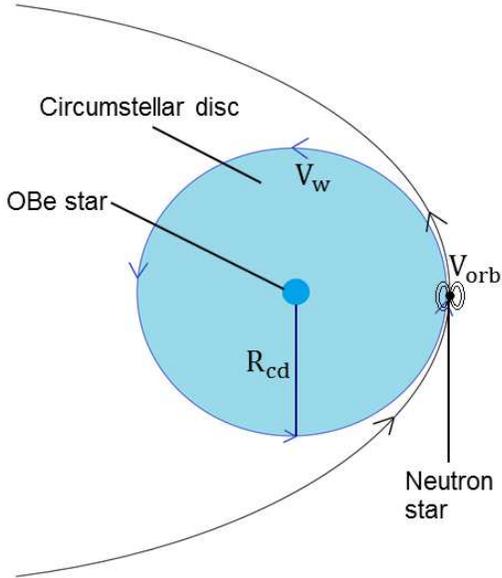}
\caption{Schematic of the orbital system of a BeXB (not to scale). In this case the OBe star's circumstellar disc is truncated by the neutron star orbit.}
\end{figure}

We determine the actual relative velocity of each system using
\begin{equation}
V\textsubscript{rel}=\sqrt{V\textsubscript{w}\textsuperscript{2}+V\textsubscript{orb}\textsuperscript{2}+2V\textsubscript{w}V\textsubscript{orb}cos\theta},
\end{equation}
where V\textsubscript{w} is the velocity of accreted material, which we calculate by determining the stellar wind velocity at the radius 
of the OBe star's circumstellar disc, R\textsubscript{cd}. V\textsubscript{orb} is the orbital velocity of the neutron star and $\theta$ is the angle at which the accreted 
material and neutron star impact, 
where $\theta$=180$^{\circ}$ indicates that the star and disc are in prograde motion. Some of these parameters are shown in Fig. 2. V\textsubscript{w} is calculated by assuming that 
the circumstellar disc is in a circular orbit using;
\begin{equation}
V\textsubscript{w}=\sqrt{\frac{GM_{OB}}{R_{cd}}},
\end{equation}
where R\textsubscript{cd} is calculated using
\begin{equation}
\log\left(\sqrt{\frac{R_{OB}}{R_{cd}}}\right)=[-0.32\times\log(-EW H\alpha)]-0.2.
\end{equation}
Here R\textsubscript{OB} is the radius of the OBe star and -EW H$\alpha$ is the equivalent width 
of H$\alpha$ lines (in Angstrom), which are given in Table 1. V\textsubscript{orb} is calculated using
\begin{equation}
V_{orb}=\sqrt{\frac{G(M+M_{OB})}{a}\frac{1+e}{1-e}},
\end{equation}
where e is the eccentricity of the system. Note that the above calculation estimates the velocity at a point and that
the Bondi radius formally extends to infinity as the relative velocity becomes
negligible. However even if the velocity difference across the Bondi radius is 
approximately V\textsubscript{w} [$\sim$140 km s\textsuperscript{-1} (M\textsubscript{OB}/15 M\textsubscript{$\odot$})\textsuperscript{1/2}(R\textsubscript{cd}/150 R\textsubscript{$\odot$})\textsuperscript{-1/2}], 
we see that V\textsubscript{w}/V\textsubscript{Crel}$\lesssim$1 when the disc is not truncated for most SXPs and 
V\textsubscript{w}/V\textsubscript{Crel}$<$0.3 when the disc is truncated for all SXPs. More accurate results could be found using numerical simulations which are beyond the scope of this paper.

The critical relative velocity for disc accretion, V\textsubscript{Crel}, and the actual relative velocity of each system, V\textsubscript{rel}, are used 
to determine which systems in our data set contain neutron stars that accrete via a disc and which accrete via a quasi-spherical wind. We also determine the minimum possible 
angle that the neutron star's orbit must be misaligned with the OBe star's circumstellar disc for disc accretion to cease $\theta$\textsubscript{Crit}, in both the truncated 
and non-truncated case, by rearranging equation (8) and using V\textsubscript{Crel} in place of V\textsubscript{rel}.

\section{Magnetic Fields}
We used three models applicable to disc accretion and two models applicable to wind accretion in order to determine the magnetic fields of the neutron star in each system. For 
spin equilibrium methods, see also Chashkina and Popov (2012).

\subsection{Ghosh and Lamb model}
The Ghosh and Lamb (1979) model is applicable to BeXB systems that contain neutron stars which accrete via a disc, whether or not they have achieved spin 
equilibrium (which entails a $\dot{P}$ of zero and is discussed further in Section 4.2).
This model predicts
\begin{equation}
  -\dot{P}=5.0\times10^{-5} \mu_{30}^{2/7} n(\omega_{s}) R_{NS6}^{6/7}\left(\frac{M}{M_\odot}\right)^{-3/7}I_{45}^{-1} (PL_{37}^{3/7})^2,
\end{equation}
where $\dot{P}$ is the rate of change of spin period measured in s yr\textsuperscript{-1}, $\mu_{30}$=$\mu$/10\textsuperscript{30} G cm\textsuperscript{3}, 
I\textsubscript{45}=I/10\textsuperscript{45} g cm\textsuperscript{2}, R\textsubscript{NS6}=R\textsubscript{NS}/10\textsuperscript{6} cm and 
L\textsubscript{37}=L/10\textsuperscript{37} erg s\textsuperscript{-1}. n($\omega$\textsubscript{s}) is the dimensionless accretion torque and 
depends on the fastness parameter $\omega\textsubscript{s}$.
For 0$<\omega$\textsubscript{s}$<$0.9,
\begin{equation} 
  n(\omega_{s})=1.39(1-(\omega_{s}[4.03(1-\omega_{s})^{0.173}-0.878]))(1-\omega_{s})^{-1} 
\end{equation}
within 5\% accuracy and 
\begin{equation}
\omega_{s}={1.35}\mu_{30}^{6/7}R_{NS6}^{-3/7}\left(\frac{M}{M_\odot}\right)^{-2/7}(PL_{37}^{3/7})^{-1}.
\end{equation}

\subsection{Kluzniak and Rappaport model}
The Kluzniak and Rappaport (2007) model, like the Ghosh and Lamb model, is also applicable to BeXB systems that contain neutron stars which accrete via a disc, 
whether or not they have achieved spin equilibrium. This model predicts
\begin{equation} 
-\dot{P}=8.2\times10^{-5}\mu_{30}^{2/7} g(\omega_{s}) R_{NS6}^{6/7}\left(\frac{M}{M_\odot}\right)^{-3/7}I_{45}^{-1} (PL_{37}^{3/7})^2,
\end{equation}
where $g$($\omega$\textsubscript{s}) is a function of the fastness parameter $\omega$\textsubscript{s} only and is of order unity; here we assume 
R\textsubscript{0}$\approx$R\textsubscript{A} for simplicity, where R\textsubscript{0} is defined in Kluzniak and Rappaport (2007). 
Note also small differences in definitions of R\textsubscript{A} and $\omega$\textsubscript{s} between here and Kluzniak and Rappaport. 
In the spin equilibrium case, when $\omega_{s}\approx$1,
\begin{equation} 
B\approx4.4\times10^{13}G\,R_{NS6}^{-3}\left(\frac{M}{M_\odot}\right)^{5/6}\dot{M}_{16}^{1/2}(P/100 s)^{7/6},
\end{equation}
where $\dot{M}_{16}$=$\dot{M}$/10\textsuperscript{16} g s\textsuperscript{-1}.

\subsection{Equilibrium period model for disc accretion}
The systems in our data set contain neutron stars which have relatively low rates of change of spin period and could be considered to be spinning close to their equilibrium period 
(see Section 5 for more details). Therefore, we also use a number of models which assume that the neutron stars in each system are close to spin equilibrium. The first of which is the 
equilibrium period model for disc accretion \citep{Davidson, Alpar}.
 
As discussed in Section 3, accretion can only occur if matter is able to penetrate the neutron star's magnetosphere which is approximately at R\textsubscript{A}. This can only 
happen if the neutron 
star, and hence its magnetosphere, are spinning slow enough; specifically, they must be spinning slower than the Keplerian velocity of matter that is corotating with the neutron star. 
The radius at which matter can corotate with the neutron star is known as the corotation radius
\begin{equation}
R_{co}=\left(\frac{GMP^{2}}{4\pi^{2}}\right)^{1/3}.
\end{equation}
For accretion to occur, R\textsubscript{co}$>$R\textsubscript{A}. This generally causes the neutron star to spin up, $\dot{P}<0$. 
If R\textsubscript{co}$<$R\textsubscript{A}, then matter is not 
accreted, but expelled by the centrifugal force of the neutron star in what is known as the propeller mechanism \citep{Shvartsman}. This causes the neutron star to spin down, 
$\dot{P}>0$.  

If a system contains a neutron star which is in spin equilibrium, then it has a $\dot{P}$ of $0$, i.e. it is neither spinning up nor spinning down. This occurs when 
R\textsubscript{co}$\approx$R\textsubscript{A} 
and so this equation can be rearranged to show the magnetic field of the neutron star in each system in terms of its X-ray luminosity - which is proportional to the mass accretion 
rate - and spin period, i.e.,
\begin{equation}
B\approx1.8\times10^{13}G\, R_{NS6}^{-3}\left(\frac{M}{M_\odot}\right)^{5/6}\dot{M}_{16}^{1/2}\left(\frac{P}{100 s}\right)^{7/6}.
\end{equation}
Thus for systems with similar X-ray luminosities, as in the BeXB considered here, neutron stars with a higher spin period will have a higher magnetic field. 

A similar result can be found with the equilibrium model for disc accretion \citep{Pringle}. This determines 
the neutron star's magnetic field by considering accelerating and decelerating torques where, in the case of a star in spin equilibrium, these torques are balanced.
The spin up torque is equal to $\dot{M}\sqrt{GM\epsilon R_A}$ where $\epsilon$ is a numerical coefficient assumed to be 0.45 \citep{Lipunov}. The spin down torque is equal to 
$\kappa$\textsubscript{t}$\mu$\textsuperscript{2}/{R\textsubscript{co}}\textsuperscript{3} where $\kappa$\textsubscript{t} is a numerical coefficient assumed to 
be 1/3 \citep{Lipunov}, i.e.,
\begin{equation}
\dot{M}\sqrt{GM\epsilon R_A}=\frac{\kappa_{t}\mu^2}{{R_{co}}^3}.
\end{equation}
This yields
\begin{equation}
B\approx1.4\times10^{13}G \left(\frac{\epsilon}{\kappa_{t}^{2}}\right)^{7/24}R_{NS6}^{-3}\left(\frac{M}{M_\odot}\right)^{5/6}\dot{M}_{16}^{1/2}\left(\frac{P}{100 s}\right)^{7/6}.
\end{equation}

\subsection{Equilibrium period model for wind accretion}
Wind accretion is possible if the system contains an OBe star with a non-truncated circumstellar disc and the neutron star's orbit is misaligned with the OBe star's 
circumstellar disc by an amount $>\theta$\textsubscript{Crit}. For this case, we use two 
models which assume accretion is occurring via a wind.
The first model is the equilibrium period model for wind accretion \citep{IllarionovS, IllarionovK} and is the same as the equilibrium model for disc accretion 
but with a 
different spin up torque. Here, the spin up torque is assumed to be $\dot{M}\eta\Omega$\textsubscript{orb}{R\textsubscript{B}\textsuperscript{2}, where 
$\Omega$\textsubscript{orb} (=2$\pi$/P\textsubscript{orb}) is the orbital frequency and $\eta$ is a numerical coefficient assumed to be 1/4 \citep{Lipunov}. 
B is then calculated using the following equation:
\begin{equation}
B\approx1\times10^{14}G \left(\frac{\eta}{\kappa_{t}}\right)^{1/2}R_{NS6}^{-3}\left(\frac{M}{M_\odot}\right)^{3/2}\dot{M}_{16}^{1/2}
\end{equation}
\begin{displaymath}
\times\left(\frac{V_{rel}}{100 km/s}\right)^{-2}\left(\frac{P/100 s}{(P_{orb}/10 d)^{1/2}}\right).
\end{displaymath}

Here V\textsubscript{rel} is measured in km s\textsuperscript{-1}, P\textsubscript{orb} is measured in days and everything else in cgs units.
Like the models for disc accretion, the magnetic field is proportional to the X-ray luminosity and spin period, but here it is also inversely proportional to the relative velocity of 
accreted matter and the orbital period of the system.

\subsection{Shakura et al. model}
The Shakura et al. model (see Postnov et al. 2011 and Shakura et al. 2012) applies to BeXB systems where accretion occurs via a wind and assumes that the neutron star is close to 
spin equilibrium. This model assumes that, in systems with slowly rotating neutron stars, matter forms an extended quasi-static shell around its magnetosphere which matter must pass 
through before being accreted. Postnov et al. (2011) find
\begin{equation}
B\approx 1\times10^{15}G\, R_{NS6}^{-3}\dot{M}_{16}^{1/3}\left(\frac{V_{rel}}{100 km/s}\right)^{-11/3}\left(\frac{P/100 s}{P_{orb}/10 d}\right)^{11/12}.
\end{equation}

\section{Results}
We determine the long-term average X-ray luminosity during outbursts, spin period and rate of change of spin period 
for 42 BeXB systems in the SMC. The results which we calculate from equations (2)-(12) are given in Table 3. A positive correlation 
is found between $\dot{P}$ and P which follows a power-law with a slope of $\sim4/3$, as is shown in Table 1 and Fig. 3; note that the Ghosh and Lamb model predicts a power 
law of 2 for slow rotators [see equation (12)]. There is also an asymmetry between the number of systems 
containing neutron stars that are spinning up and down, with 
27 systems containing neutron stars that are spinning up on average and 15 systems containing neutron stars which are spinning down. Such a large proportion of spin down sources 
may indicate that these systems are close to spin equilibrium.

Fig. 4 shows a correlation between the neutron star spin period and the orbital period of the BeXB in our data set as would be expected from the Corbet 
relation \citep{Corbet}. There is no apparent relationship between luminosity and either of these factors. However care should be taken as the instrument limitations of RXTE 
prevent the detection of luminosities below $\sim$10\textsuperscript{36} erg s\textsuperscript{-1} (at the SMC distance). Given the uncertainty in relative velocities, we are unable to determine
whether a correlation between velocity and luminosity exists, which would
have provided an argument for very low relative velocities.

Fig. 5 shows the ratio of the relative velocity which we calculate using equation (8) and the critical relative velocity which we calculate using equations (2), (7) and either (3) or (4)
against spin period. Equation (3) assumes that the neutron star and its magnetosphere are completely engulfed by the 
circumstellar disc during periastron and equation (4) assumes that, due to truncation of the OBe star's circumstellar disc, only half of the neutron 
star and its magnetosphere are exposed to the circumstellar disc at a time. In order for disc accretion to occur this ratio must be $<$1. If this ratio is $>$1, then 
accretion via wind becomes possible. It is clear from Fig. 5 that, whether the neutron star and its magnetosphere are completely engulfed by the 
circumstellar disc during periastron or not, all prograde systems are expected to contain neutron stars which accrete via a disc. 

Fig. 5 shows that the systems in 
our data set are likely to contain neutron stars which are disc accreting. Therefore we consider models which assume disc accretion to be more appropriate. We also consider the 
disc accretion models which take $\dot{P}$ into account to be more accurate. This applies to the Kluzniak and Rappaport and Ghosh and Lamb models. Fig. 6 shows the 
magnetic fields inferred from these models (using the measured P, $\dot{P}$ and L for each BeXB). Results from the Ghosh and Lamb model are also given in Table 3. 
Both models predict two possible fields, where the higher magnetic fields are close to the magnetic fields predicted by spin equilibrium models which are shown in Fig. 7.
Figs 6 and 7 show that the higher magnetic fields predicted by the Kluzniak and Rappaport and Ghosh and Lamb models, and all models which assume disc accretion and spin equilibrium, 
predict magnetic fields over the quantum critical level of 4.4$\times$10\textsuperscript{13} G for all systems containing neutron stars with spin periods over about 100 s. 
This is over half of the systems in our data set. 
The lower fields predicted by the Kluzniak and Rappaport and Ghosh and Lamb models are all 
well below this value and are similar to those of neutron stars in LMXB. Unlike systems containing a neutron star which is close to spin equilibrium, the magnetic field in this 
case does not appear to depend on spin period.

Whilst disc accretion seems more likely, we also show in Table 3 the angle at which neutron star's orbit must be misaligned with the OBe star's circumstellar disc for disc 
accretion to cease in the non-truncated case. It is possible that this could happen - the OBe star could be tilted by the supernova that created the neutron star \citep{Lai} or 
the OBe star's circumstellar disc could be tilted due to radiation-induced warping \citep{Pringle96} - and so we also determine the magnetic field using models for wind accretion 
assuming that these systems are in retrograde motion. If the circumstellar disc is truncated then wind 
accretion is not possible at any angle.

Fig. 7 shows that the models for wind accretion, specifically the equilibrium period model and the Shakura et al. model, predict lower magnetic fields than the models for disc accretion, 
assuming the systems contain a neutron star which is close to spin equilibrium, 
but still have systems containing neutron stars with fields over the quantum critical level. Incidentally, if these systems are wind accreting and in prograde motion 
then the predicted magnetic fields of their neutron stars would almost all be greater than the quantum critical level, with fields predicted to be as 
high as 10\textsuperscript{20} G.

Fig. 8 shows the magnetic field calculated by the Ghosh and Lamb model - and assuming the neutron stars in these systems are close to spin equilibrium - alongside magnetars, 
and neutron stars in Galactic 
BeXB that have had their magnetic fields determined from cyclotron resonance scattering features (CRSF; see below), plotted against spin period. 
If we were to plot the lower values from the Ghosh and Lamb model, then they would be situated far below the bottom 
of the plot and away from all other known sources (for further discussion of the systems in our data set in relation to other neutron
star populations, see \cite{Ho}).

\begin{figure*} 
\centering
\includegraphics[scale=0.55,angle=90]{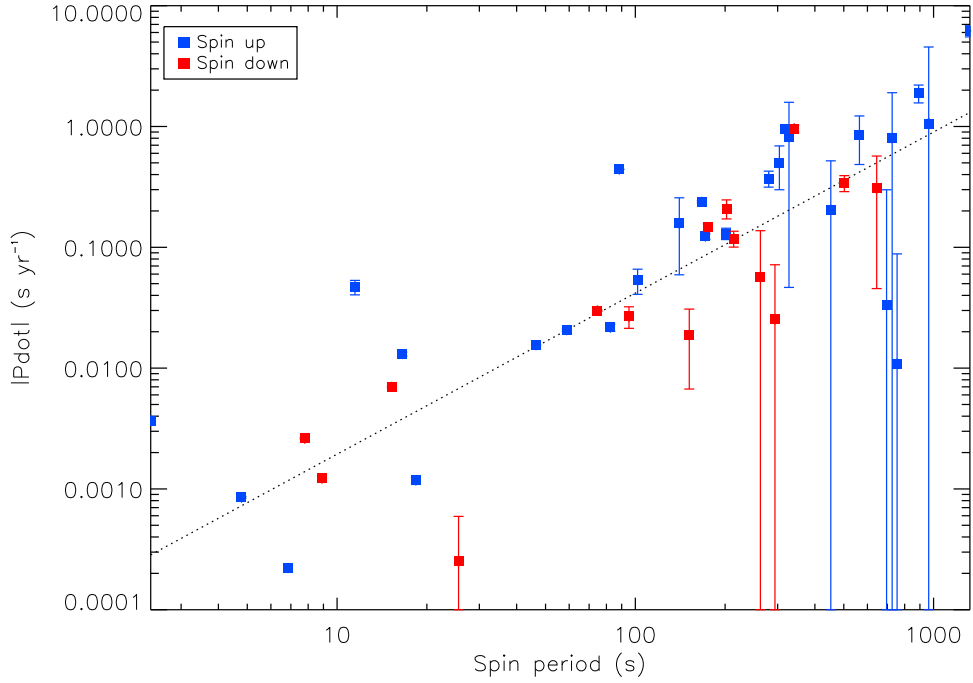}
\caption{The long-term average rate of change of spin period $\dot{P}$, against spin period P, for neutron stars in the 42 BeXB in our data set: 
blue for $\dot{P}<0$ (spin up) and red for $\dot{P}>0$ (spin down). The dotted line indicates a correlation of $\dot{P}\propto$ P\textsuperscript{4/3}.}
\end{figure*}

\begin{figure*} 
\centering
\includegraphics[scale=0.55,angle=0]{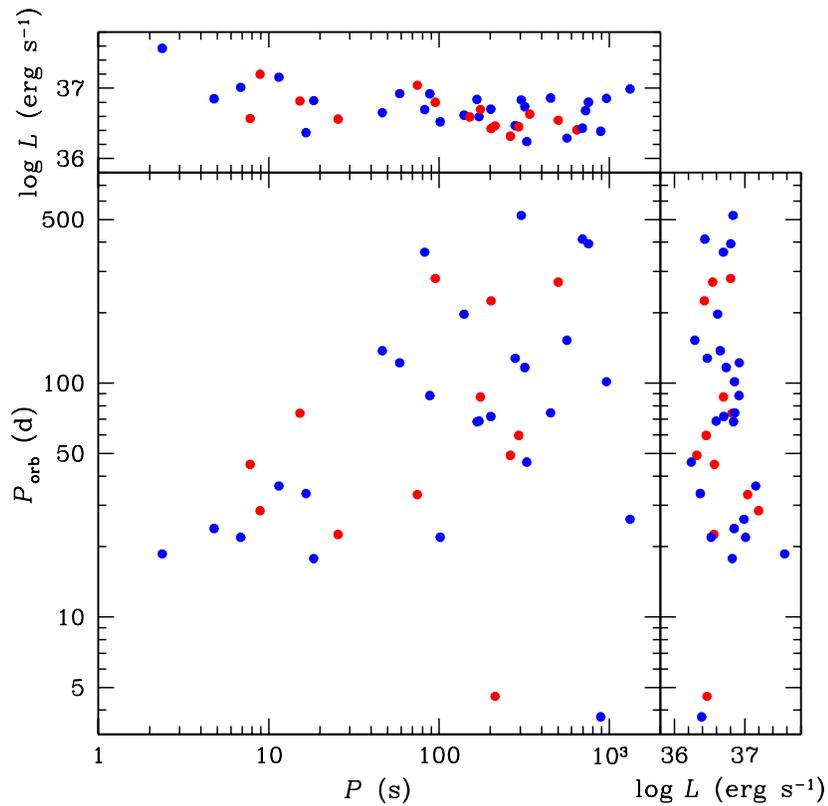}
\caption{Corbet diagram (P\textsubscript{orb} versus P)) for the BeXB in our data set. Circles indicate neutron stars that are spinning up (blue) and spinning down (red). 
Error bars are not shown but are mostly smaller than the symbols.}
\end{figure*}

\begin{figure*} 
\centering
\includegraphics[scale=0.55,angle=90]{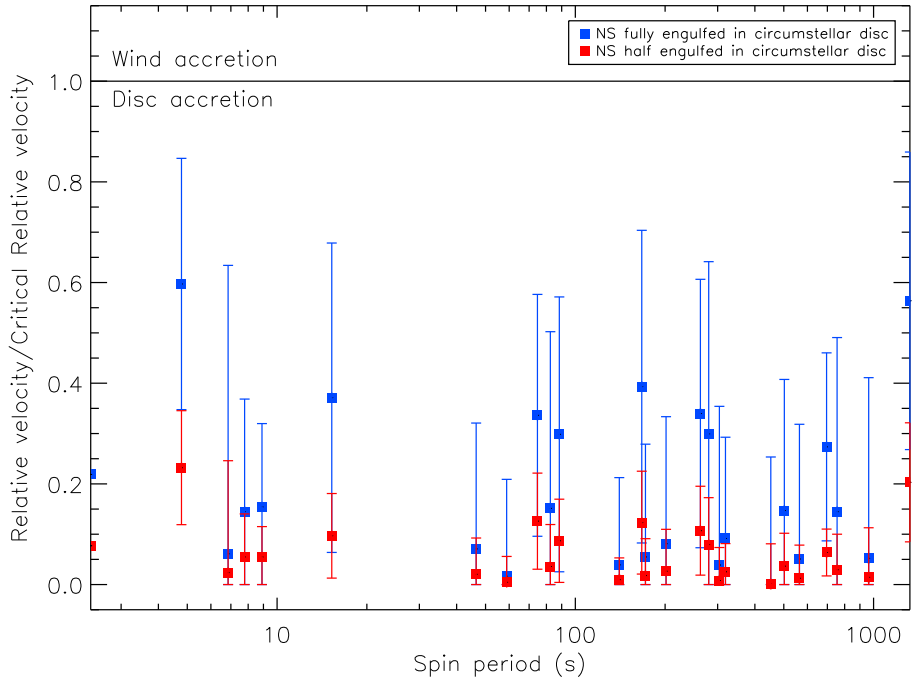}
\caption{The ratio of relative velocity [equation (8)] and critical relative velocity [blue for equation (3) and red for equation (4)] versus spin period. 
Disc accretion occurs when this ratio $<$1 and wind accretion when this ratio $>$1.}
\end{figure*}

\begin{figure*}
\centering
\includegraphics[scale=0.55,angle=90]{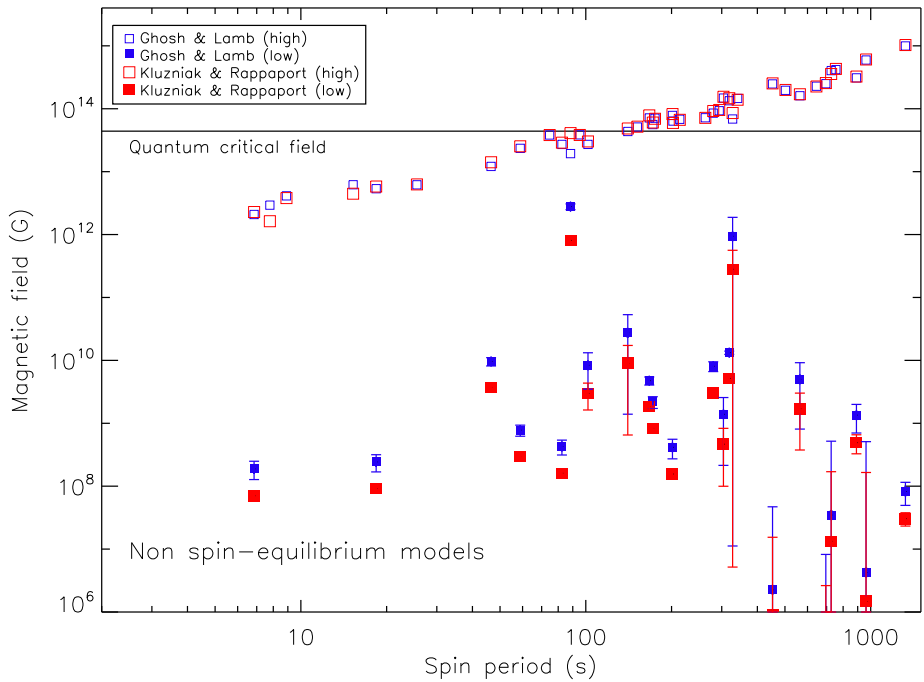}
\caption{Neutron star magnetic field versus spin period determined using the Ghosh and Lamb (1979) and Kluzniak and Rappaport (2007) non-spin equilibrium 
models (see sections 4.1 and 4.2, respectively). The fact that there are two possible values, referred to as high and low, is discussed in Section 5. 
Errors bars for the higher values are not shown but are mostly the size of the symbols.}
\end{figure*}

\begin{figure*}
\centering
\includegraphics[scale=0.55,angle=90]{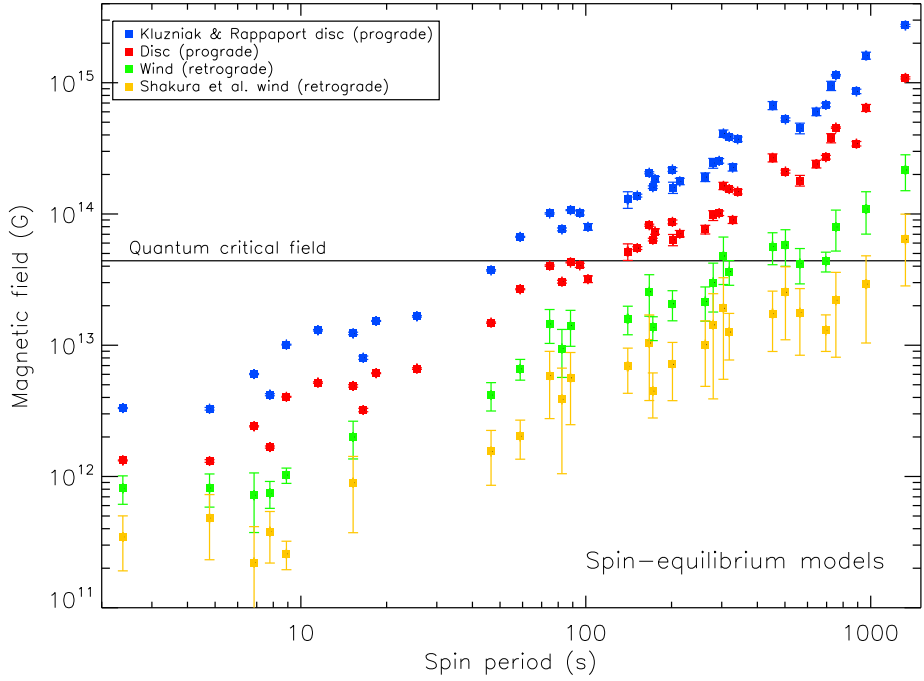}
\caption{Neutron star magnetic field versus spin period determined using the spin equilibrium models discussed in Sections 4.2-4.5}
\end{figure*}

\begin{figure*} 
\centering
\includegraphics[scale=0.55,angle=90]{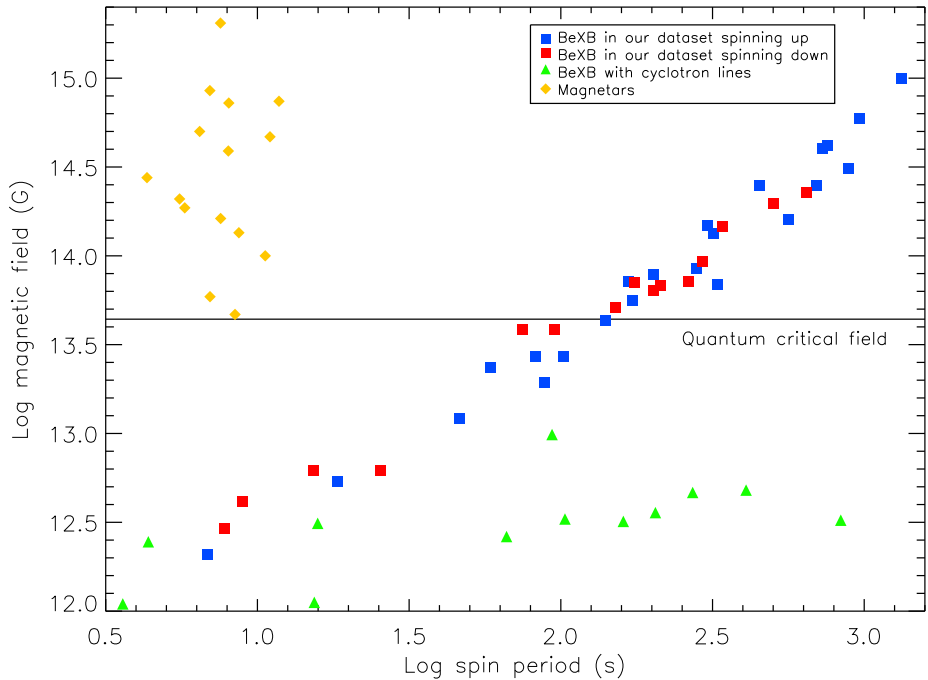}
\caption{Magnetic field for the neutron stars in the BeXB in our data set which we calculate using the Ghosh and Lamb model and assuming these systems contain neutron stars which are 
close to spin equilibrium (blue for spin up and red for spin down). Also shown are BeXB whose magnetic fields are measured 
using cyclotron resonance scattering features (green) - where B and P are from; \citep{san, Mak, De, Hei, Mih, Shr, Ken, Hei2, Ts, Klo, Doroa, Cob} - and magnetars 
(yellow) where P and $\dot{P}$ are from Manchester et al. (2005) and B is determined from the standard 
B=3.2$\times$10\textsuperscript{19} G (P$\dot{P}$)\textsuperscript{1/2} relation.}
\end{figure*}

\begin{table*}
 \centering
\begin{tabular}{@{}cccccccc@{}}
  \hline
BeXB&R\textsubscript{cd}/R\textsubscript{OB}&V\textsubscript{Crel}&V\textsubscript{Crel}&V\textsubscript{rel}&$\theta$\textsubscript{Crit}&B&B\\
&&(km/s)&(km/s)&(km/s)&(degrees)&(10\textsuperscript{12}G)&(10\textsuperscript{10}G)\\
&&&(truncated disc)&&&&\\
\\
  \hline
SXP2.37	&	9.4	$\pm$	0.5	&	242	$\pm$	27	&	687	$\pm$	191	&	53	$\pm$	52	&	68	$\pm$	20	&				&				\\
SXP4.78	&	28.2	$\pm$	0.5	&	219	$\pm$	25	&	563	$\pm$	155	&	131	$\pm$	53	&	60	$\pm$	46	&				&				\\
SXP6.85	&	6	$\pm$	5	&	224	$\pm$	25	&	577	$\pm$	159	&	14	$\pm$	128	&	49	$\pm$	37	&	2.1	$\pm$	0.4	&	0.019	$\pm$	0.006	\\
SXP7.78	&	14	$\pm$	1	&	187	$\pm$	21	&	496	$\pm$	136	&	27	$\pm$	42	&	62	$\pm$	22	&	2.9	$\pm$	0.5	&				\\
SXP8.80	&	7.1	$\pm$	0.4	&	215	$\pm$	24	&	607	$\pm$	169	&	19	$\pm$	35	&	47	$\pm$	16	&	4.1	$\pm$	0.7	&				\\
SXP11.5	&				&	203	$\pm$	23	&	615	$\pm$	172	&				&				&				&				\\
SXP15.3	&	19.8	$\pm$	0.8	&	165	$\pm$	19	&	632	$\pm$	178	&	61	$\pm$	50	&	59	$\pm$	36	&	6	$\pm$	1	&				\\
SXP16.6	&				&				&				&				&				&				&				\\
SXP18.3	&				&	240	$\pm$	28	&	554	$\pm$	152	&				&				&	5	$\pm$	1	&	0.024	$\pm$	0.007	\\
SXP25.5	&		 		&				&				&				&				&	6	$\pm$	1	&				\\
SXP46.6	&	18.1	$\pm$	0.4	&	144	$\pm$	16	&	502	$\pm$	141	&	10	$\pm$	36	&	58	$\pm$	23	&	12	$\pm$	2	&	1.0	$\pm$	0.1	\\
SXP59.0	&	18.9	$\pm$	0.7	&	149	$\pm$	16	&	557	$\pm$	157	&	2	$\pm$	29	&	56	$\pm$	18	&	23	$\pm$	4	&	0.08	$\pm$	0.02	\\
SXP74.7	&	16	$\pm$	1	&	219	$\pm$	25	&	583	$\pm$	161	&	74	$\pm$	52	&	72	$\pm$	27	&	38	$\pm$	7	&				\\
SXP82.4	&	20.2	$\pm$	0.6	&	120	$\pm$	14	&	507	$\pm$	144	&	18	$\pm$	42	&	68	$\pm$	34	&	27	$\pm$	6	&	0.04	$\pm$	0.01	\\
SXP91.1	&	21	$\pm$	1	&	163	$\pm$	18	&	559	$\pm$	157	&	49	$\pm$	44	&	66	$\pm$	28	&	19	$\pm$	4	&	278	$\pm$	29	\\
SXP95.2	&				&				&				&				&				&	38	$\pm$	8	&				\\
SXP101	&	9	$\pm$	4	&				&				&				&				&	27	$\pm$	7	&	0.8	$\pm$	0.5	\\
SXP140	&	30	$\pm$	1	&	135	$\pm$	15	&	542	$\pm$	153	&	5	$\pm$	23	&	70	$\pm$	17	&	43	$\pm$	19	&	3	$\pm$	3	\\
SXP152	&	16	$\pm$	1	&				&				&				&				&	51	$\pm$	11	&				\\
SXP169	&	22	$\pm$	1	&	171	$\pm$	20	&	546	$\pm$	153	&	67	$\pm$	53	&	65	$\pm$	37	&	71	$\pm$	14	&	0.47	$\pm$	0.07	\\
SXP172	&	14.2	$\pm$	0.8	&	166	$\pm$	19	&	508	$\pm$	142	&	9	$\pm$	37	&	54	$\pm$	21	&	56	$\pm$	11	&	0.22	$\pm$	0.05	\\
SXP175	&				&	157	$\pm$	17	&	512	$\pm$	143	&				&				&	71	$\pm$	16	&				\\
SXP202A	&	16.0	$\pm$	3.0	&	169	$\pm$	19	&	514	$\pm$	144	&	14	$\pm$	43	&	60	$\pm$	24	&	78	$\pm$	17	&	0.04	$\pm$	0.01	\\
SXP202B	&				&	128	$\pm$	16	&	536	$\pm$	153	&				&				&	64	$\pm$	17	&				\\
SXP214	&				&	309	$\pm$	35	&	525	$\pm$	135	&				&				&	68	$\pm$	16	&				\\
SXP264	&	22.2	$\pm$	0.8	&	177	$\pm$	20	&	561	$\pm$	157	&	60	$\pm$	47	&	61	$\pm$	30	&	72	$\pm$	19	&				\\
SXP280	&	27	$\pm$	1	&	144	$\pm$	17	&	543	$\pm$	154	&	43	$\pm$	49	&	68	$\pm$	37	&	85	$\pm$	24	&	0.8	$\pm$	0.1	\\
SXP293	&				&	178	$\pm$	20	&	609	$\pm$	171	&				&				&	93	$\pm$	22	&				\\
SXP304	&	38.2	$\pm$	2.2	&	110	$\pm$	13	&	529	$\pm$	151	&	4	$\pm$	35	&	76	$\pm$	29	&	148	$\pm$	37	&	0.1	$\pm$	0.1	\\
SXP323	&	22.6	$\pm$	0.5	&	151	$\pm$	17	&	547	$\pm$	154	&	14	$\pm$	30	&	63	$\pm$	19	&	134	$\pm$	28	&	1.32	$\pm$	0.08	\\
SXP327	&				&		$\pm$		&	509	$\pm$	142	&				&				&	69	$\pm$	34	&	94	$\pm$	94	\\
SXP348	&	24	$\pm$	1	&	153	$\pm$	17	&	577	$\pm$	167	&	27	$\pm$	29	&	63	$\pm$	19	&	146	$\pm$	31	&				\\
SXP455	&	14	$\pm$	1	&	173	$\pm$	20	&	541	$\pm$	152	&	0	$\pm$	44	&	63	$\pm$	23	&	248	$\pm$	63	&	0.0002	$\pm$	0.0045	\\
SXP504	&	32	$\pm$	2	&	123	$\pm$	14	&	496	$\pm$	141	&	18	$\pm$	32	&	72	$\pm$	24	&	196	$\pm$	40	&				\\
SXP565	&	26	$\pm$	1	&	138	$\pm$	16	&	560	$\pm$	159	&	7	$\pm$	37	&	64	$\pm$	25	&	161	$\pm$	47	&	0.5	$\pm$	0.4	\\
SXP645	&				&				&				&				&				&	228	$\pm$	59	&				\\
SXP701	&	25	$\pm$	2	&	109	$\pm$	12	&	467	$\pm$	132	&	30	$\pm$	20	&	53	$\pm$	18	&	250	$\pm$	53	&	0.0000003	$\pm$	0.0008222	\\
SXP726	&				&				&				&				&		$\pm$		&	404	$\pm$	41	&	0.003	$\pm$	0.048	\\
SXP756	&	21	$\pm$	2	&	114	$\pm$	13	&	562	$\pm$	160	&	17	$\pm$	39	&	59	$\pm$	36	&	419	$\pm$	79	&				\\
SXP893	&				&				&				&				&				&	310	$\pm$	69	&	0.13	$\pm$	0.06	\\
SXP967	&	13	$\pm$	4	&	156	$\pm$	18	&	569	$\pm$	161	&	8	$\pm$	56	&	58	$\pm$	38	&	595	$\pm$	150	&	0.0004	$\pm$	0.0506	\\
SXP1323	&	15.5	$\pm$	0.9	&	213	$\pm$	24	&	591	$\pm$	164	&	120	$\pm$	61	&	54	$\pm$	28	&	996	$\pm$	196	&	0.008	$\pm$	0.003	\\
  \hline
\end{tabular}
\caption{Ratio of the radius of circumstellar disc R\textsubscript{cd}, over the radius of the OBe star in each system R\textsubscript{OB} [see equation (10)], and  
the critical relative velocity for disc accretion which we determine using equations (2)-(7). Also shown are the relative velocities of each system which we determine using equation (8), 
where for disc accretion to occur, the relative velocity must be below the critical velocity, and the angle at which the neutron star's orbit must be misaligned with the OBe 
star's circumstellar disc for 
disc accretion to cease in the non-truncated case, which we determine using equations (2)-(6) and (10). The last two columns show the magnetic fields which we determine using the 
Ghosh and Lamb model.}
\end{table*}

\section{Discussion and conclusions}
We determine the long-term average spin period, rate of change of spin period and X-ray luminosity during outbursts for 
42 BeXB in the SMC. All systems are expected to contain neutron stars which accrete from a disc, assuming that their orbital axis is not misaligned with the orbit of the OBe star's 
circumstellar disc by more than the values of $\theta$\textsubscript{Crit} given in Table 3. If the neutron stars in these 
systems are close to spin equilibrium, then the magnetic field of over half, and all with spin periods over 
about 100 s, are over the quantum critical field B\textsubscript{Q}=4.4$\times$10\textsuperscript{13} G. 
Note that similarly high estimates for the magnetic fields of 
neutron stars in binaries have been made before (see for example \cite{Lipunov, Li1, Bozzo1, Rei, Klus1}). If the systems containing neutron stars that are spinning up are not close 
to spin equilibrium, then some are predicted to have a much lower field. Both of these possibilities are unexpected when compared to the 
magnetic fields of neutron stars in Galactic BeXB measured using CRSF (see below).

Our work extends well-beyond that of Chashkina \& Popov (2012) who derived magnetic
fields of BeXB using some of the spin equilibrium models used here
(see Section 4.1, 4.3$-$4.5), as well as others, but without accounting for
the observed $\dot{P}$.
They find similar results but arrive at different conclusions, in particular
they favour the Shakura et al. wind accretion model (see Section 4.5), which
produces magnetic fields that better match their population synthesis
calculations (including magnetic field decay).
As we showed in Section 5 (see Fig. 5), we find that it is more likely that the
BeXB in our data set are accreting via a disc, rather than via a wind.
This was derived (see Section 3) using the same disc versus wind accretion
criteria as in Chashkina \& Popov (2012), except we account for the pulsar
interacting with the circumstellar disc of the OBe star near periastron
passage (this has not been accounted for in previous works, see, e.g.,
Illarionov \& Sunyaev 1975; Shapiro \& Lightman 1976; Wang 1981) and we
evaluate the criteria using the measured system parameters for each BeXB.
We also note that the population synthesis calculations performed in
Chashkina \& Popov (2012) assume a Ohmic magnetic field decay time-scale
of $10^6\mbox{ yr}$, as indicated by Pons et al. (2009).
Our results suggest that the field decay time-scale is longer and therefore
may still match population synthesis models with a longer time-scale.

12 Galactic BeXB contain neutron stars which have had their magnetic fields measured using cyclotron features. All have fields between 
10\textsuperscript{12}-10\textsuperscript{13} G (see Fig. 8) 
and they do not show the same correlation between magnetic field and spin period that would be expected for systems containing a neutron star in spin equilibrium [see equations (16), 
(18), (20), (21) or (22)]. The lower fields for our data set, 
which we calculate from the Ghosh and Lamb model, are far lower than the magnetic fields of neutron stars in Galactic BeXB whose fields have been determined 
from cyclotron features ($\sim$10\textsuperscript{6}-10\textsuperscript{10} G). The higher fields, which we also calculate using the Ghosh and Lamb model, are mostly 
much higher ($\sim$10\textsuperscript{12}-10\textsuperscript{15} G) than the magnetic fields of neutron stars in Galactic BeXB which have been determined from cyclotron features. 
The values of the magnetic field determined for cyclotron line sources are 
closest to those predicted by the Shakura et al. model for systems in our data set that contain neutron stars which are wind accreting, orbiting OBe stars with a non-truncated circumstellar disc 
and in retrograde motion. However it is unlikely that this applies to most of our systems.

Another way to predict magnetic fields from our data set that are below the quantum critical level, and closer to the magnetic fields of neutron stars in Galactic BeXB, 
is to assume that the systems in our data set containing neutron stars with spin periods over 100 s have eccentricities larger than $\approx0.8$. This would allow for wind 
accretion in prograde systems. The fields predicted from prograde quasi-spherical wind accretion with these eccentricities are below the quantum critical level. 
However Townsend et al. (2011a) show that 
no known BeXB system containing neutron stars with spin periods above 100 s have eccentricities above 0.5. There are four known BeXB with eccentricities above 0.8: three 
have spin periods below 1 s and one has a spin period of 94.3 s. Four of the cyclotron line sources have known eccentricities - two of which have spin periods above 100 
s - and all are below 0.5.

\begin{table*}
 \centering
\begin{tabular}{@{}ccccccccccc@{}}
  \hline
BeXB & $P$ & $\dot{P}$ & $L$ & $B$ from CRSF& $B$ from GL&Height from GL&Height from B+12\\
& (s) & (s yr$^{-1}$) & ($10^{37}$ erg s$^{-1}$) &(10\textsuperscript{12} G)&(10\textsuperscript{12} G)&(km)&(km)\\
  \hline
GRO J1008-57	&	94	&	0.25	&	0.4	&	9.9	&	39	&	5.8	&	0.3	\\
A0535+26	&	103	&	-0.03	&	0.1	&	4.3	&	14 or 0.03	&	4.8	&	1.7	\\
RX J0440.9+4431	&	205	&	0.21	&	0.4	&	3.2	&	73	&	18.3	&	0.3	\\
1A1118-616	&	408	&	-14.52	&	1.4	&	4.8	&	230 or 2.2	&	26.2	&	0.1	\\
X Per	&	837	&	0.11	&	0.004	&	3.3	&	42	&	13.5	&	8.7	\\
  \hline
\end{tabular}
\caption{Galactic BeXB containing neutron stars with measured cyclotron resonance scattering feature (CRSF) and spin period time derivative (see text for references). 
Also shown are magnetic fields determined 
using CRSF and the Ghosh and Lamb (GL) model, CRSF accretion column heights if the field predicted by the GL model is the surface field, and heights determined using the model described 
in Becker et al. (2012) (B+12).}
\end{table*}

For the sources whose magnetic field has been measured by CRSF, we can use
their known values of P, L and $\dot{P}$ and the models
discussed in Section 4 to cross-check this magnetic field.
This is currently possible for five sources:
GRO J1008-57 \citep{Shr}, A0535+26 \citep{Ken,Mai}, 
RX J0440.9+4431 \citep{Ts}, 1A1118-616 \citep{Doroa, Nes} and X Per \citep{Cob}, whose magnetic fields are given in Table 4. Assuming that the neutron stars in these systems are disc 
accreting and close to spin equilibrium, then all models predict higher fields than those 
determined by the cyclotron features. The model which predicts magnetic fields closest to those determined by CRSF is the Shakura et al. model 
for systems containing a neutron star which is wind accreting from a non-truncated circumstellar disc and in retrograde motion, though it seems unlikely that this applies to all 
12 of the CRSF systems. The magnetic fields predicted by the Ghosh and Lamb model are shown in Table 4. 
If the magnetic fields predicted by the Ghosh and Lamb model are taken to be the magnetic field at the surface of the neutron star and the CRSF gives the magnetic field 
of the accretion column, then the column would have to be
$4-30$~km above the surface (assuming $B\propto r^{-3}$), where the height for each source is shown in Table 4. 
A similar difference in field determination has previously been noted for SGXB GX 301-2 \citep{Dorob}. It has also been observed that cyclotron lines may not be a true indicator of 
the surface magnetic field of neutron stars since different values can be measured at different times, with these values sometimes changing rapidly \citep{Rey, Sta}. 

Becker et al. (2012) show that different heights of the accretion column can be determined depending on whether or not the X-ray luminosity is above a critical value 
($\sim\mbox{a few}\times 10^{37}\mbox{ erg s$^{-1}$}$). The BeXB containing neutron stars that have had their fields measured with cyclotron features mostly have X-ray 
luminosities which are lower than this,
and therefore the height of the accretion column should be inversely proportional to the X-ray luminosity. The heights calculated using the equations given by Becker et al. are 
between $\sim0.1$ and 2 km from the surface except in the case of X Per where the height is $\sim9$ km from the surface (see Table~4). In order to 
obtain the required height to reconcile the two magnetic field determinations for each source, the X-ray luminosity of the neutron star must be less than 
$\sim$10\textsuperscript{34}~erg~s\textsuperscript{-1}. However 
Poutanen et al. (2013) suggest that this issue may be more complex, with the CRSF from a single source changing due to the fact that the CRSF can originate 
from radiation that is produced by the accretion column 
and reflected from the neutron star's surface rather than from an accretion column that is changing heights. Although it is currently unclear why the values 
predicted using the Ghosh and Lamb model are so different to those determined by cyclotron features, future work in this area would help resolve the matter.

The wide range of magnetic fields may be caused by different mechanisms for forming neutron stars e.g. as a result of an electron capture supernova
(\cite{Nomoto, Podsia}) or accretion induced
collapse (\cite{Nomoto, Taam, NomotoK}). If over half the neutron stars in systems in our data set do have fields over the quantum critical value then this could mean that magnetic field decay occurs more 
slowly than previously thought \citep{Pons}. It may also mean that half the isolated neutron star population also have fields this high and are currently 
not observed due to selection effects.

\section*{Acknowledgements}
We would like to thank the anonymous referee whose helpful suggestions have improved the quality of the paper. 
HK acknowledges a studentship from the Science and Technology Facilities Council (STFC) in the United Kingdom.
WCGH acknowledges support from STFC.
LJT acknowledges support from the University of Southampton Mayflower scholarship.

\appendix
\section{Derivation of angular momentum of accreting matter}
For a neutron star accreting from a companion, two cases have previously been considered: 
accretion via Roche lobe overflow or accretion via a stellar wind. For Roche lobe overflow, accreting matter
has high angular momentum and therefore easily forms an accretion disc around the neutron
star. In the case of a stellar wind, matter has low angular momentum and therefore accretion
likely occurs spherically \citep{IllarionovS, shapirolightman76}. We consider a third case: accretion from a circumstellar
disc around an OBe main sequence companion at periastron passage. We follow closely the derivation from Shapiro \& Lightman (1976) (see also \citealt{wang81}) of the
angular momentum of the accreting matter, bearing in mind that the work
of Shapiro \& Lightman is for matter from a radially outflowing wind,
while our case is a wind that 
(1) is moving tangentially to the direction of the companion and 
(2) has a Kelperian velocity gradient with distance from the companion.

When the neutron star enters the circumstellar disc, the star forms an accretion cylinder
of radius R\textsubscript{B}. If there is no density or velocity gradient in the wind, then there is 
no net angular momentum transferred due to symmetry, i.e. spin up on one side
and spin down on the other. However, since the density and velocity decrease as a function
of distance, e.g.,
\begin{equation}
\rho=\rho_{0}R_{cd}^{-n_{\rho}},
\end{equation}
there will be a net angular momentum change due to accretion.

Let us consider a cross-section of the accretion cylinder, which defines the \emph{xy}-plane, with
radius R\textsubscript{B}. The angular momentum passing through this plane \emph{dx dy} is
\begin{equation}
dJ = (\rho\ dx\ dy\ V_{rel}\ dt)V_{rel}\ y=\rho yV_{rel}^{2}\ dx\ dy\ dt,
\end{equation}
where \emph{y} is the radial distance from the cylinder axis. We examine the
first-order density and velocity perturbation about the periastron separation a, i.e.,
\begin{equation}
\begin{split}
\rho(x,y)\approx\rho(a)+\left.\frac{d\rho}{dR_{cd}}\right|_{a} y=\rho(a)\left(1-n_{\rho}\frac{y}{a}\right)
\end{split}
\end{equation}
\begin{equation}
\begin{split}
V_{rel}(x,y)\approx V_{rel}(a)+\left.\frac{dV_{rel}}{dR_{cd}}\right|_{a} y=V_{rel}(a)\left(1-\frac{y}{2a}\right).                         
\end{split}
\end{equation}
Note that the $dV\textsubscript{rel}/dR\textsubscript{cd}$ term accounts for both the gradient in disc velocity and neutron star
orbital velocity. Substituting back into the angular momentum equation we obtain
\begin{equation}
\frac{dJ}{dt}=\rho(a)yV_{rel}(a)^{2}\ dx\ dy\left[1-\left(n_{\rho}+\frac{1}{2}\right)\frac{y}{a}\right].
\end{equation}

The net angular momentum transferred per unit mass is then found by integrating \emph{dJ/dt} 
over the accretion cylinder and dividing by the mass accretion rate where $\dot{M}$=$\pi R\textsubscript{B}\textsuperscript{2}\rho V\textsubscript{rel}$, i.e.,
\begin{equation}
\begin{split}
J&=\frac{\rho V_{rel}^2}{\pi R_B^2 \rho V_{rel}}\int y[1-(n_{\rho}+1/2)(y/a)]\ dx\ dy \\
&=-\frac{1}{4}(n_{\rho}+1/2)V_{rel}\frac{R_B^2}{a}.
\end{split}
\end{equation}

We also consider the case where the neutron star truncates the radial extent of the
circumstellar disc. If matter only occupies the hemisphere closest to the companion, then we find
\begin{equation}
\begin{split}
J_t&=\frac{V_{rel}}{\pi R_B^2}\int_{\pi}^{2\pi}\int_0^{R_B}[1-(n_{\rho}+1/2)(R_{cd}\sin\theta/ a )R_{cd}^2\sin\theta drd\theta\\
&=-V_{rel}R_B\left[\frac{2}{3\pi}+\frac{1}{8}(n_{\rho}+1/2)\frac{R_B}{a}\right].
\end{split}
\end{equation}
\label{lastpage}

\section{Plots of P and L vs MJD}
Figs B1-B42 show plots of spin period (upper panel) and luminosity (lower panel) as functions of MJD for the 42 systems in our data set. 
The weighted line of best-fit, used to determine the long-term average $\dot{P}$, is calculated using MPFITEXPR (see Section 2).
\FloatBarrier
\begin{figure}
\centering
\includegraphics[scale=0.32,angle=90]{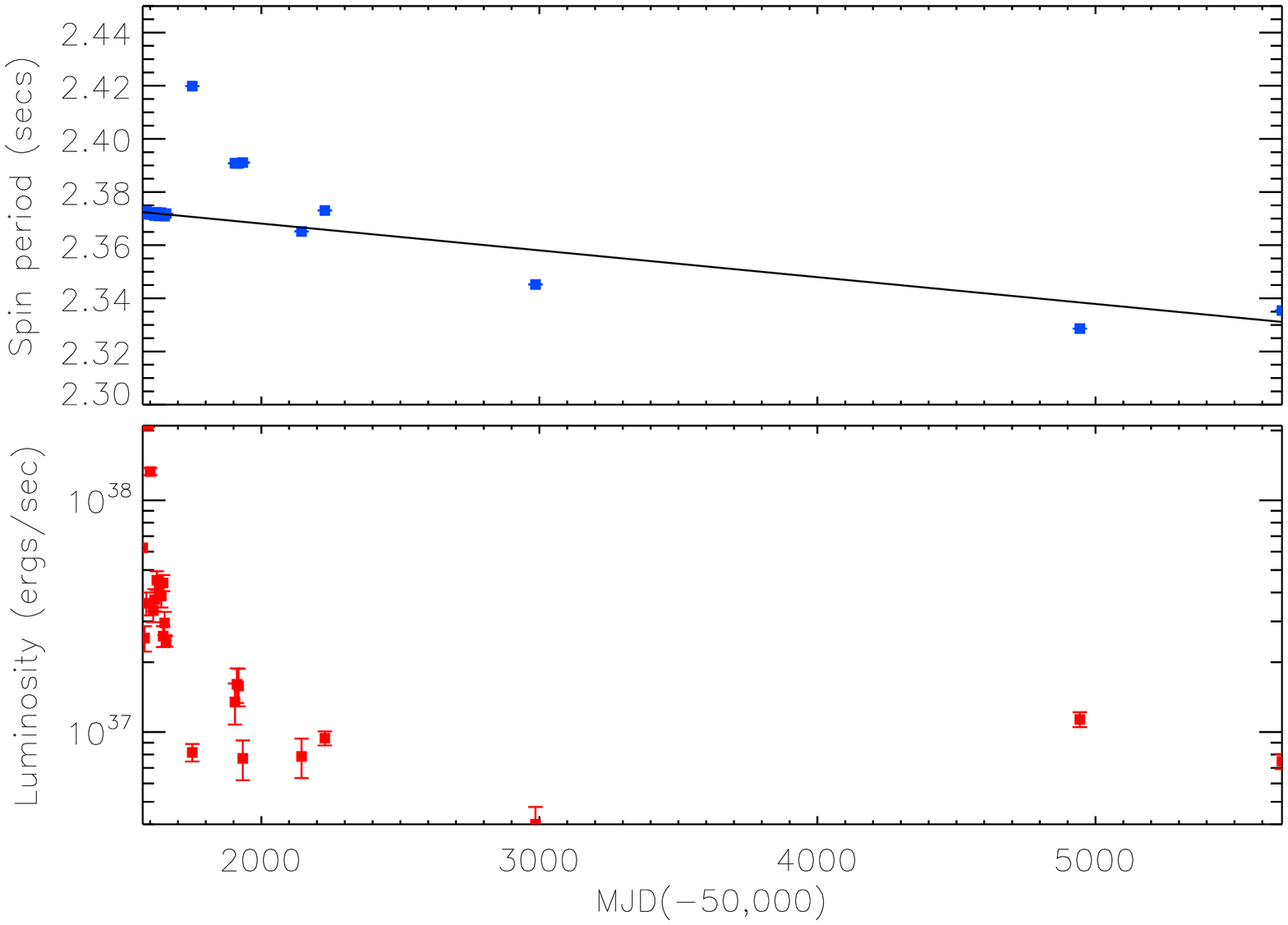}
\caption{The upper panel shows spin period as a function of MJD and the lower panel shows luminosity as a function of MJD for the source SXP2.37. The line in the upper panel shows the best-fitting $\dot{P}$.}
\end{figure}

\begin{figure} 
\centering
\includegraphics[scale=0.32,angle=90]{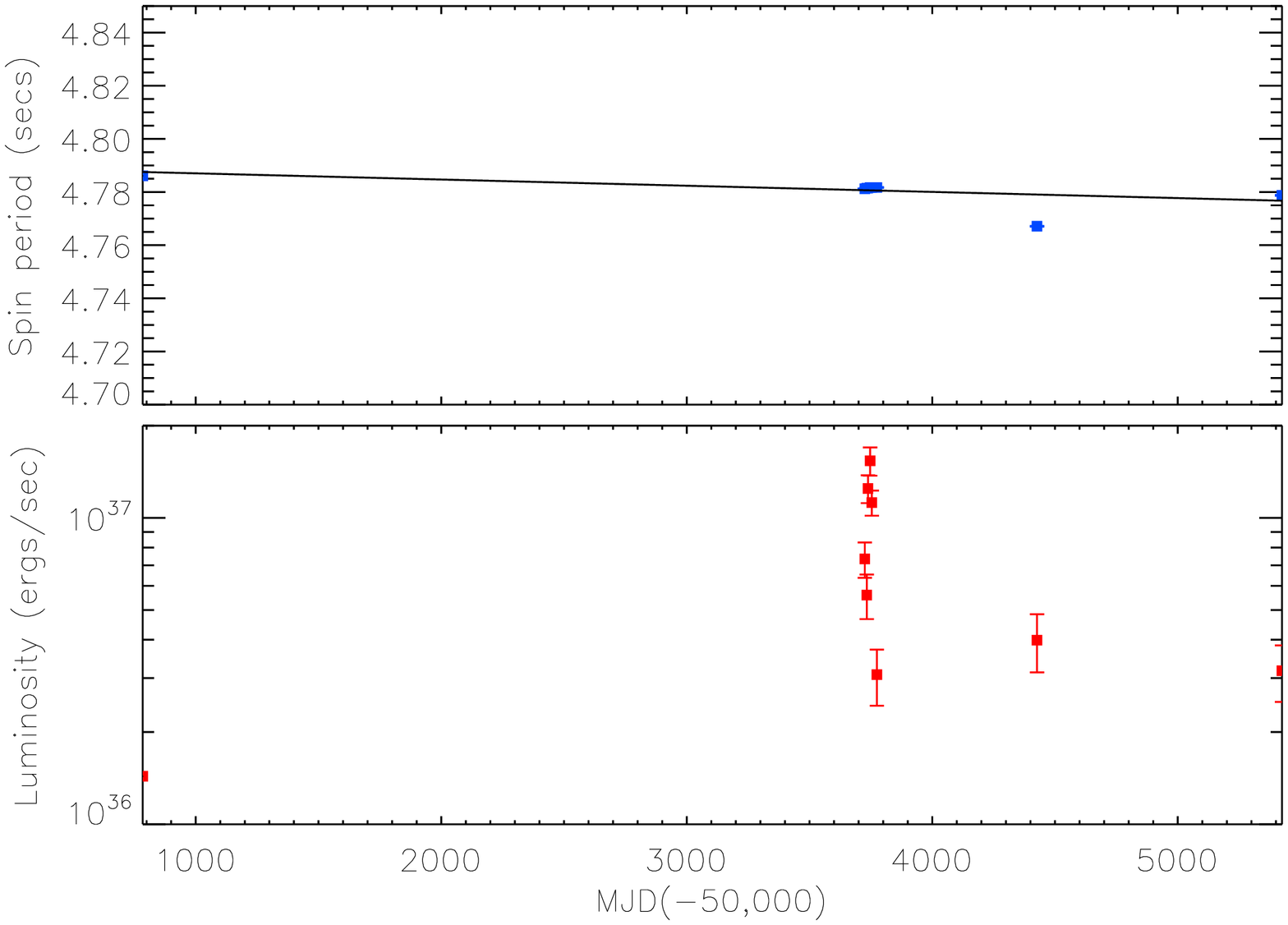}
\caption{The upper panel shows spin period as a function of MJD and the lower panel shows luminosity as a function of MJD for the source SXP4.78. The line in the upper panel shows the best-fitting $\dot{P}$.}
\end{figure}

\begin{figure} 
\centering
\includegraphics[scale=0.32,angle=90]{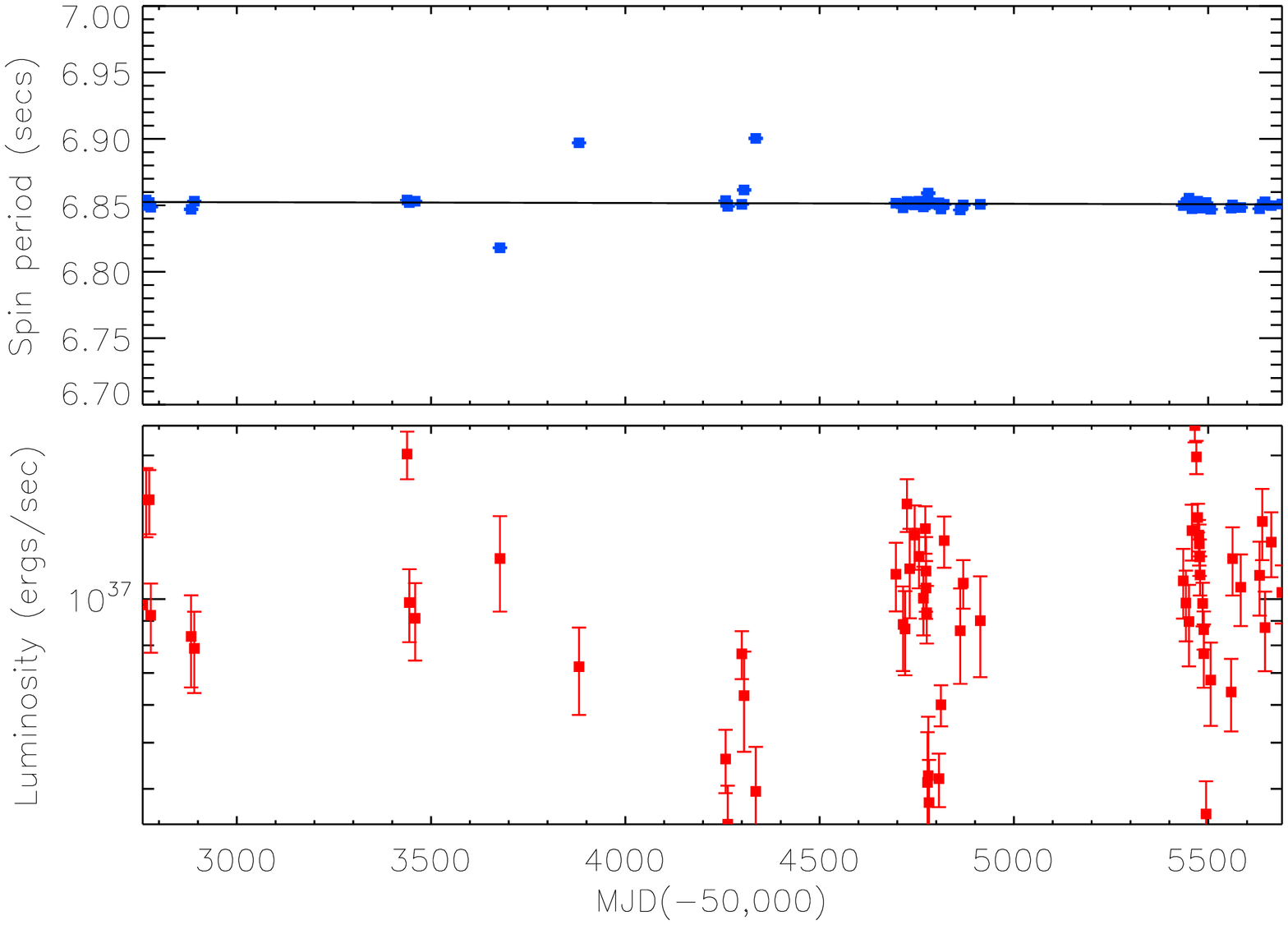}
\caption{The upper panel shows spin period as a function of MJD and the lower panel shows luminosity as a function of MJD for the source SXP6.85. The line in the upper panel shows the best-fitting $\dot{P}$.}
\end{figure}

\begin{figure} 
\centering
\includegraphics[scale=0.32,angle=90]{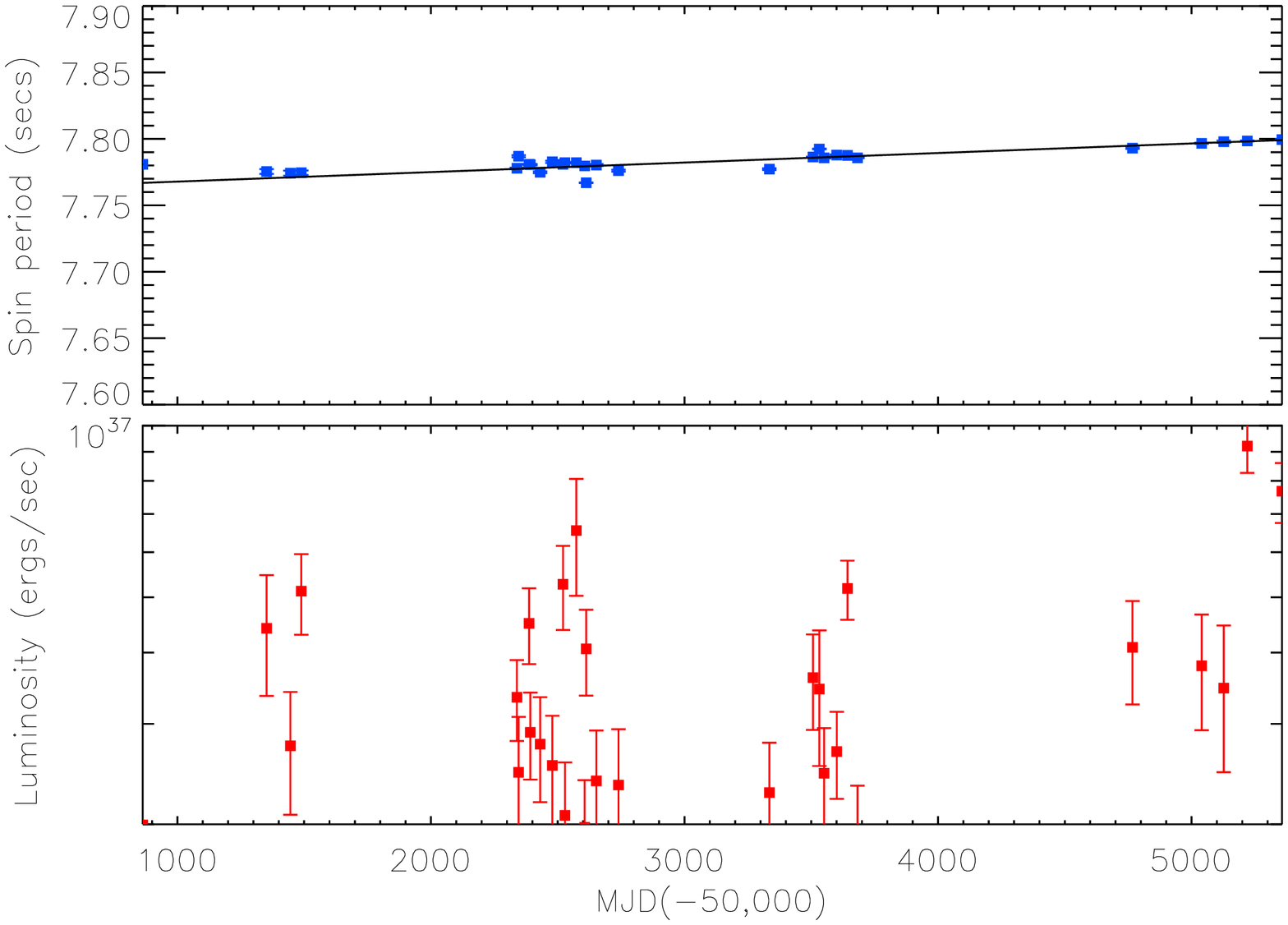}
\caption{The upper panel shows spin period as a function of MJD and the lower panel shows luminosity as a function of MJD for the source SXP7.78. The line in the upper panel shows the best-fitting $\dot{P}$.}
\end{figure}

\begin{figure} 
\centering
\includegraphics[scale=0.32,angle=90]{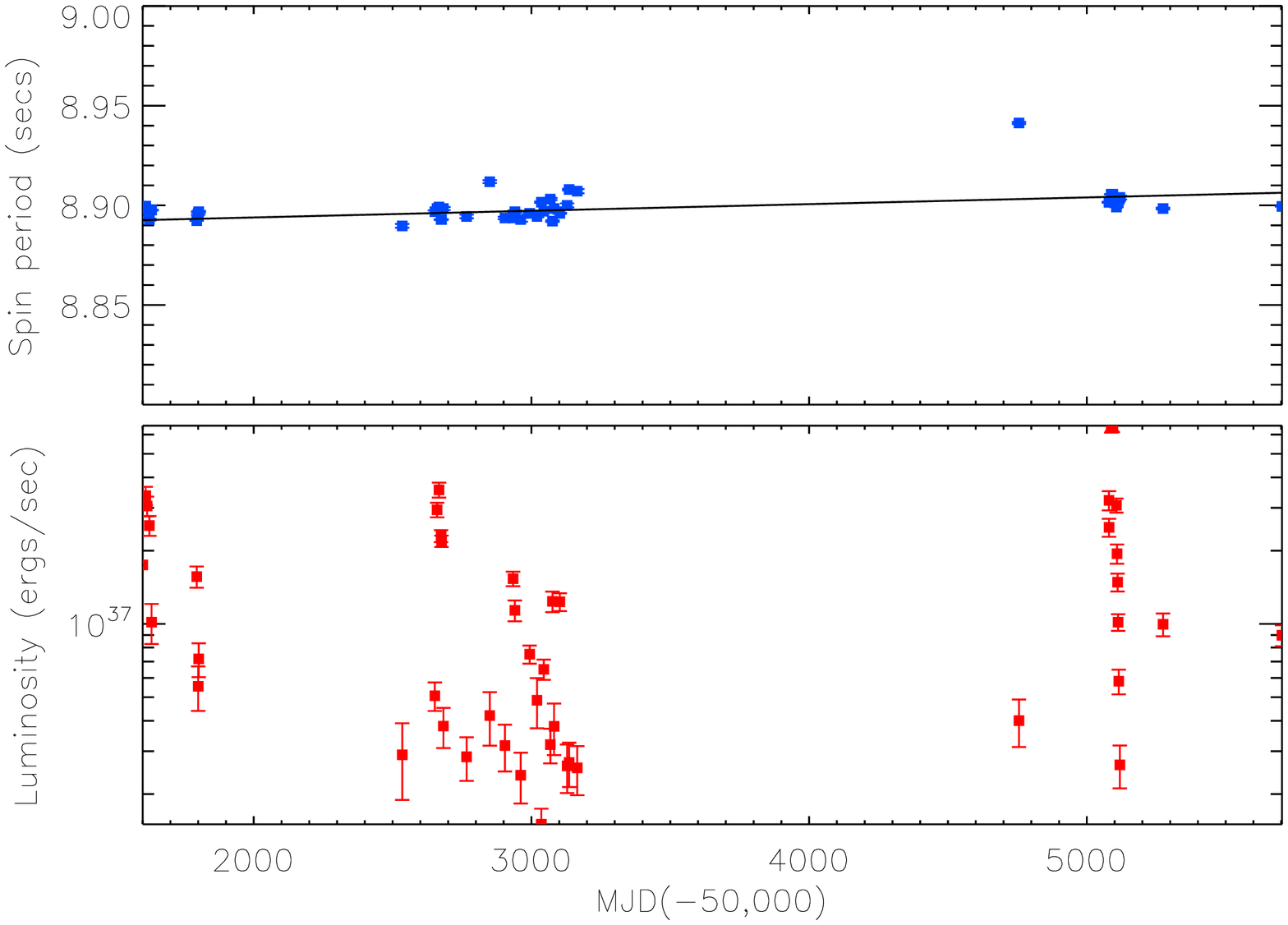}
\caption{The upper panel shows spin period as a function of MJD and the lower panel shows luminosity as a function of MJD for the source SXP8.80. The line in the upper panel shows the best-fitting $\dot{P}$.}
\end{figure}

\begin{figure} 
\centering
\includegraphics[scale=0.32,angle=90]{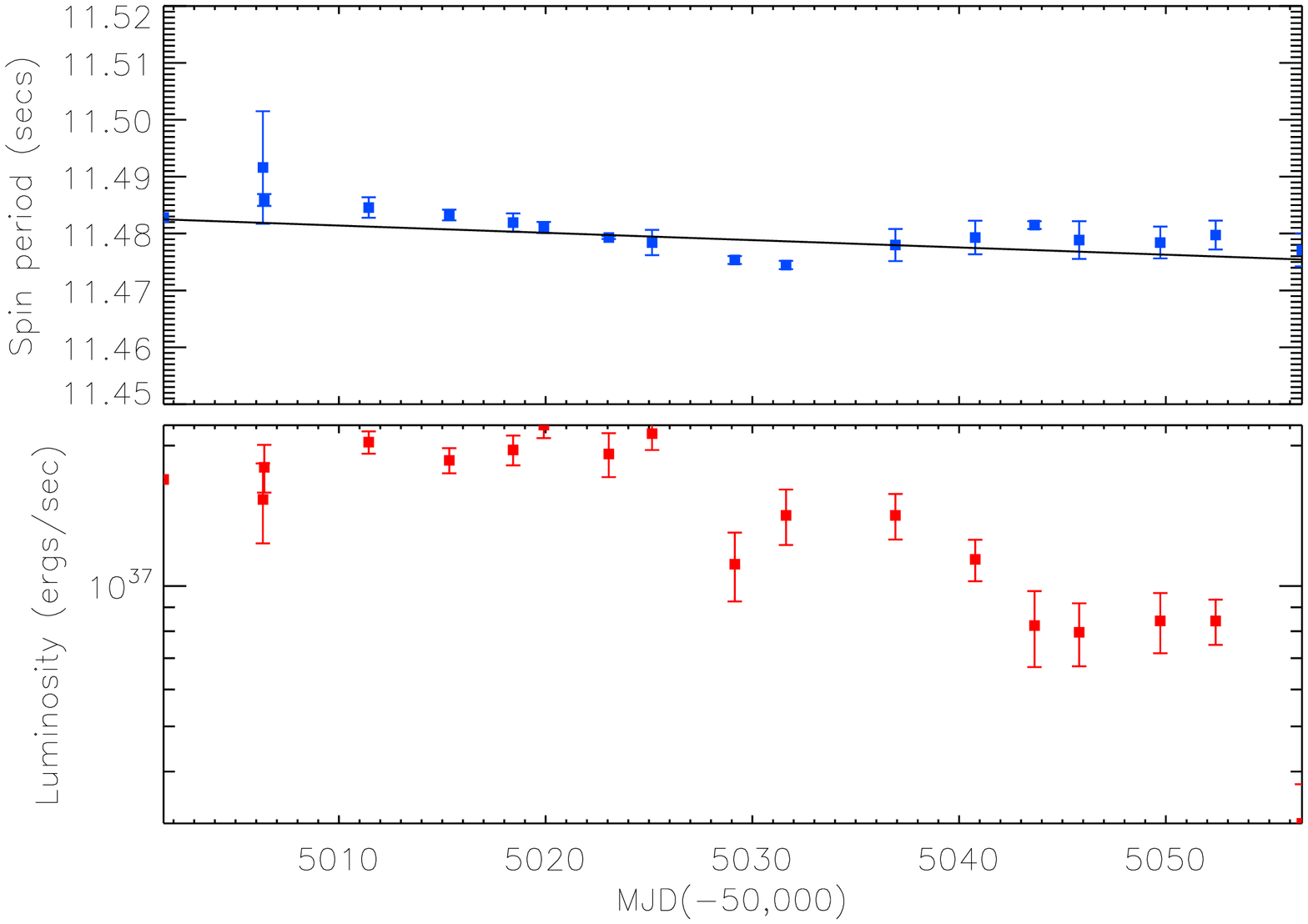}
\caption{The upper panel shows spin period as a function of MJD and the lower panel shows luminosity as a function of MJD for the source SXP11.5. The line in the upper panel shows the best-fitting $\dot{P}$.}
\end{figure}
\FloatBarrier
\begin{figure}
\centering
\includegraphics[scale=0.32,angle=90]{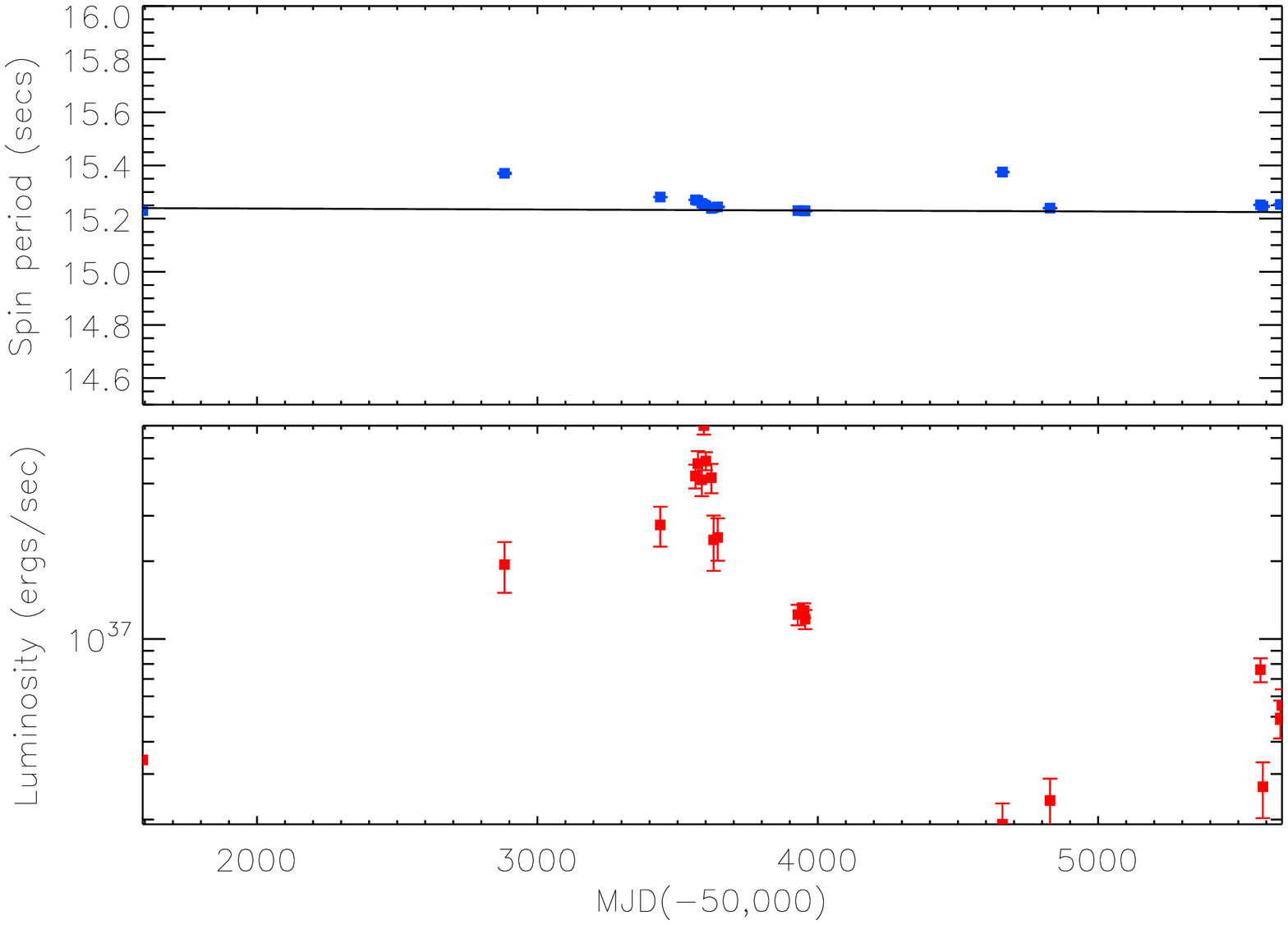}
\caption{The upper panel shows spin period as a function of MJD and the lower panel shows luminosity as a function of MJD for the source SXP15.3. The line in the upper panel shows the best-fitting $\dot{P}$.}
\end{figure}

\begin{figure} 
\centering
\includegraphics[scale=0.32,angle=90]{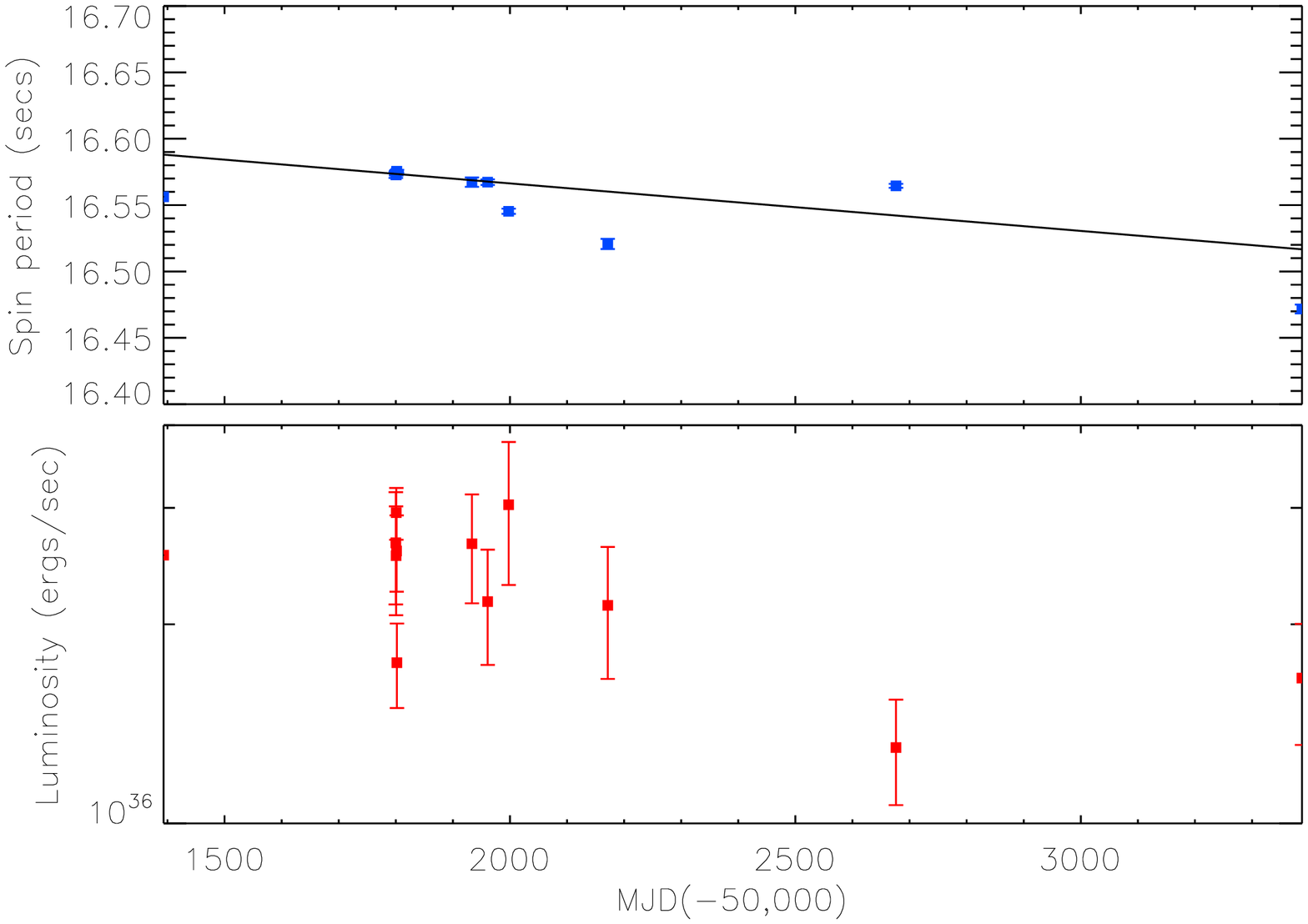}
\caption{The upper panel shows spin period as a function of MJD and the lower panel shows luminosity as a function of MJD for the source SXP16.6. The line in the upper panel shows the best-fitting $\dot{P}$.}
\end{figure}

\begin{figure} 
\centering
\includegraphics[scale=0.32,angle=90]{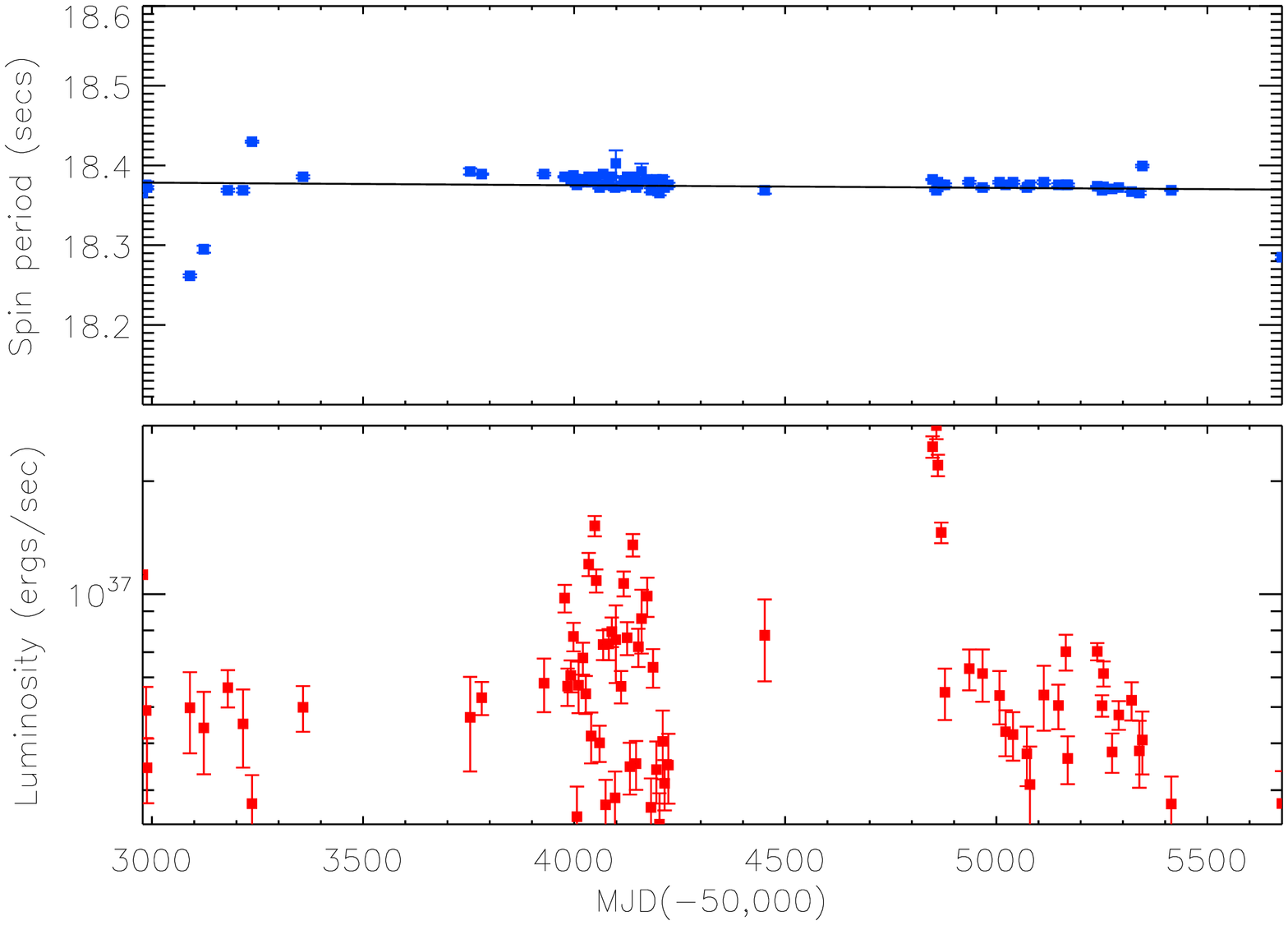}
\caption{The upper panel shows spin period as a function of MJD and the lower panel shows luminosity as a function of MJD for the source SXP18.3. The line in the upper panel shows the best-fitting $\dot{P}$.}
\end{figure}

\begin{figure} 
\centering
\includegraphics[scale=0.32,angle=90]{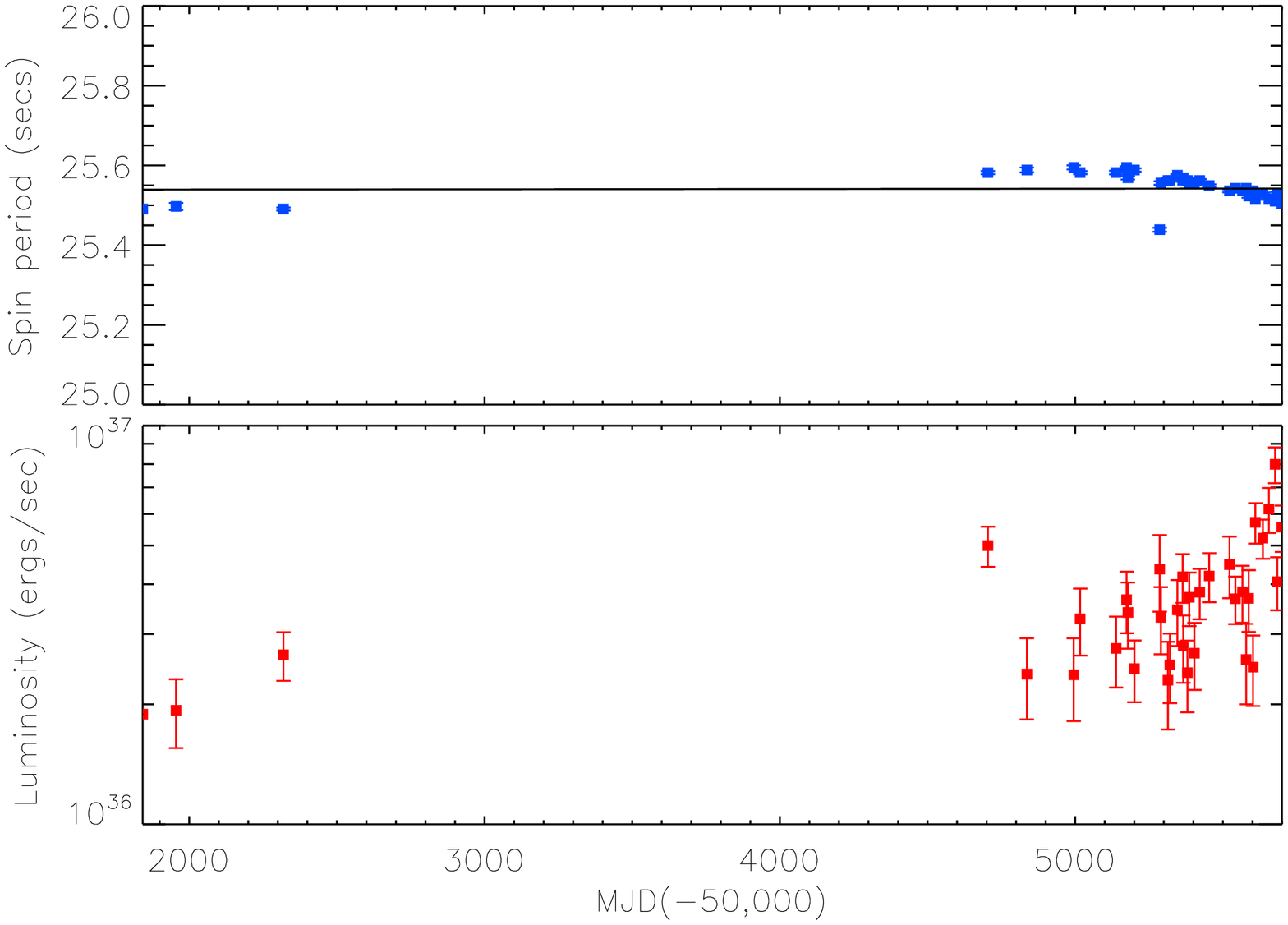}
\caption{The upper panel shows spin period as a function of MJD and the lower panel shows luminosity as a function of MJD for the source SXP25.5. The line in the upper panel shows the best-fitting $\dot{P}$.}
\end{figure}

\begin{figure} 
\centering
\includegraphics[scale=0.32,angle=90]{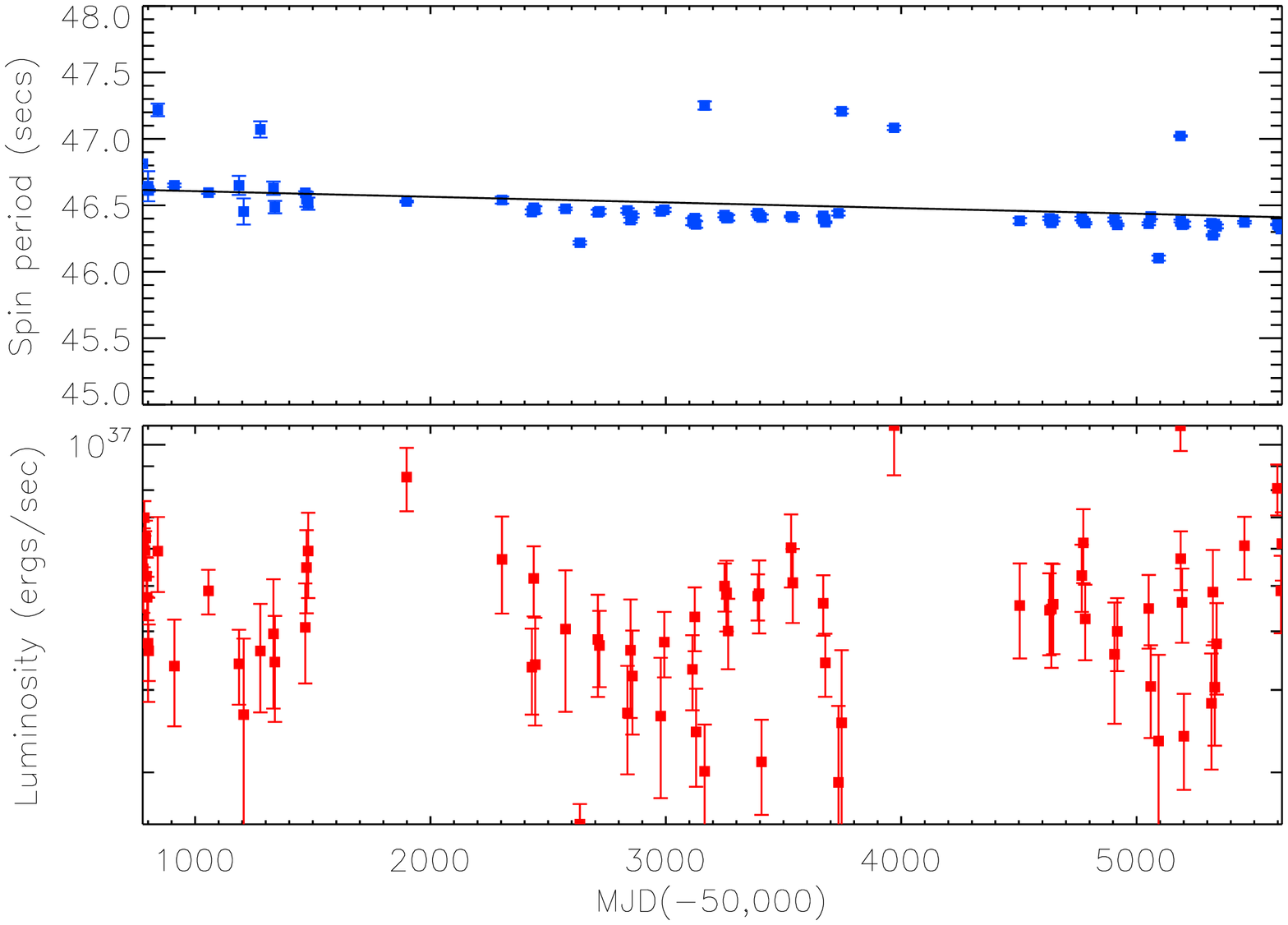}
\caption{The upper panel shows spin period as a function of MJD and the lower panel shows luminosity as a function of MJD for the source SXP46.6. The line in the upper panel shows the best-fitting $\dot{P}$.}
\end{figure}

\begin{figure} 
\centering
\includegraphics[scale=0.32,angle=90]{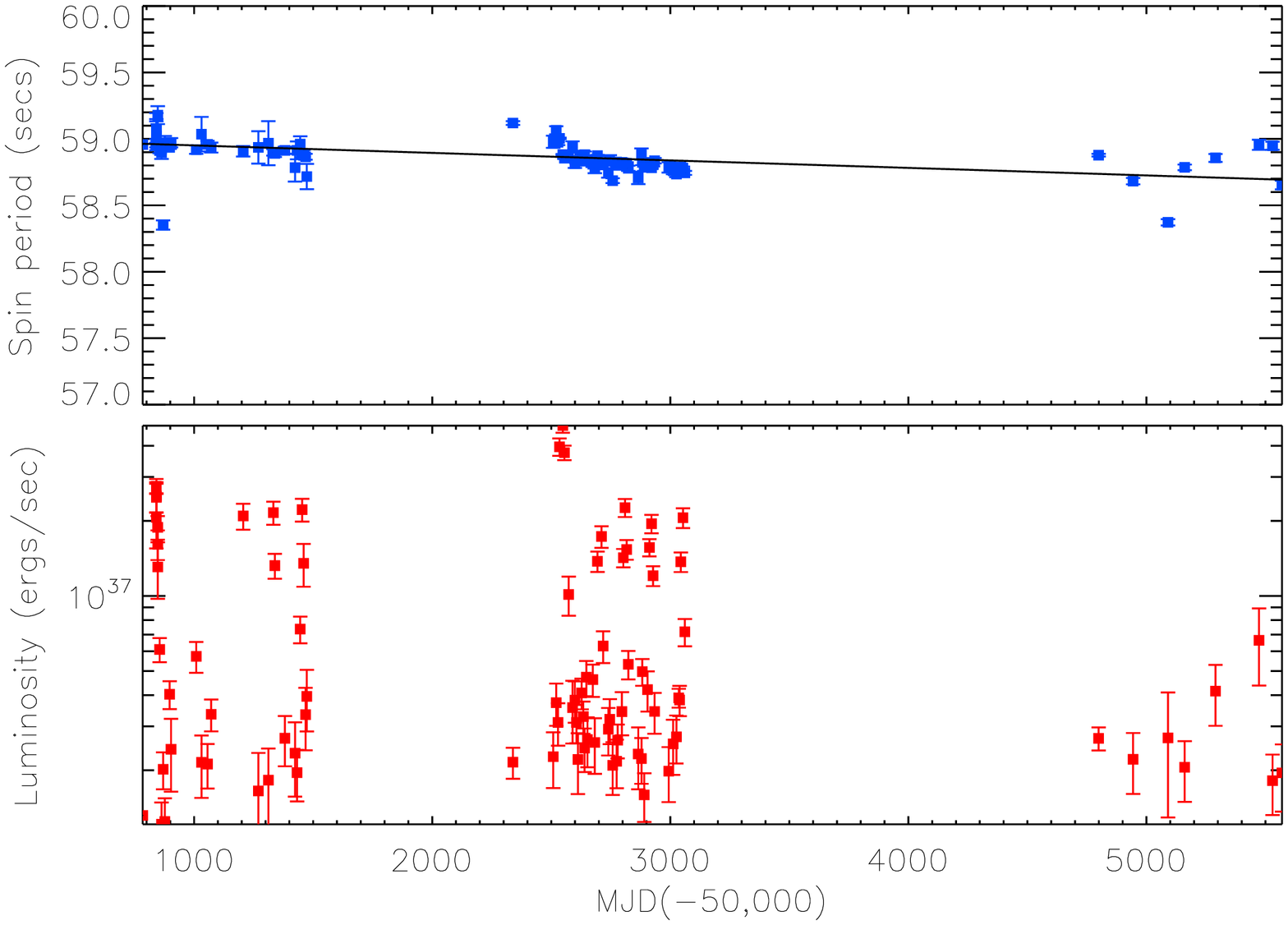}
\caption{The upper panel shows spin period as a function of MJD and the lower panel shows luminosity as a function of MJD for the source SXP59.0. The line in the upper panel shows the best-fitting $\dot{P}$.}
\end{figure}
\FloatBarrier
\begin{figure} 
\centering
\includegraphics[scale=0.32,angle=90]{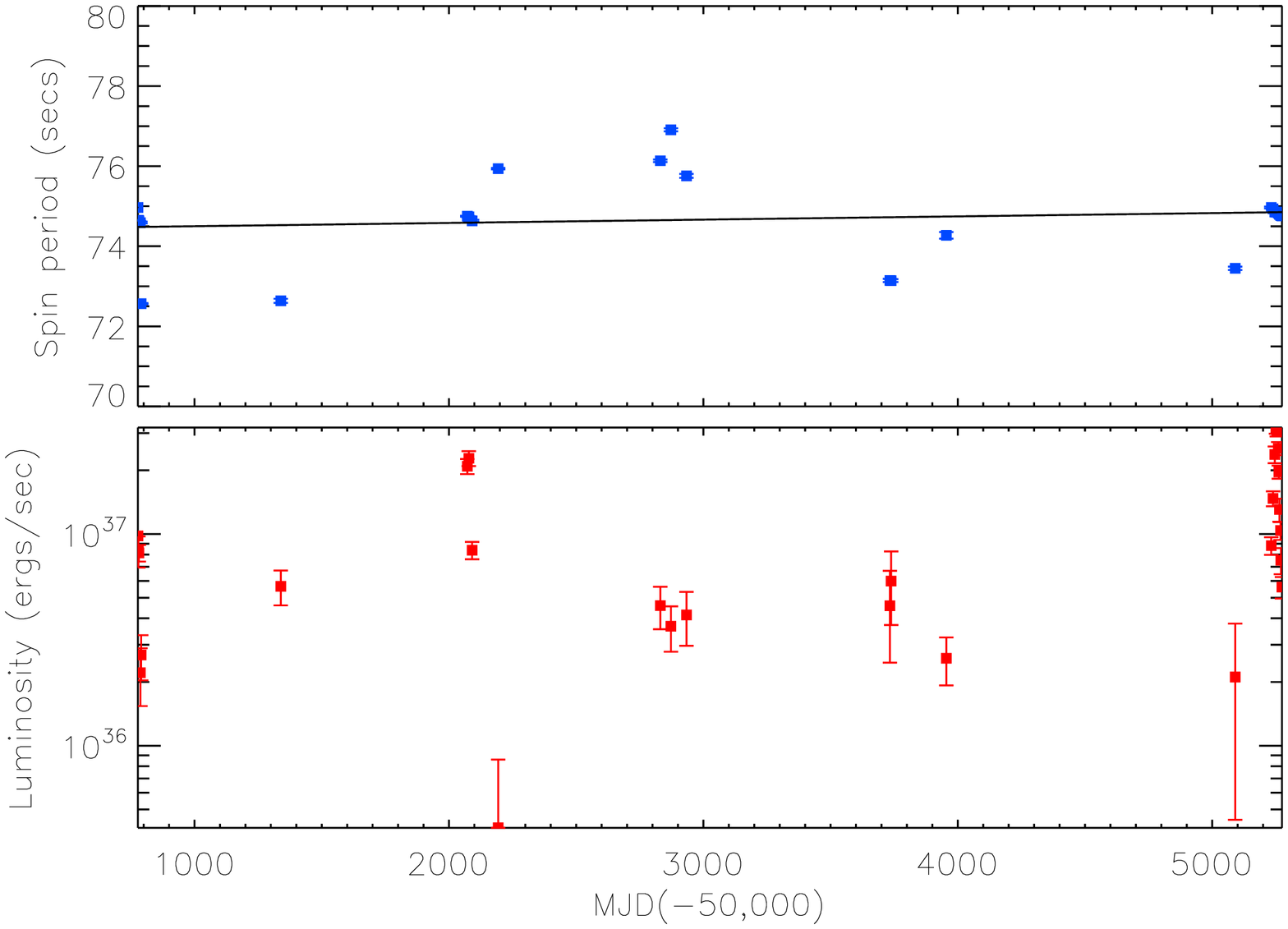}
\caption{The upper panel shows spin period as a function of MJD and the lower panel shows luminosity as a function of MJD for the source SXP74.7. The line in the upper panel shows the best-fitting $\dot{P}$.}
\end{figure}

\begin{figure} 
\centering
\includegraphics[scale=0.32,angle=90]{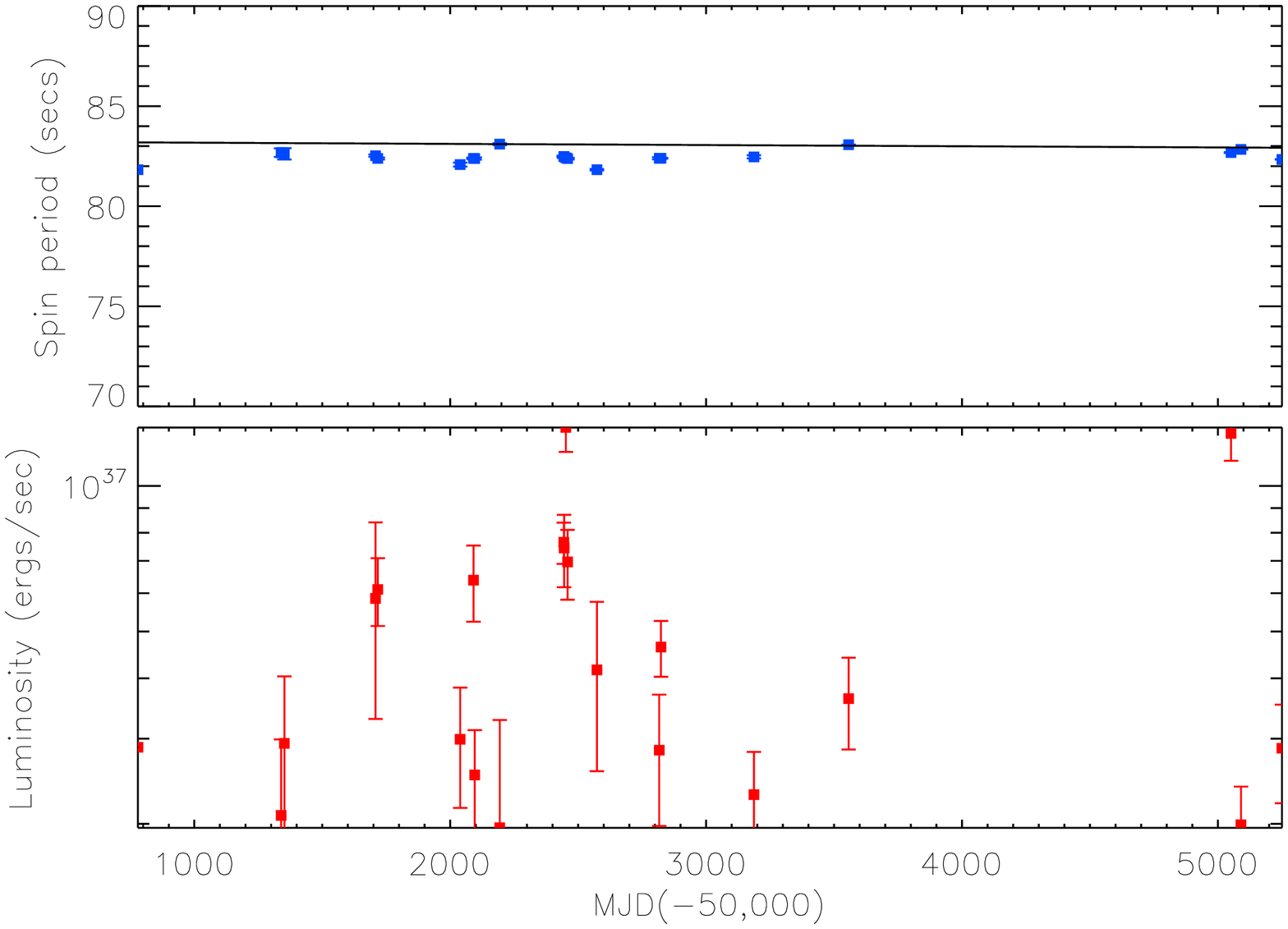}
\caption{The upper panel shows spin period as a function of MJD and the lower panel shows luminosity as a function of MJD for the source SXP82.4. The line in the upper panel shows the best-fitting $\dot{P}$.}
\end{figure}

\begin{figure} 
\centering
\includegraphics[scale=0.32,angle=90]{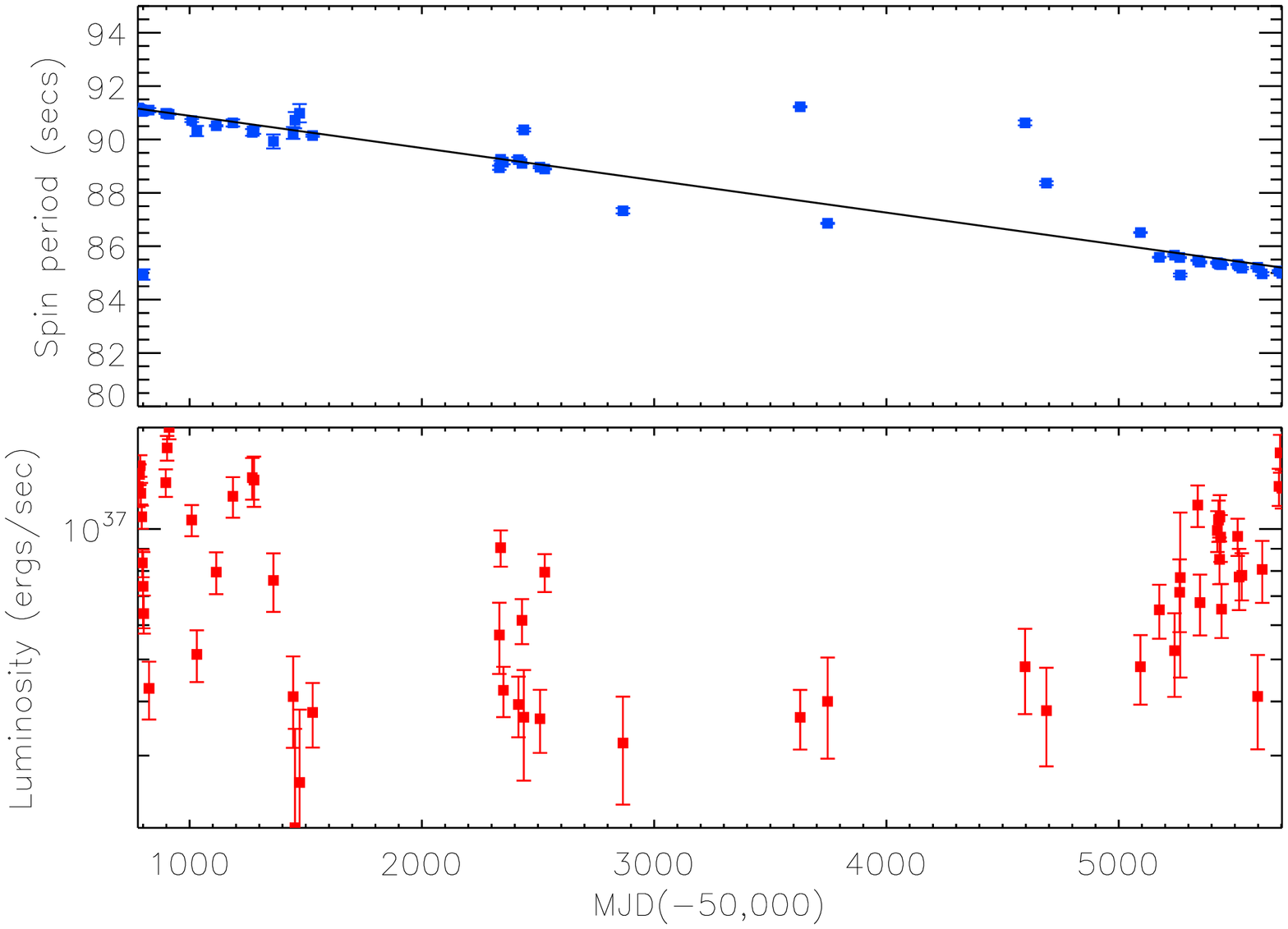}
\caption{The upper panel shows spin period as a function of MJD and the lower panel shows luminosity as a function of MJD for the source SXP91.1. The line in the upper panel shows the best-fitting $\dot{P}$.}
\end{figure}

\begin{figure} 
\centering
\includegraphics[scale=0.32,angle=90]{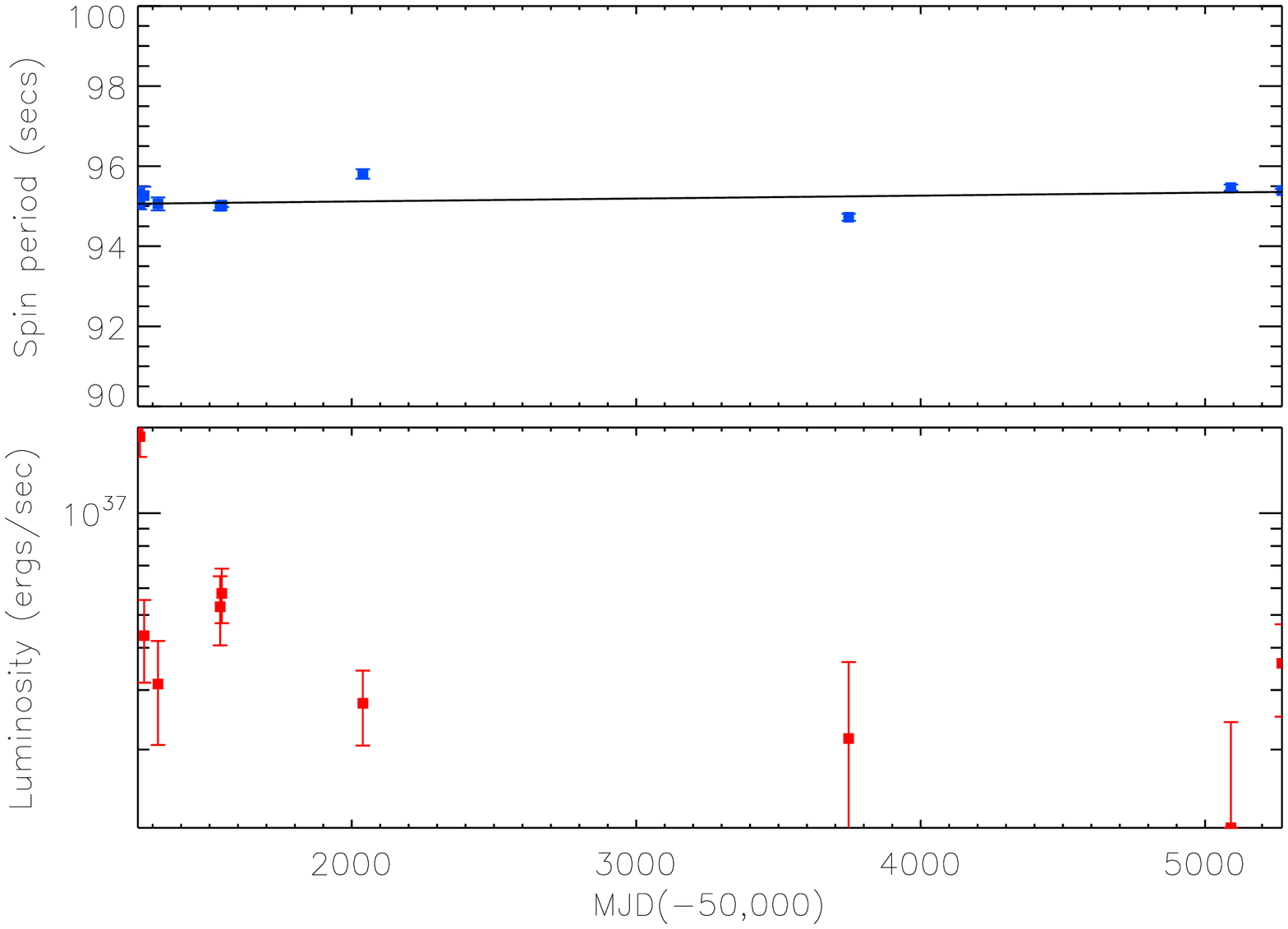}
\caption{The upper panel shows spin period as a function of MJD and the lower panel shows luminosity as a function of MJD for the source SXP95.2. The line in the upper panel shows the best-fitting $\dot{P}$.}
\end{figure}

\begin{figure} 
\centering
\includegraphics[scale=0.32,angle=90]{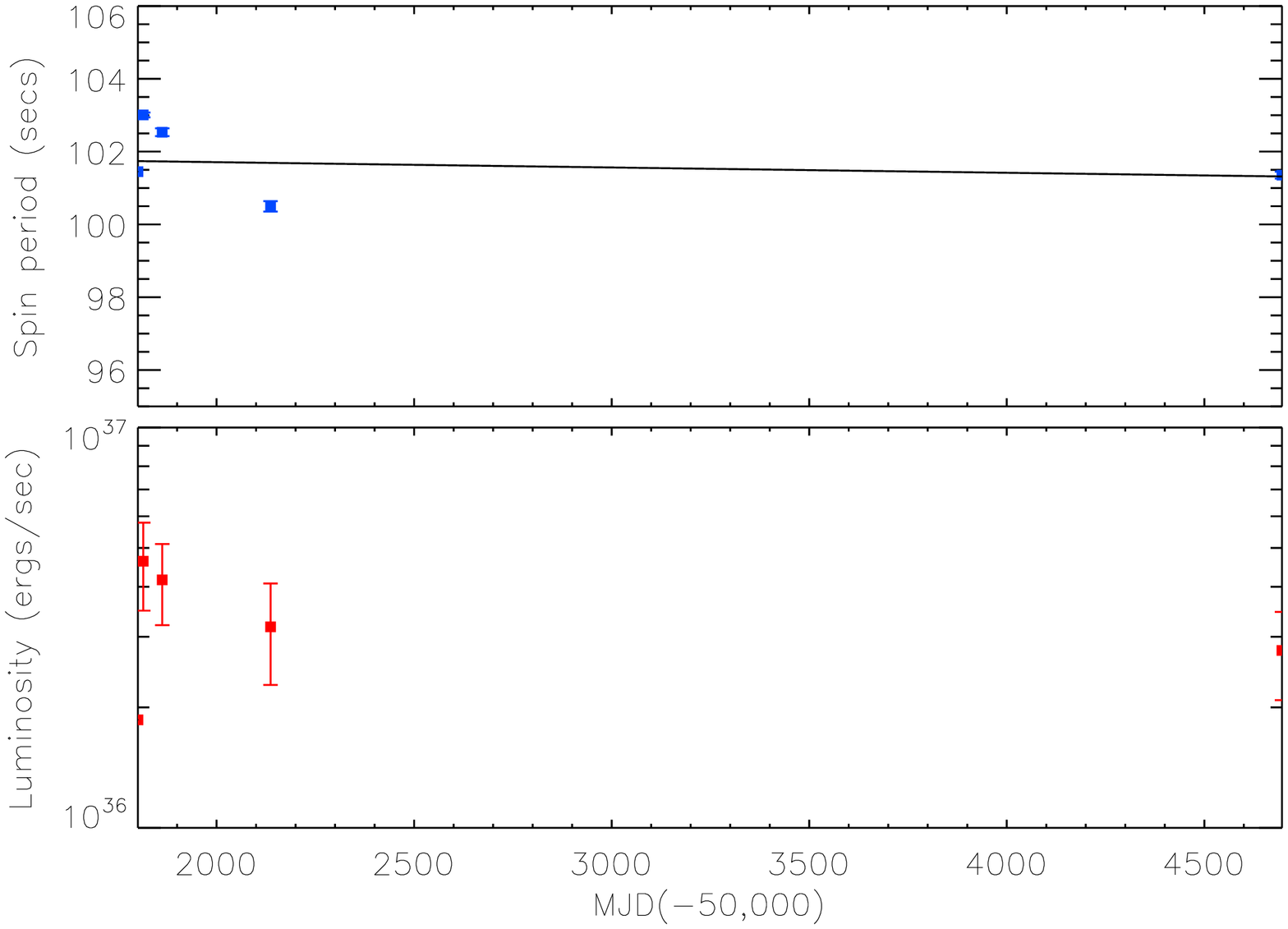}
\caption{The upper panel shows spin period as a function of MJD and the lower panel shows luminosity as a function of MJD for the source SXP101. The line in the upper panel shows the best-fitting $\dot{P}$.}
\end{figure}

\begin{figure} 
\centering
\includegraphics[scale=0.32,angle=90]{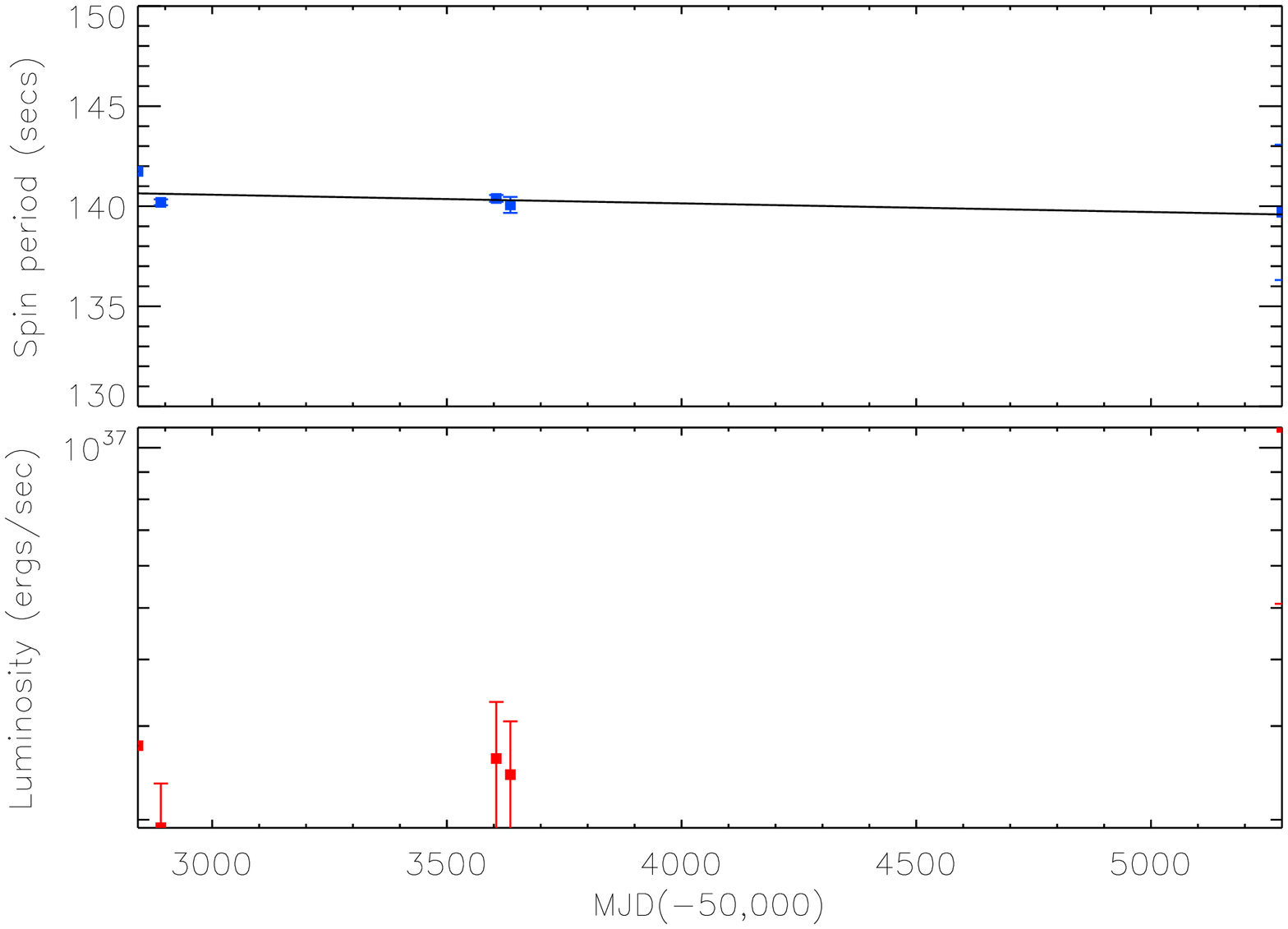}
\caption{The upper panel shows spin period as a function of MJD and the lower panel shows luminosity as a function of MJD for the source SXP140. The line in the upper panel shows the best-fitting $\dot{P}$.}
\end{figure}
\FloatBarrier
\begin{figure} 
\centering
\includegraphics[scale=0.32,angle=90]{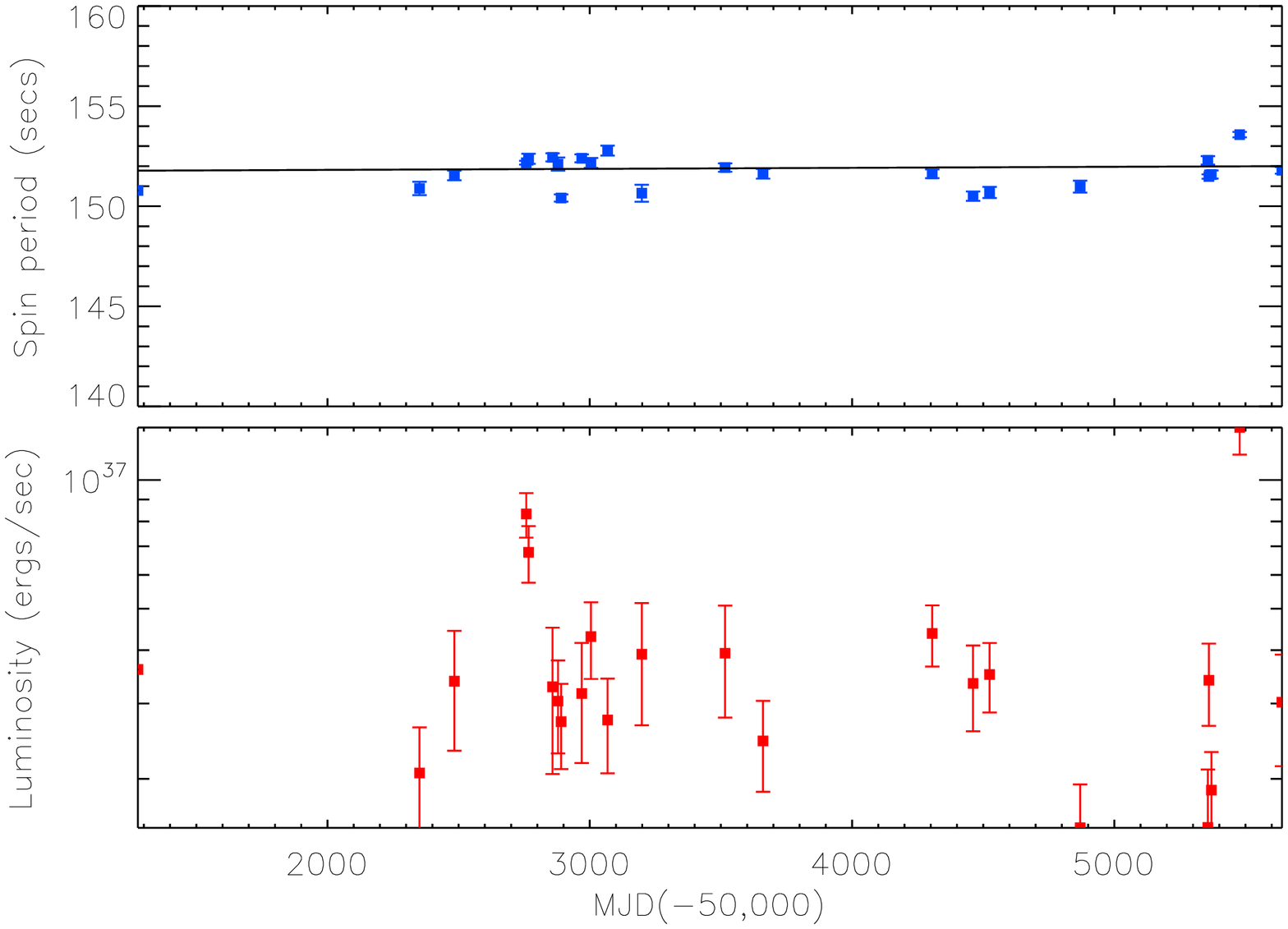}
\caption{The upper panel shows spin period as a function of MJD and the lower panel shows luminosity as a function of MJD for the source SXP152. The line in the upper panel shows the best-fitting $\dot{P}$.}
\end{figure}

\begin{figure} 
\centering
\includegraphics[scale=0.32,angle=90]{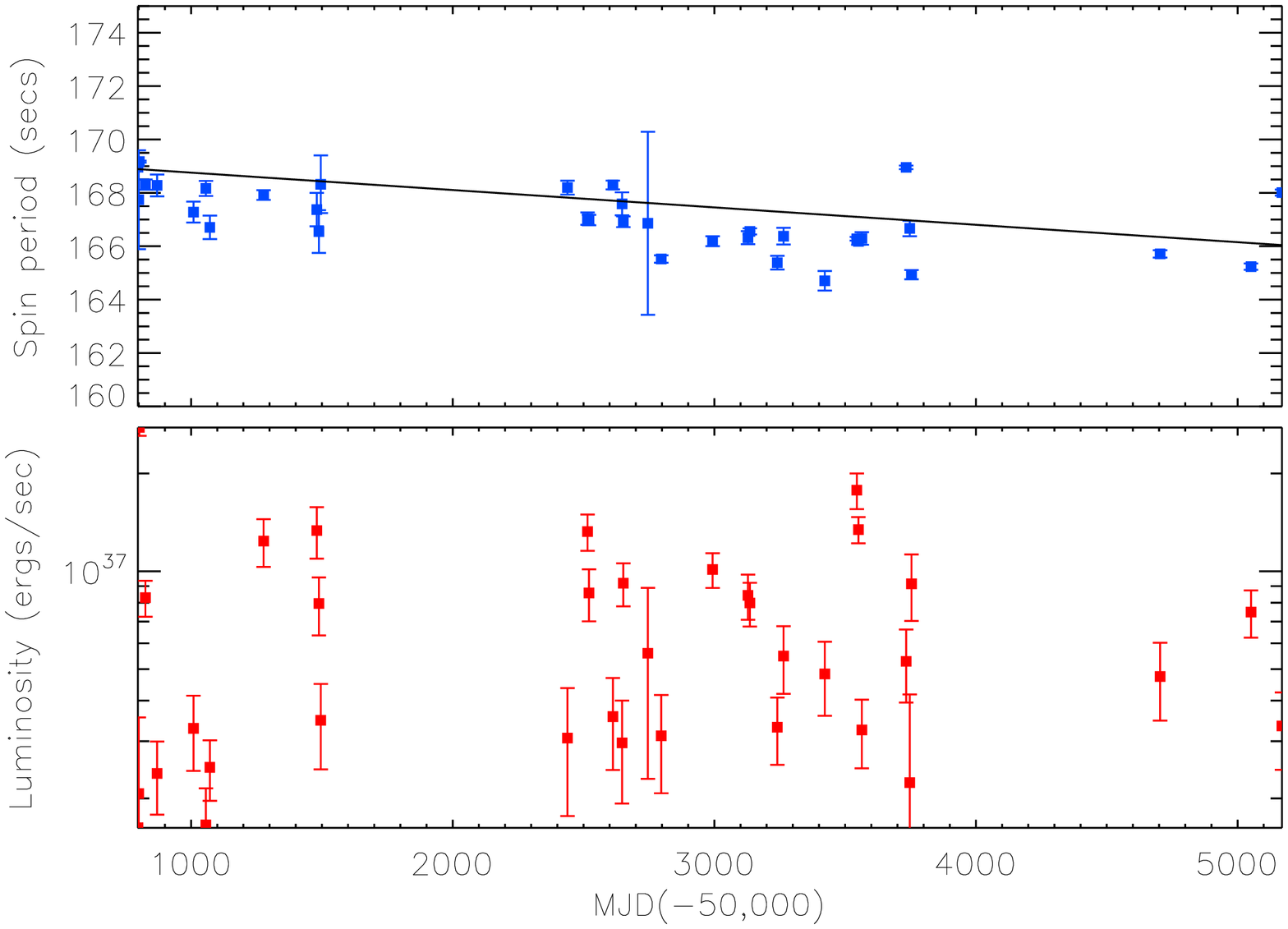}
\caption{The upper panel shows spin period as a function of MJD and the lower panel shows luminosity as a function of MJD for the source SXP169. The line in the upper panel shows the best-fitting $\dot{P}$.}
\end{figure}

\begin{figure} 
\centering
\includegraphics[scale=0.32,angle=90]{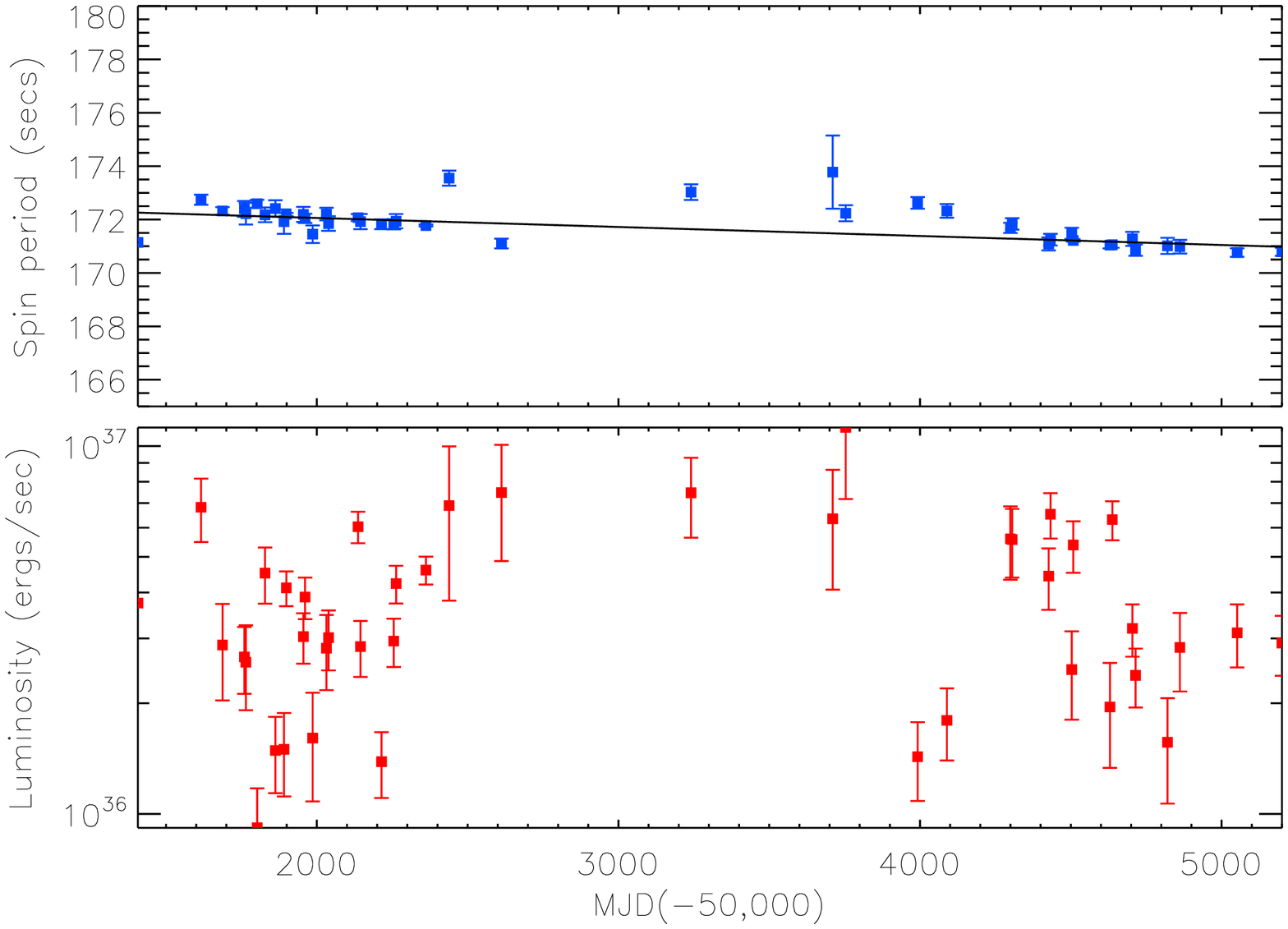}
\caption{The upper panel shows spin period as a function of MJD and the lower panel shows luminosity as a function of MJD for the source SXP172. The line in the upper panel shows the best-fitting $\dot{P}$.}
\end{figure}

\begin{figure} 
\centering
\includegraphics[scale=0.32,angle=90]{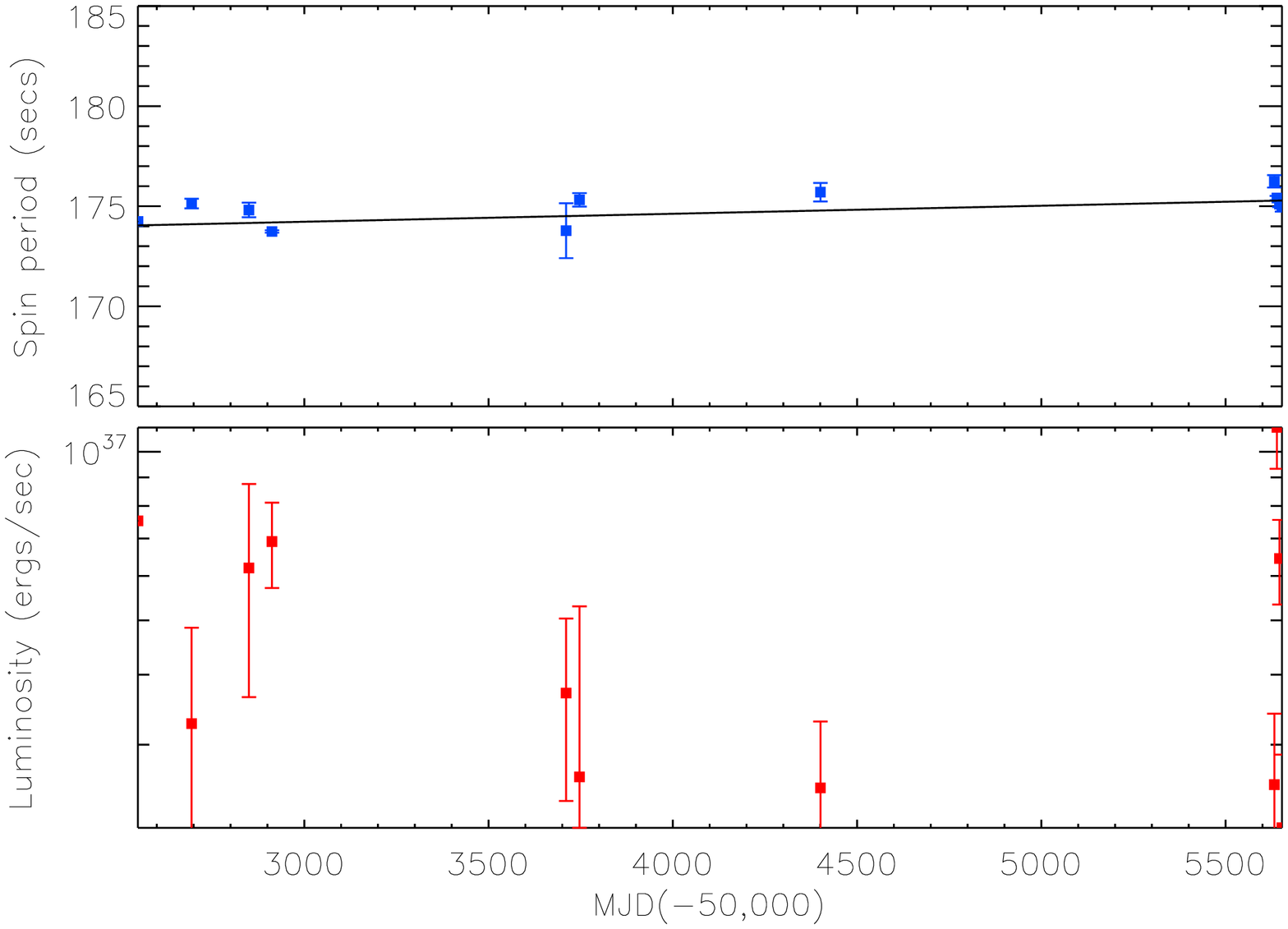}
\caption{The upper panel shows spin period as a function of MJD and the lower panel shows luminosity as a function of MJD for the source SXP175. The line in the upper panel shows the best-fitting $\dot{P}$.}
\end{figure}

\begin{figure} 
\centering
\includegraphics[scale=0.32,angle=90]{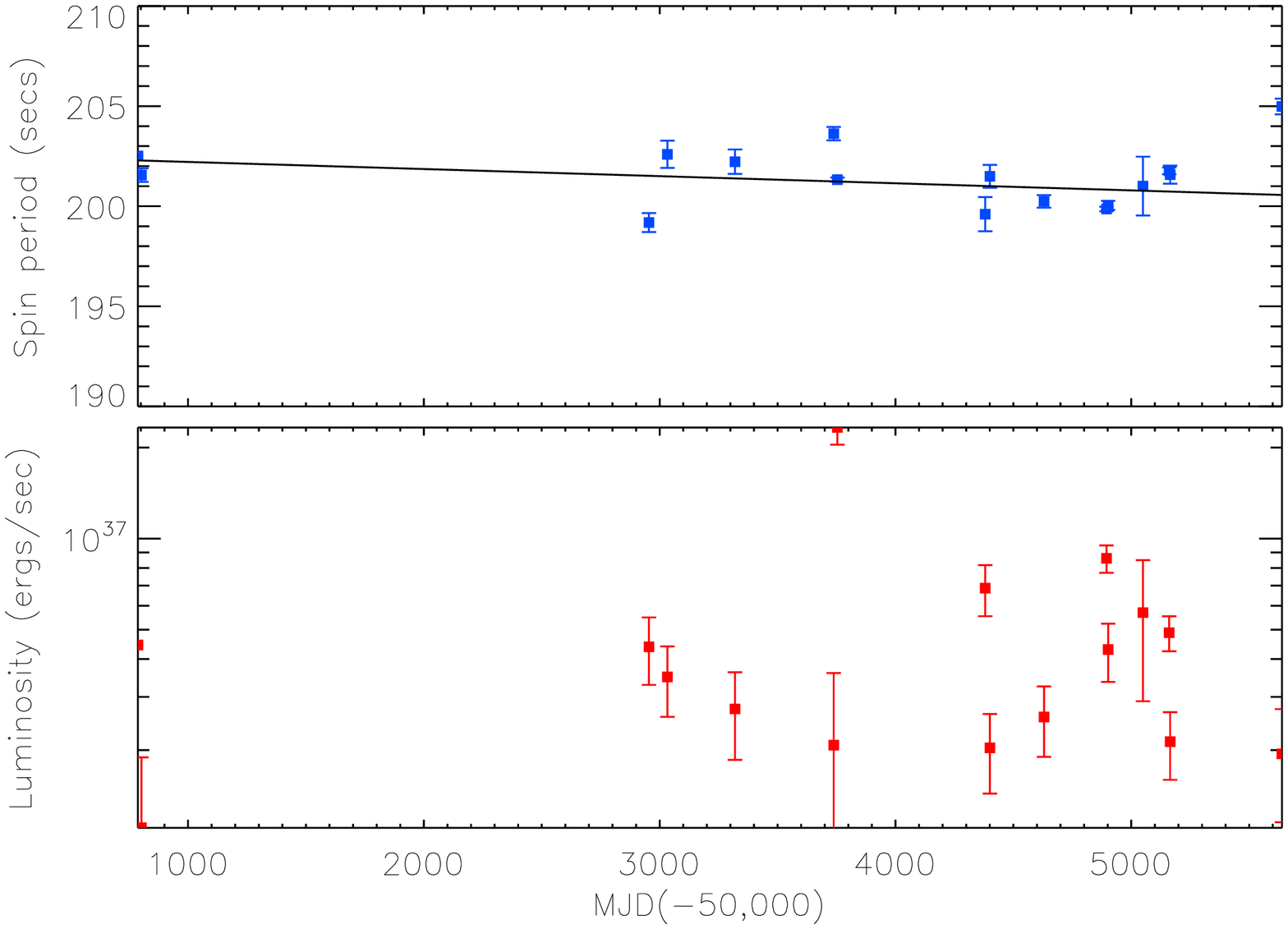}
\caption{The upper panel shows spin period as a function of MJD and the lower panel shows luminosity as a function of MJD for the source SXP202A. The line in the upper panel shows the best-fitting $\dot{P}$.}
\end{figure}

\begin{figure} 
\centering
\includegraphics[scale=0.32,angle=90]{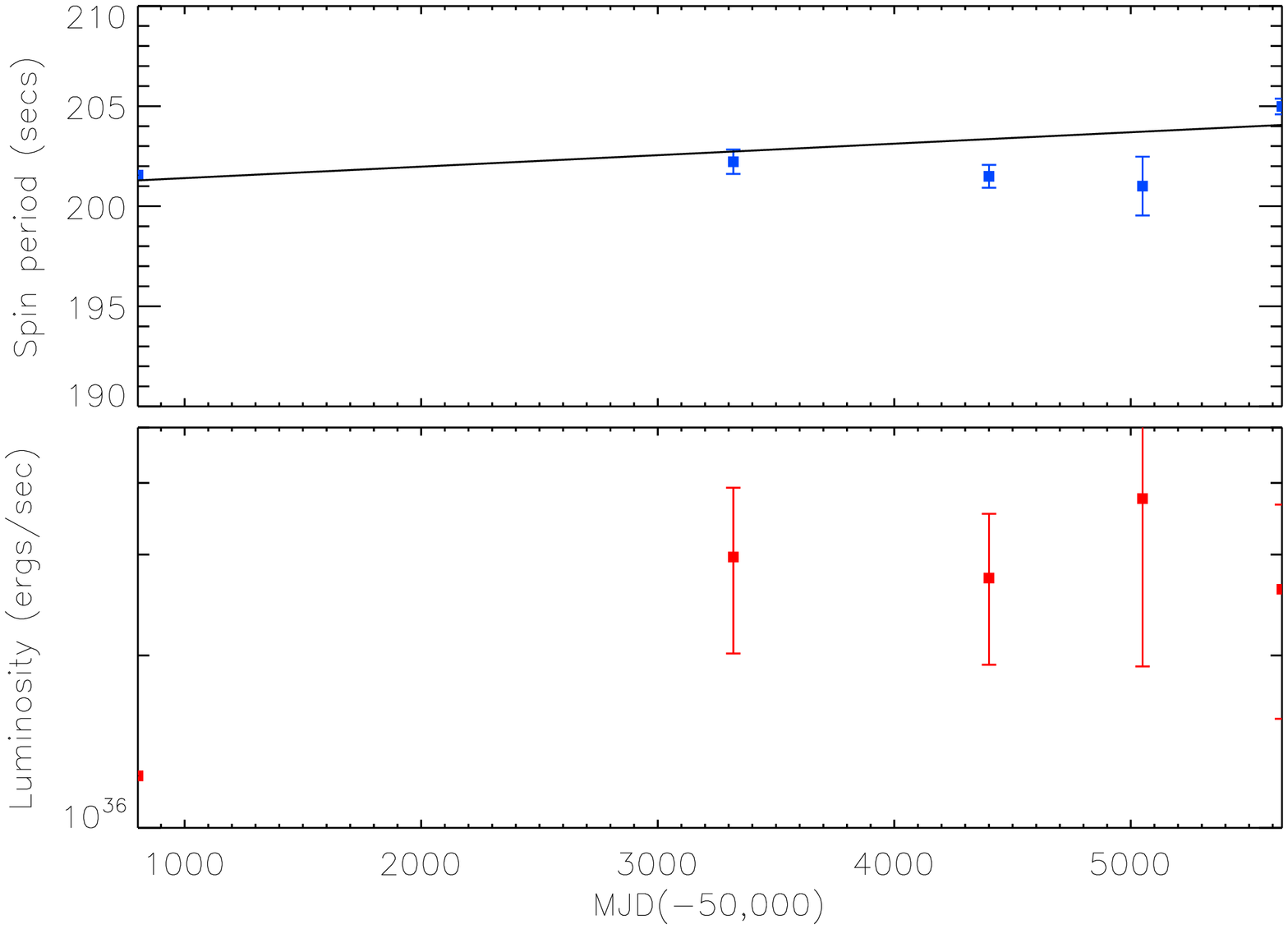}
\caption{The upper panel shows spin period as a function of MJD and the lower panel shows luminosity as a function of MJD for the source SXP202B. The line in the upper panel shows the best-fitting $\dot{P}$.}
\end{figure}
\FloatBarrier
\begin{figure} 
\centering
\includegraphics[scale=0.32,angle=90]{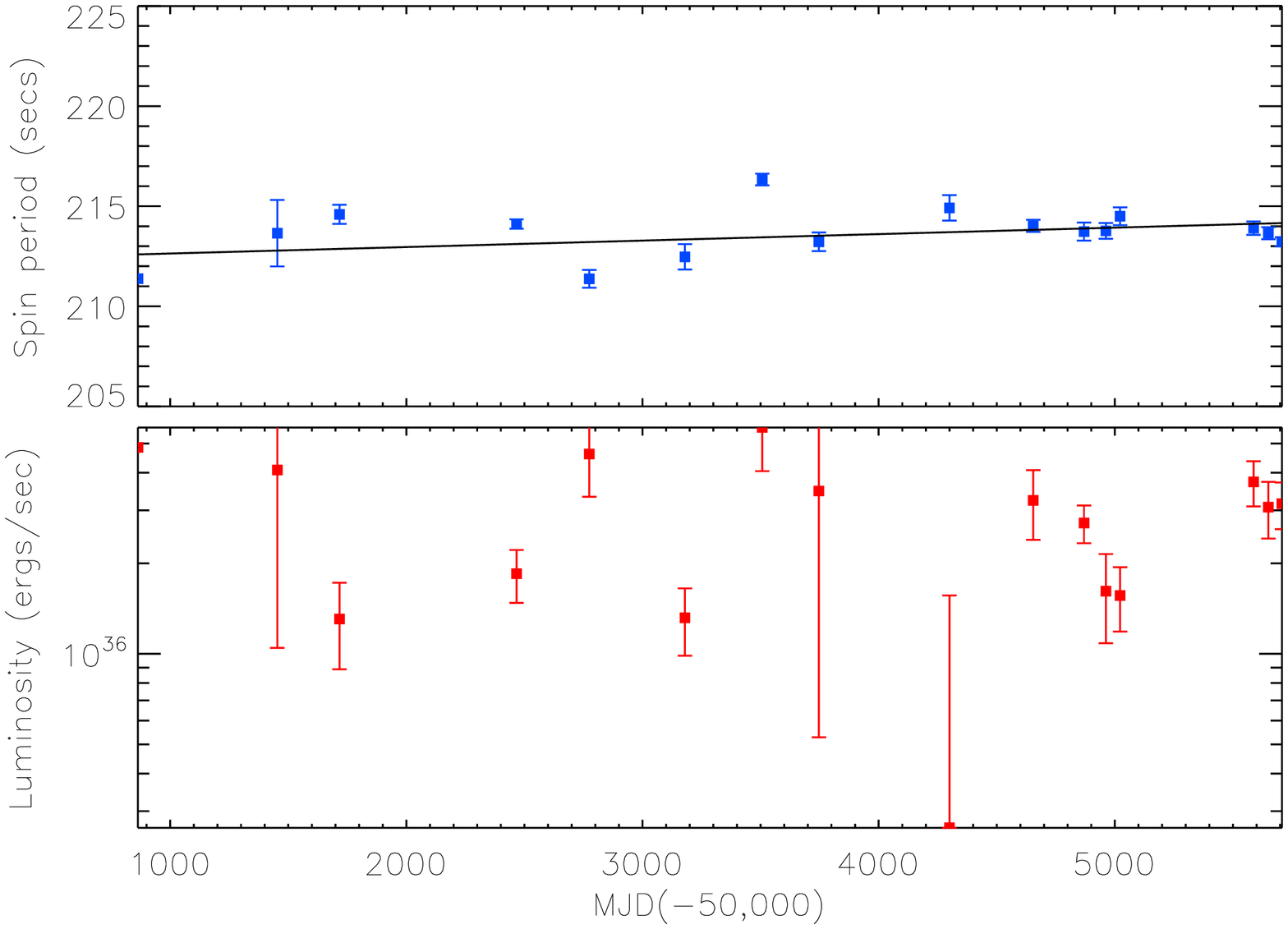}
\caption{The upper panel shows spin period as a function of MJD and the lower panel shows luminosity as a function of MJD for the source SXP214. The line in the upper panel shows the best-fitting $\dot{P}$.}
\end{figure}

\begin{figure} 
\centering
\includegraphics[scale=0.32,angle=90]{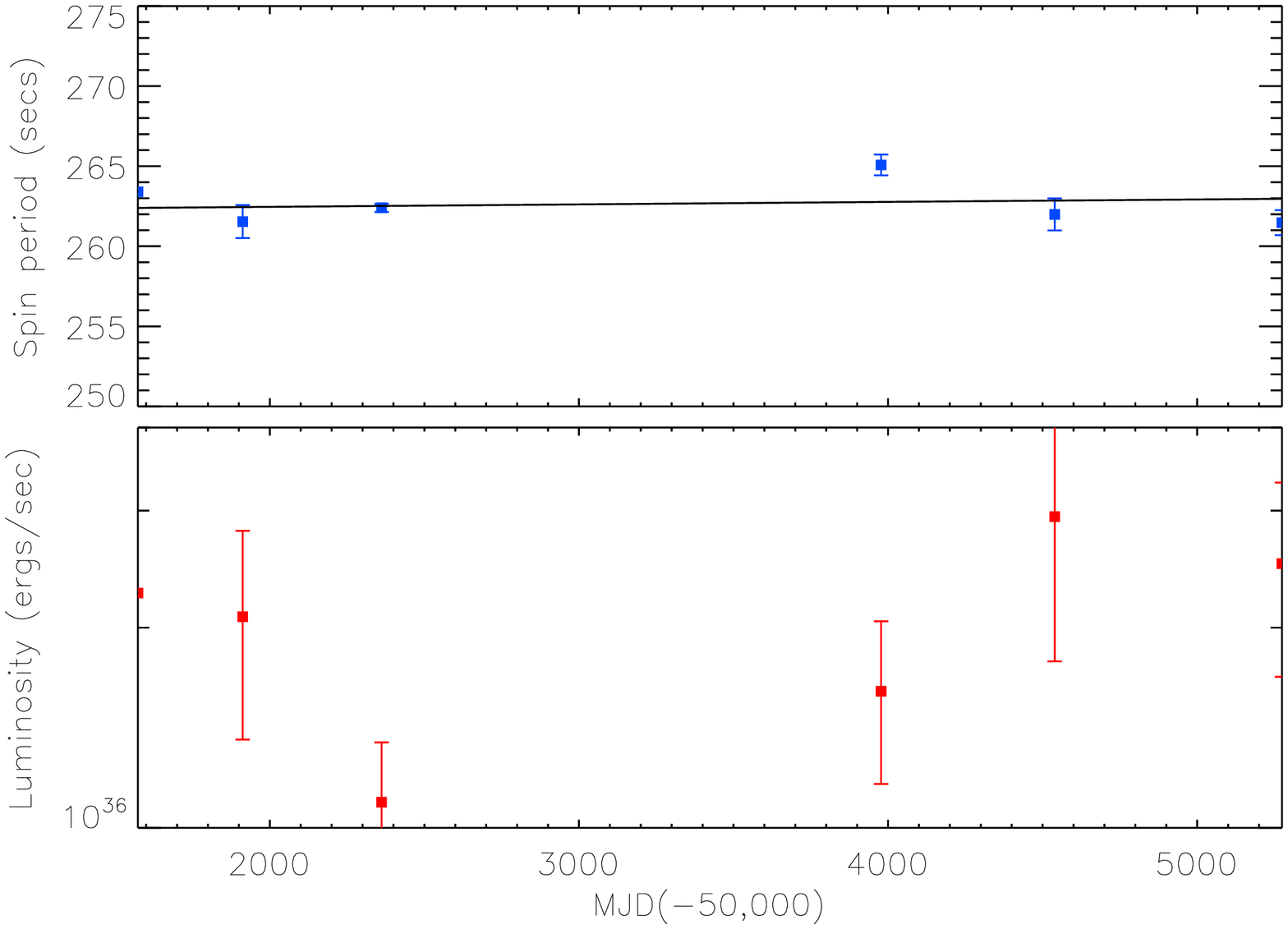}
\caption{The upper panel shows spin period as a function of MJD and the lower panel shows luminosity as a function of MJD for the source SXP264. The line in the upper panel shows the best-fitting $\dot{P}$.}
\end{figure}

\begin{figure} 
\centering
\includegraphics[scale=0.32,angle=90]{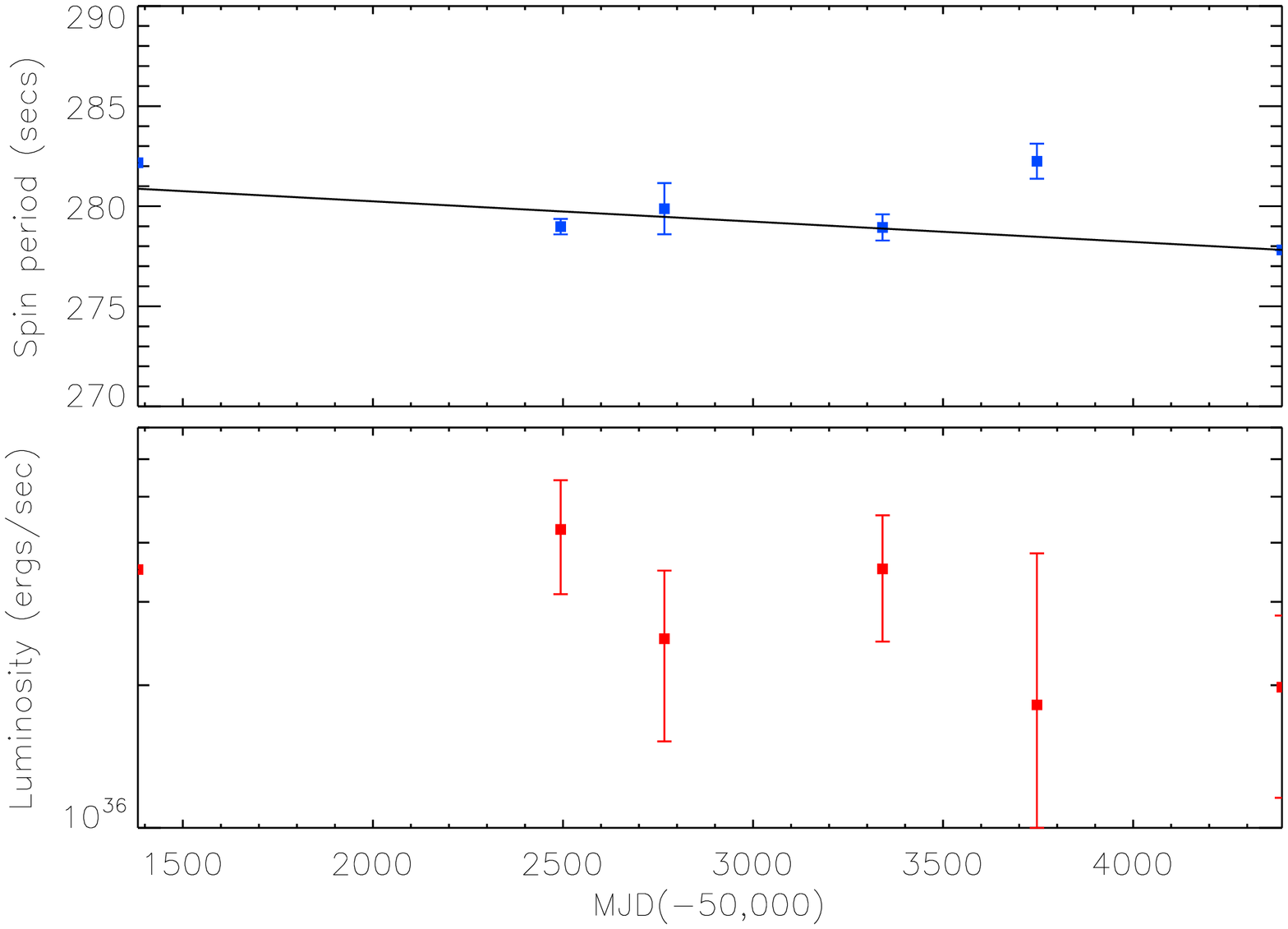}
\caption{The upper panel shows spin period as a function of MJD and the lower panel shows luminosity as a function of MJD for the source SXP280. The line in the upper panel shows the best-fitting $\dot{P}$.}
\end{figure}

\begin{figure} 
\centering
\includegraphics[scale=0.32,angle=90]{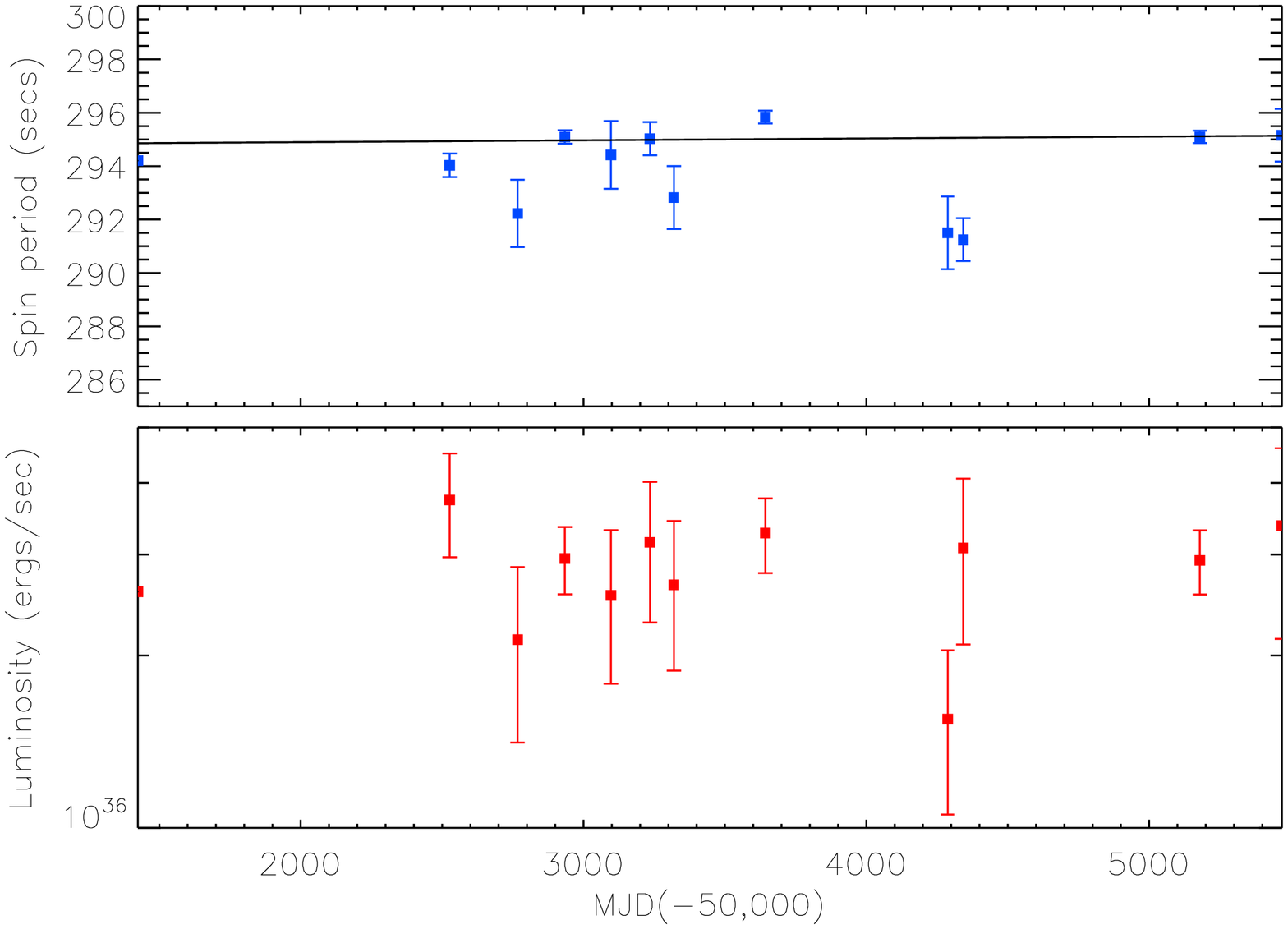}
\caption{The upper panel shows spin period as a function of MJD and the lower panel shows luminosity as a function of MJD for the source SXP293. The line in the upper panel shows the best-fitting $\dot{P}$.}
\end{figure}

\begin{figure} 
\centering
\includegraphics[scale=0.32,angle=90]{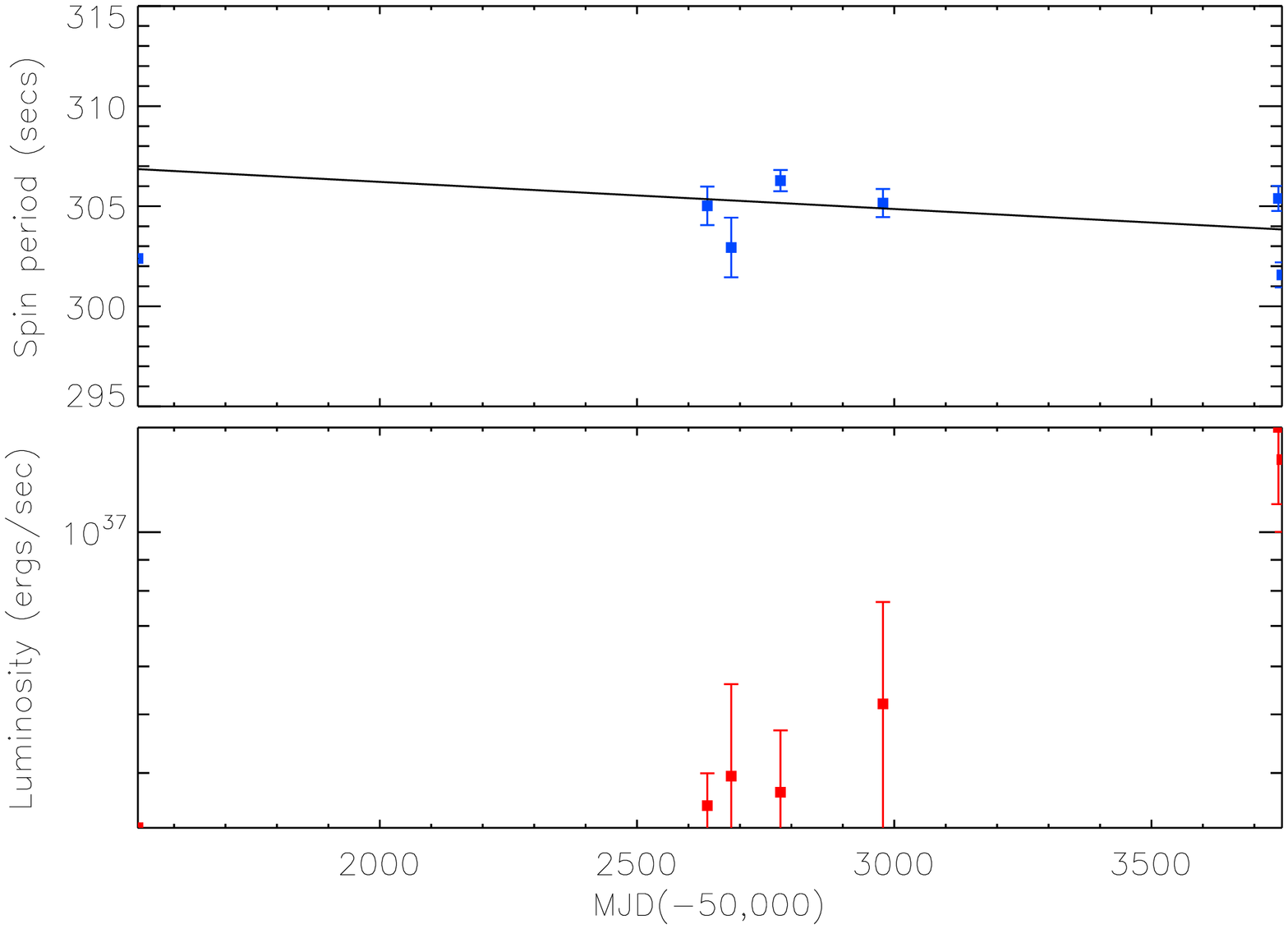}
\caption{The upper panel shows spin period as a function of MJD and the lower panel shows luminosity as a function of MJD for the source SXP304. The line in the upper panel shows the best-fitting $\dot{P}$.}
\end{figure}

\begin{figure} 
\centering
\includegraphics[scale=0.32,angle=90]{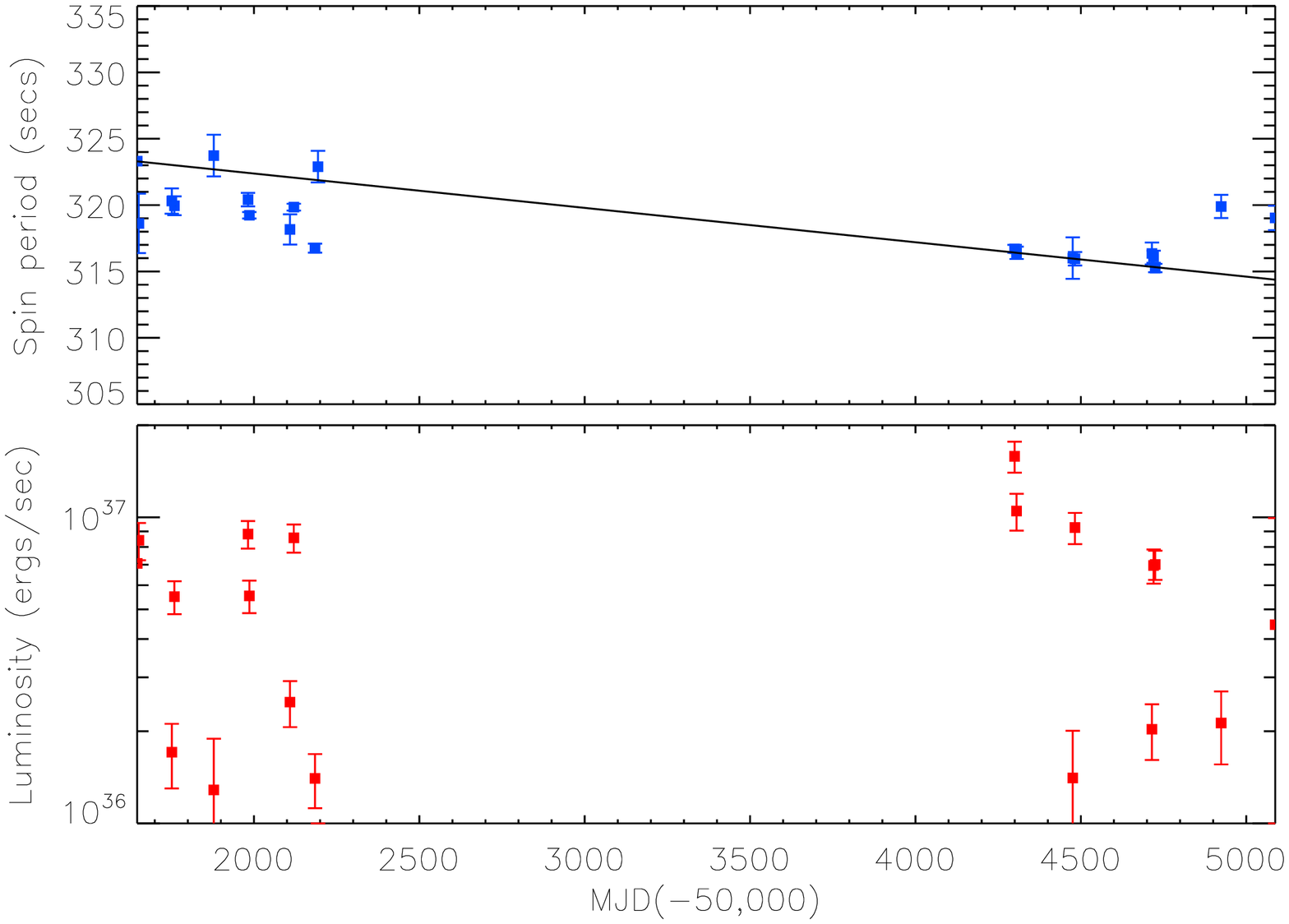}
\caption{The upper panel shows spin period as a function of MJD and the lower panel shows luminosity as a function of MJD for the source SXP323. The line in the upper panel shows the best-fitting $\dot{P}$.}
\end{figure}
\FloatBarrier
\begin{figure} 
\centering
\includegraphics[scale=0.32,angle=90]{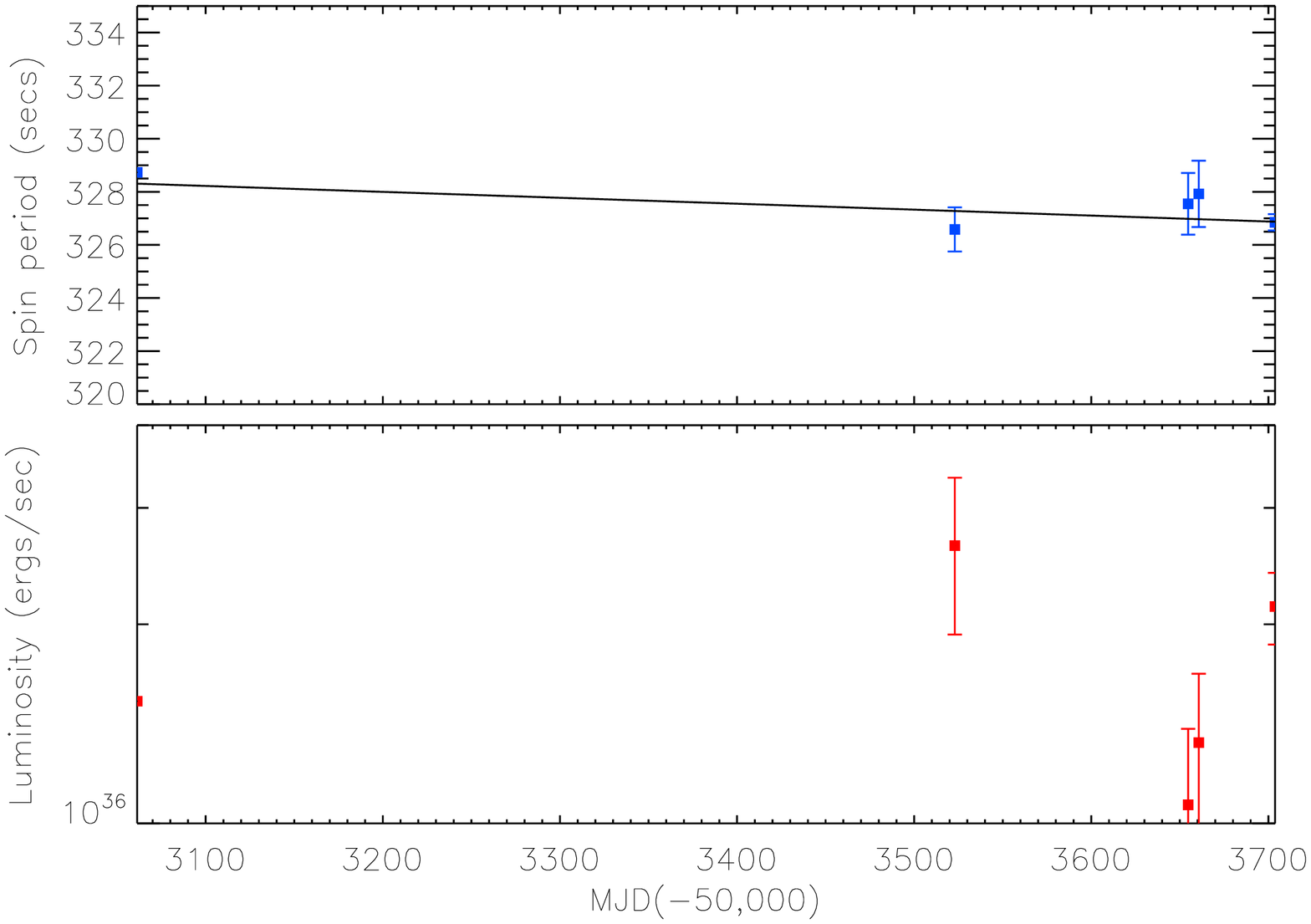}
\caption{The upper panel shows spin period as a function of MJD and the lower panel shows luminosity as a function of MJD for the source SXP327. The line in the upper panel shows the best-fitting $\dot{P}$.}
\end{figure}

\begin{figure} 
\centering
\includegraphics[scale=0.32,angle=90]{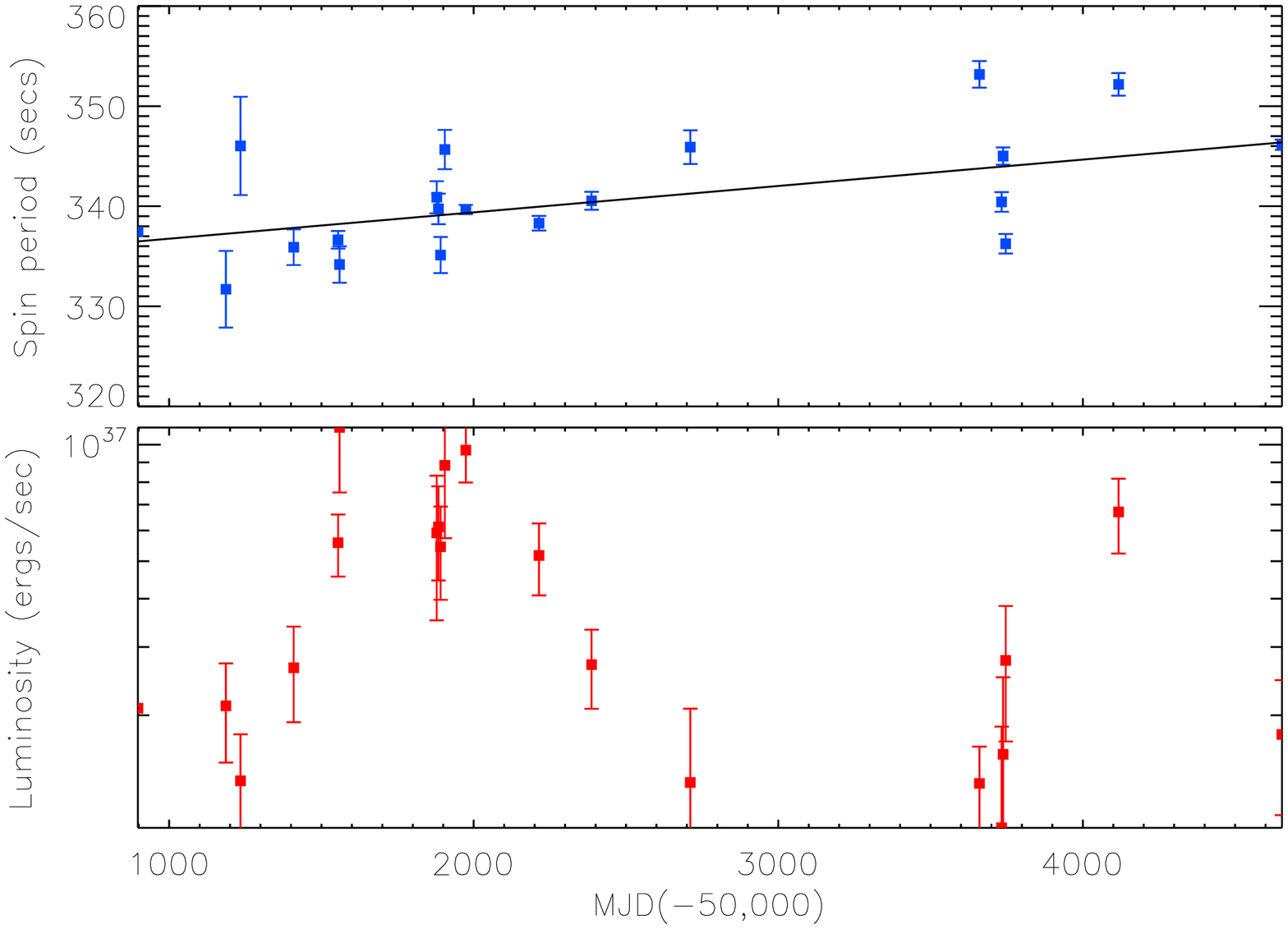}
\caption{The upper panel shows spin period as a function of MJD and the lower panel shows luminosity as a function of MJD for the source SXP342. The line in the upper panel shows the best-fitting $\dot{P}$.}
\end{figure}

\begin{figure} 
\centering
\includegraphics[scale=0.32,angle=90]{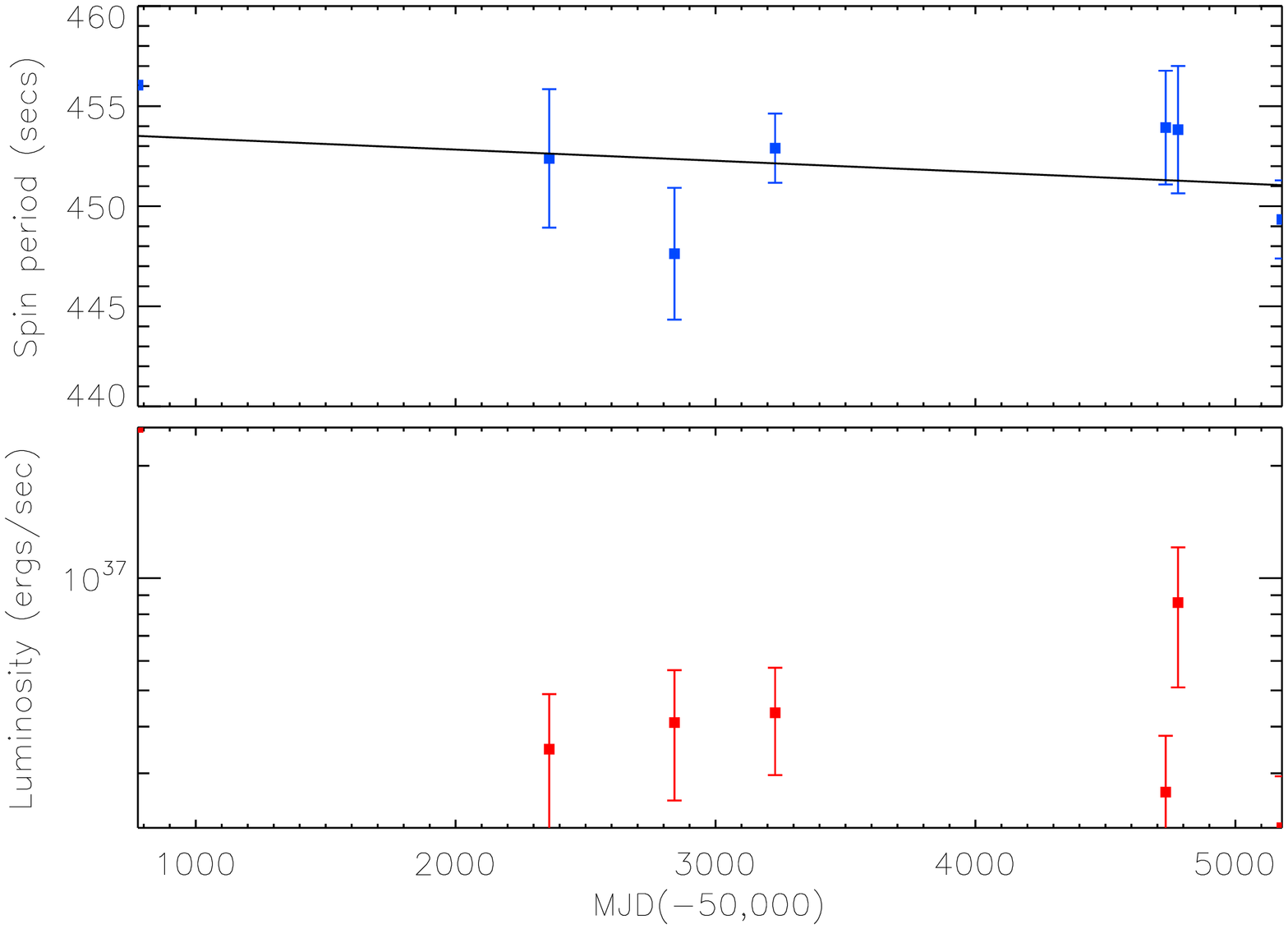}
\caption{The upper panel shows spin period as a function of MJD and the lower panel shows luminosity as a function of MJD for the source SXP455. The line in the upper panel shows the best-fitting $\dot{P}$.}
\end{figure}

\begin{figure} 
\centering
\includegraphics[scale=0.32,angle=90]{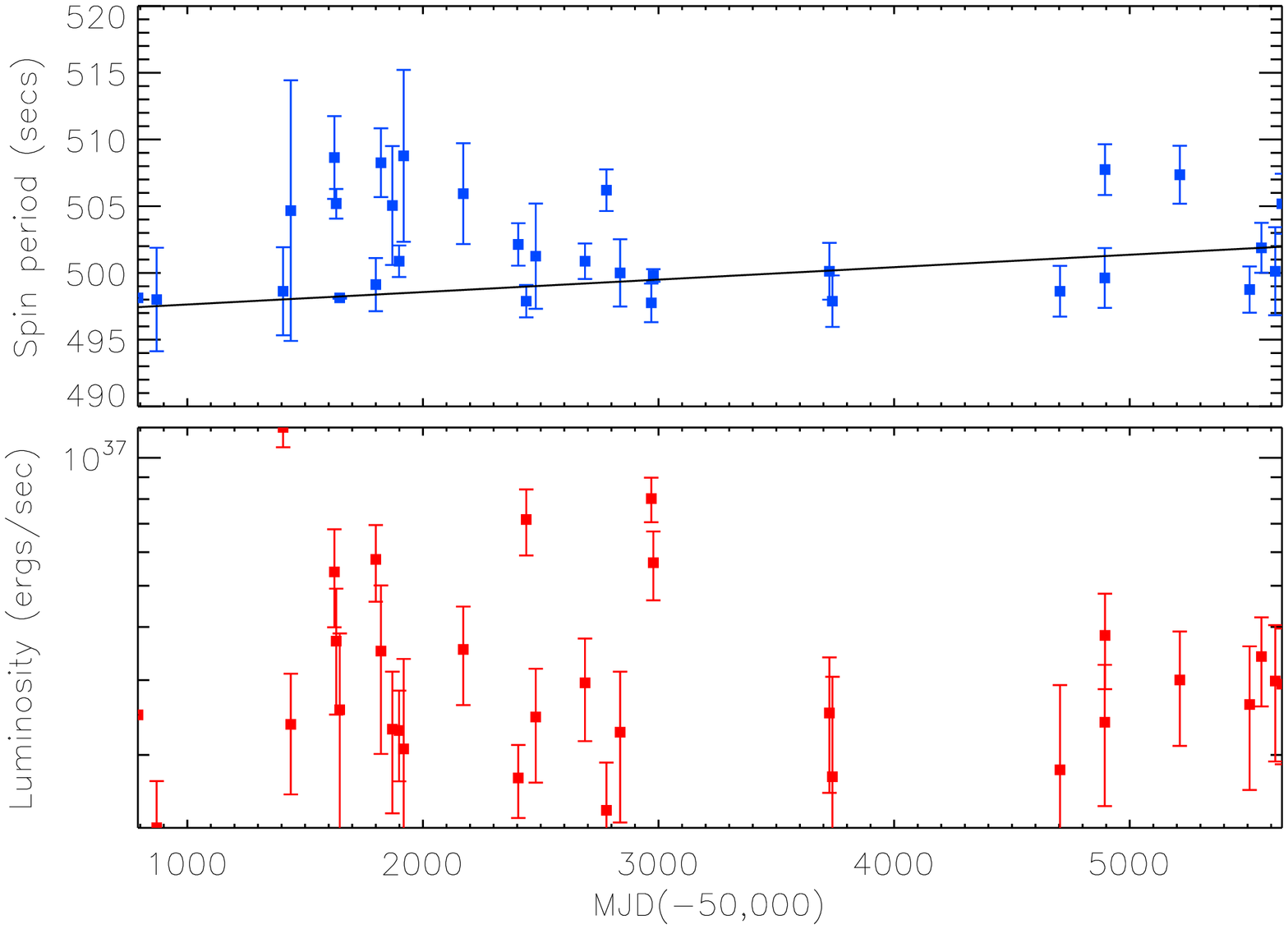}
\caption{The upper panel shows spin period as a function of MJD and the lower panel shows luminosity as a function of MJD for the source SXP504. The line in the upper panel shows the best-fitting $\dot{P}$.}
\end{figure}

\begin{figure} 
\centering
\includegraphics[scale=0.32,angle=90]{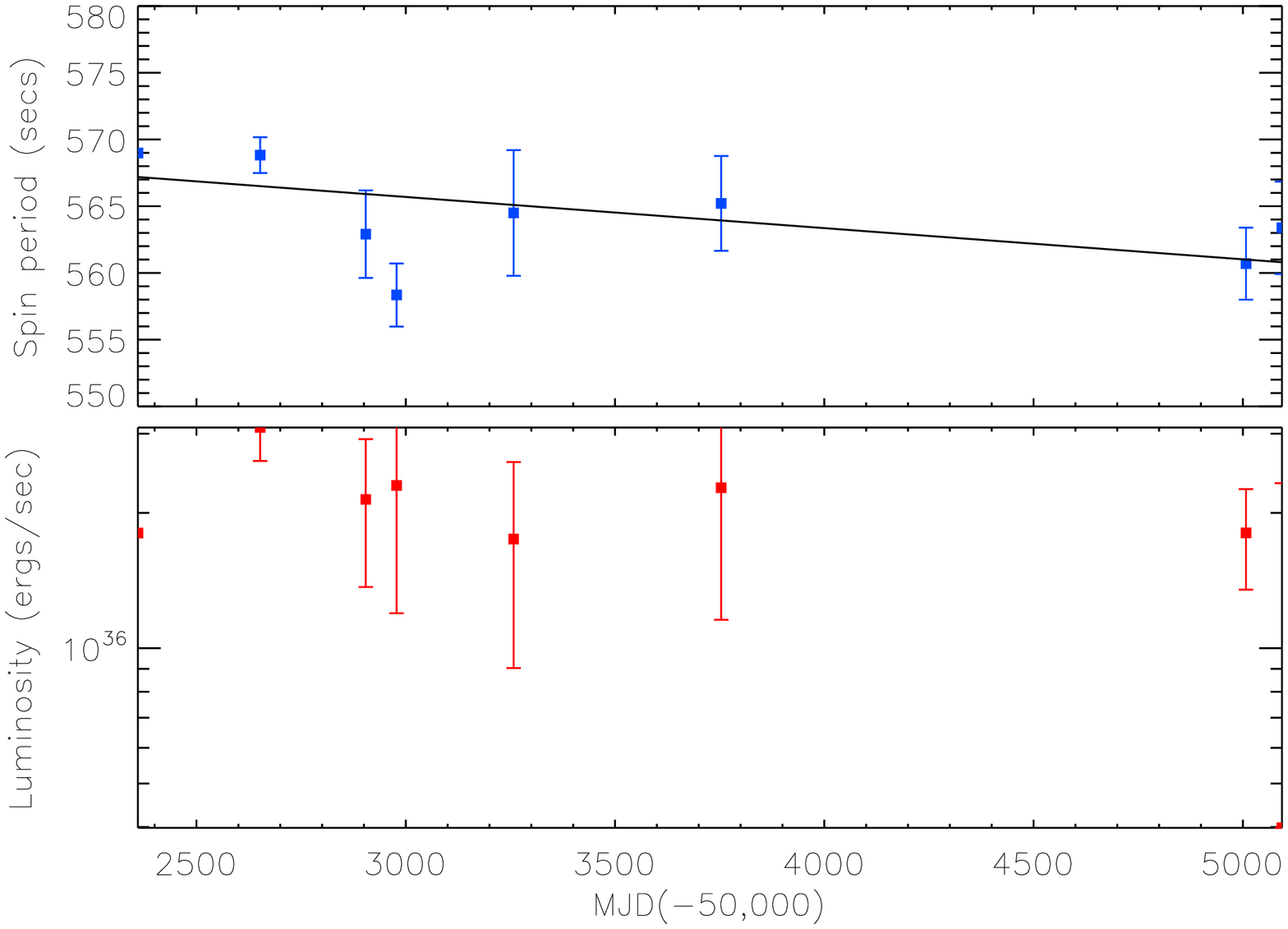}
\caption{The upper panel shows spin period as a function of MJD and the lower panel shows luminosity as a function of MJD for the source SXP565. The line in the upper panel shows the best-fitting $\dot{P}$.}
\end{figure}

\begin{figure} 
\centering
\includegraphics[scale=0.32,angle=90]{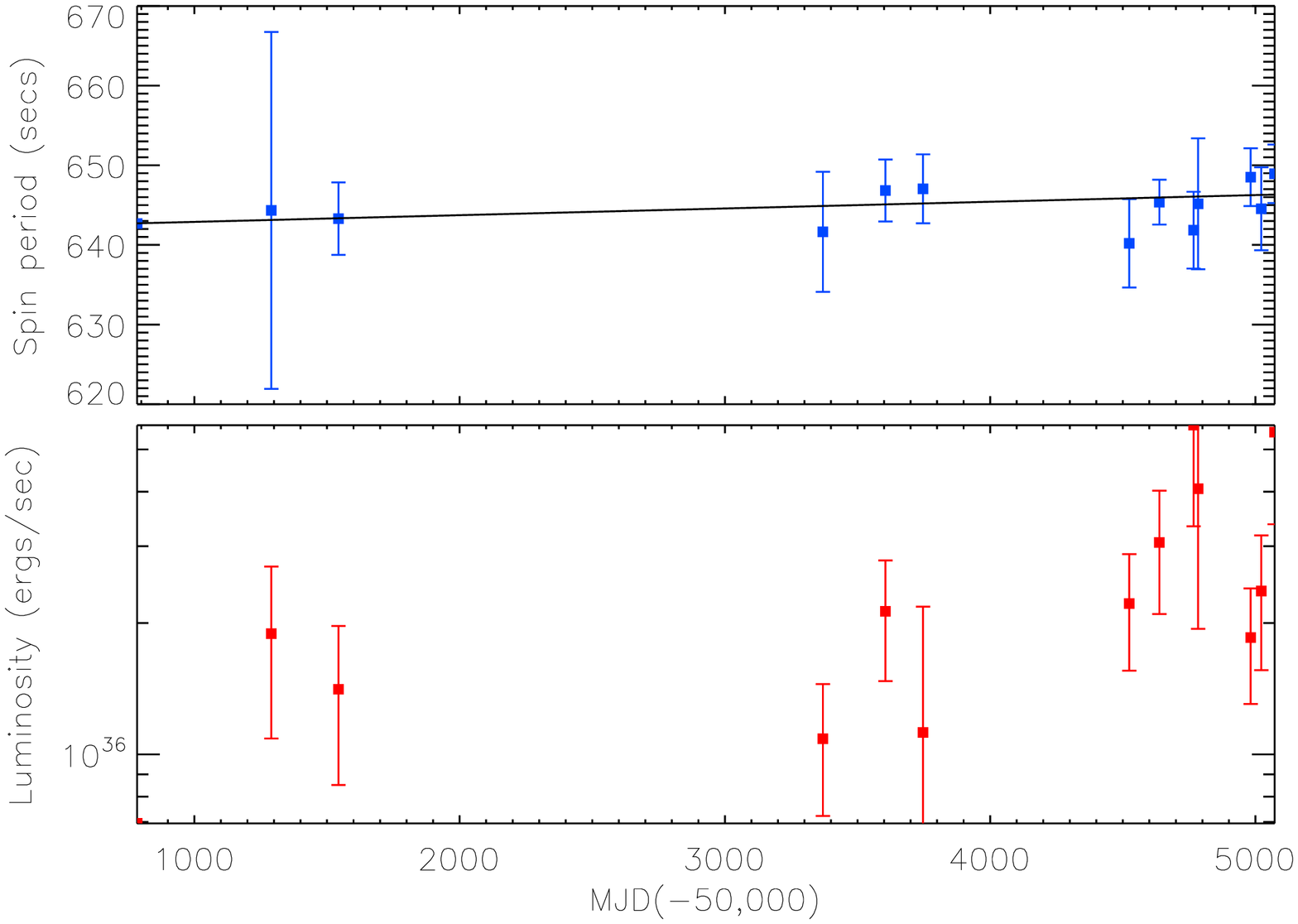}
\caption{The upper panel shows spin period as a function of MJD and the lower panel shows luminosity as a function of MJD for the source SXP645. The line in the upper panel shows the best-fitting $\dot{P}$.}
\end{figure}
\FloatBarrier
\begin{figure}
\centering
\includegraphics[scale=0.32,angle=90]{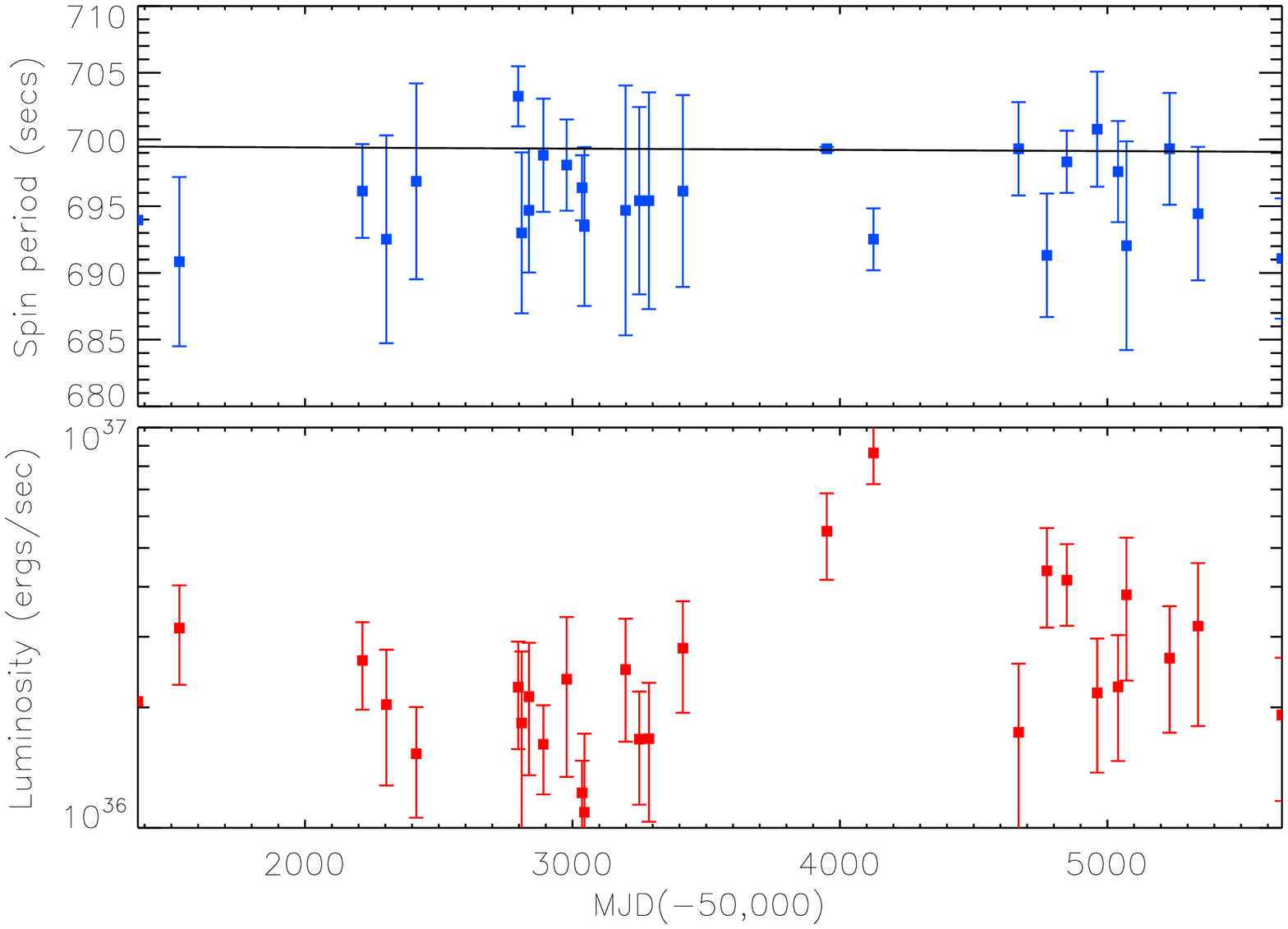}
\caption{The upper panel shows spin period as a function of MJD and the lower panel shows luminosity as a function of MJD for the source SXP701. The line in the upper panel shows the best-fitting $\dot{P}$.}
\end{figure}

\begin{figure} 
\centering
\includegraphics[scale=0.32,angle=90]{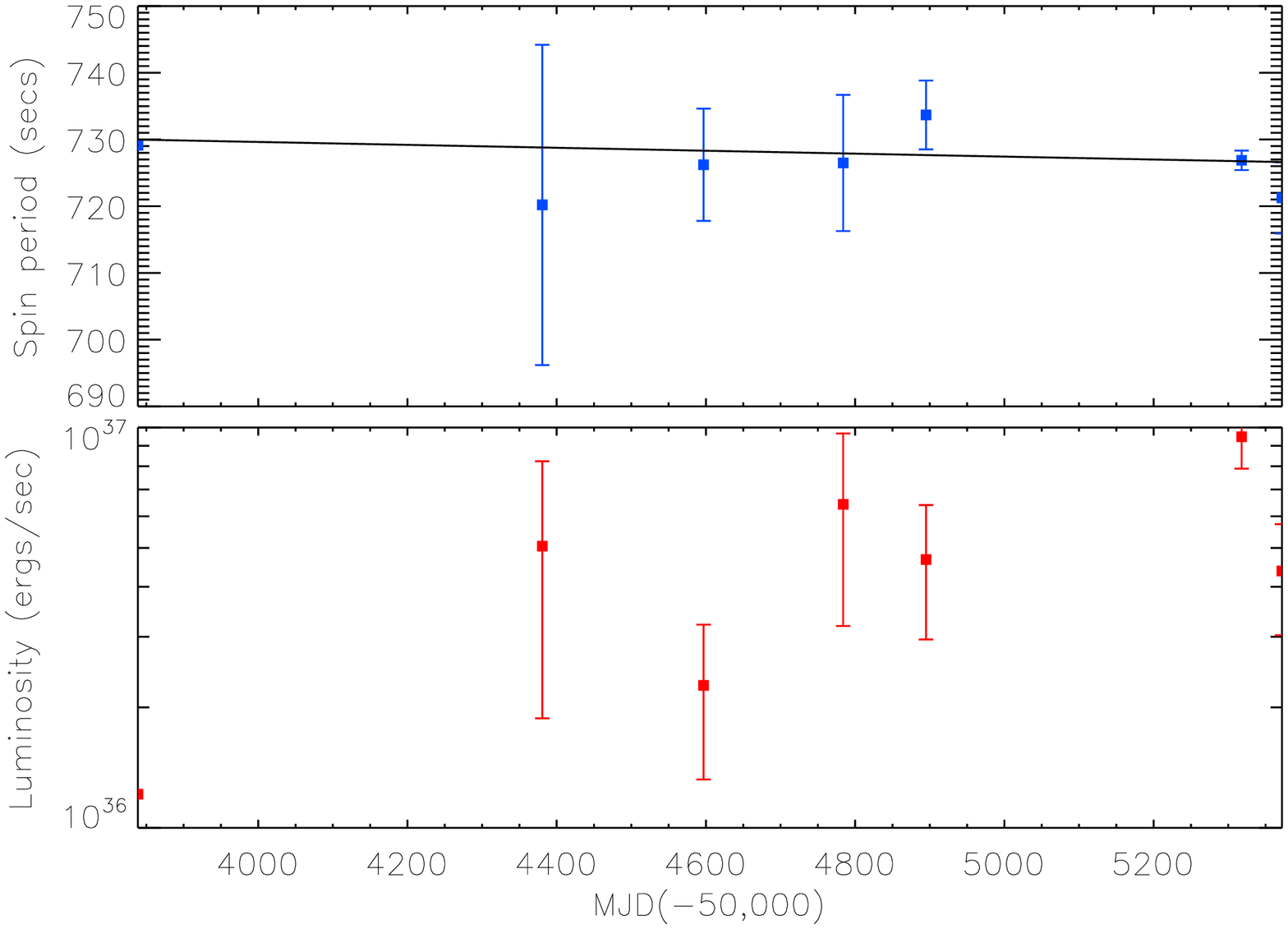}
\caption{The upper panel shows spin period as a function of MJD and the lower panel shows luminosity as a function of MJD for the source SXP726. The line in the upper panel shows the best-fitting $\dot{P}$.}
\end{figure}

\begin{figure} 
\centering
\includegraphics[scale=0.32,angle=90]{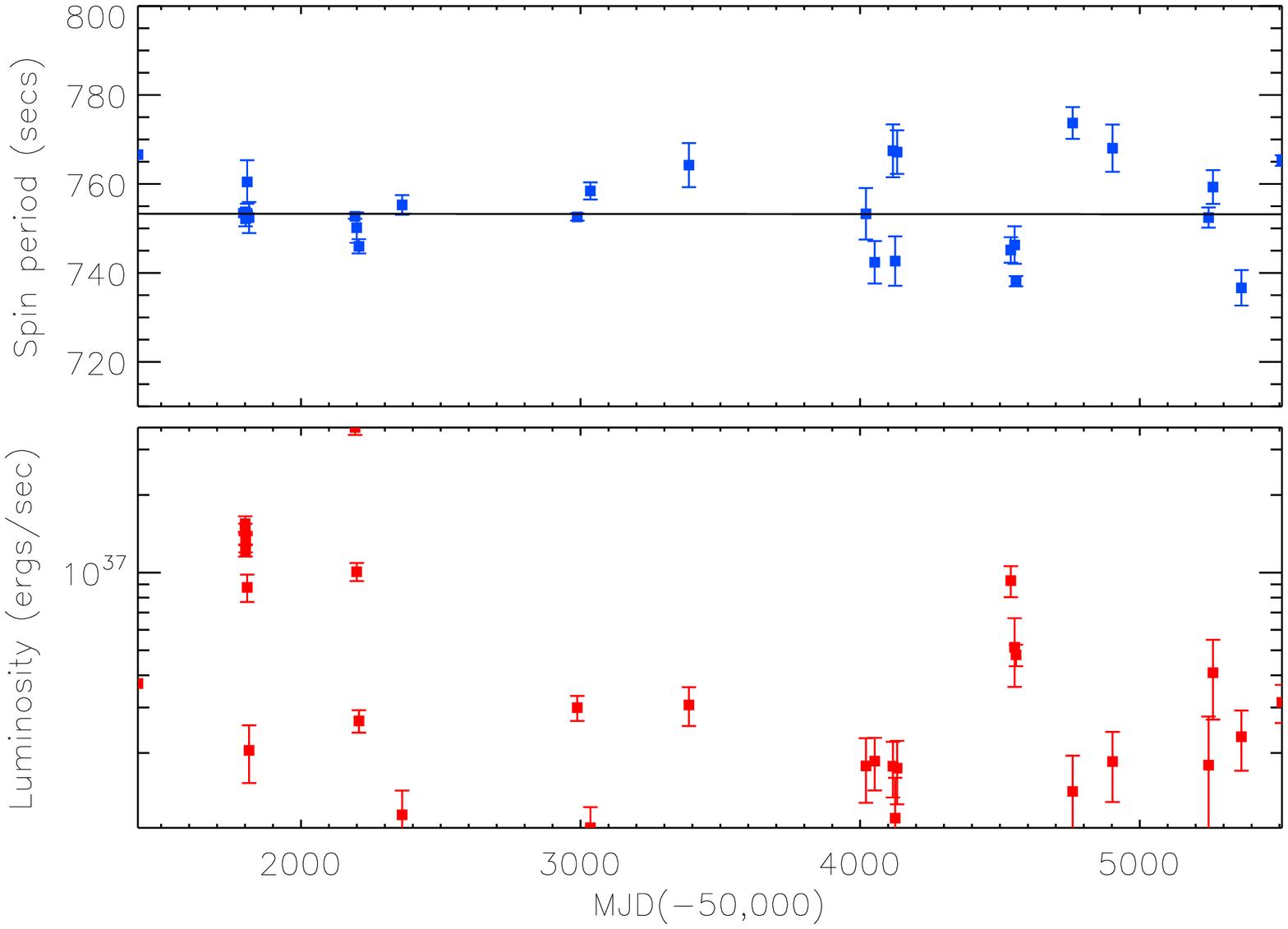}
\caption{The upper panel shows spin period as a function of MJD and the lower panel shows luminosity as a function of MJD for the source SXP756. The line in the upper panel shows the best-fitting $\dot{P}$.}
\end{figure}

\begin{figure} 
\centering
\includegraphics[scale=0.32,angle=90]{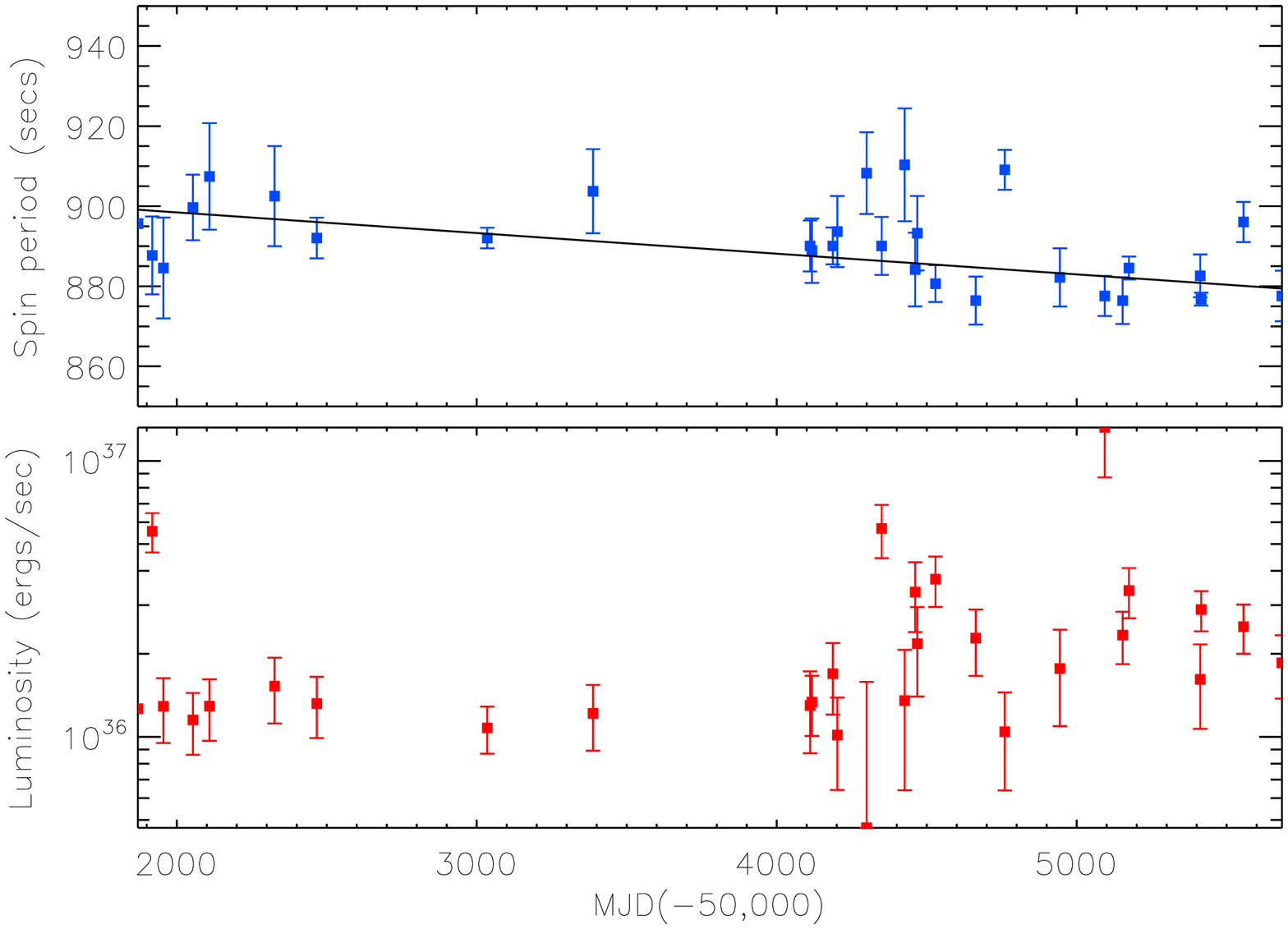}
\caption{The upper panel shows spin period as a function of MJD and the lower panel shows luminosity as a function of MJD for the source SXP893. The line in the upper panel shows the best-fitting $\dot{P}$.}
\end{figure}

\begin{figure} 
\centering
\includegraphics[scale=0.32,angle=90]{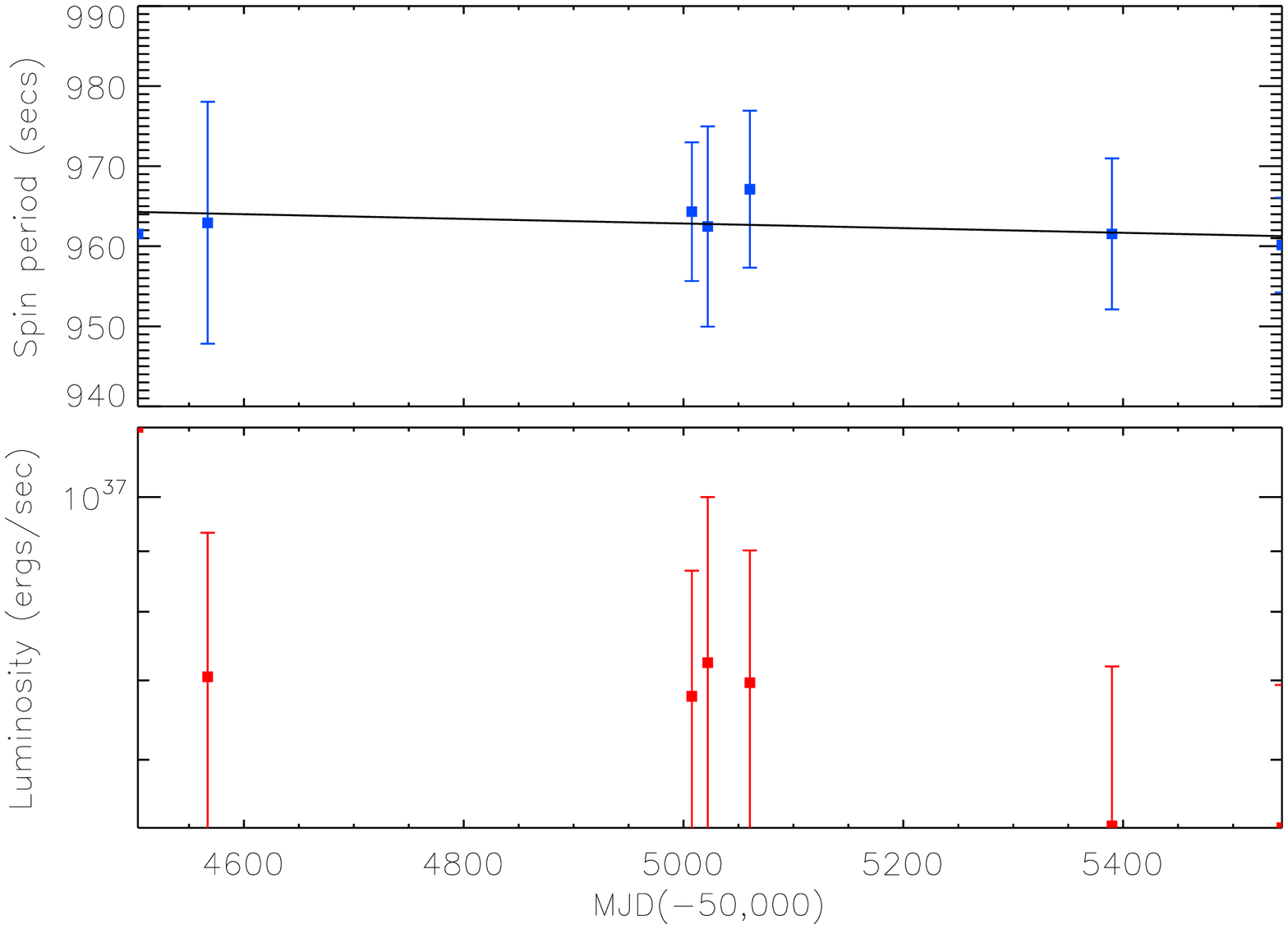}
\caption{The upper panel shows spin period as a function of MJD and the lower panel shows luminosity as a function of MJD for the source SXP967. The line in the upper panel shows the best-fitting $\dot{P}$.}
\end{figure}

\begin{figure} 
\centering
\includegraphics[scale=0.32,angle=90]{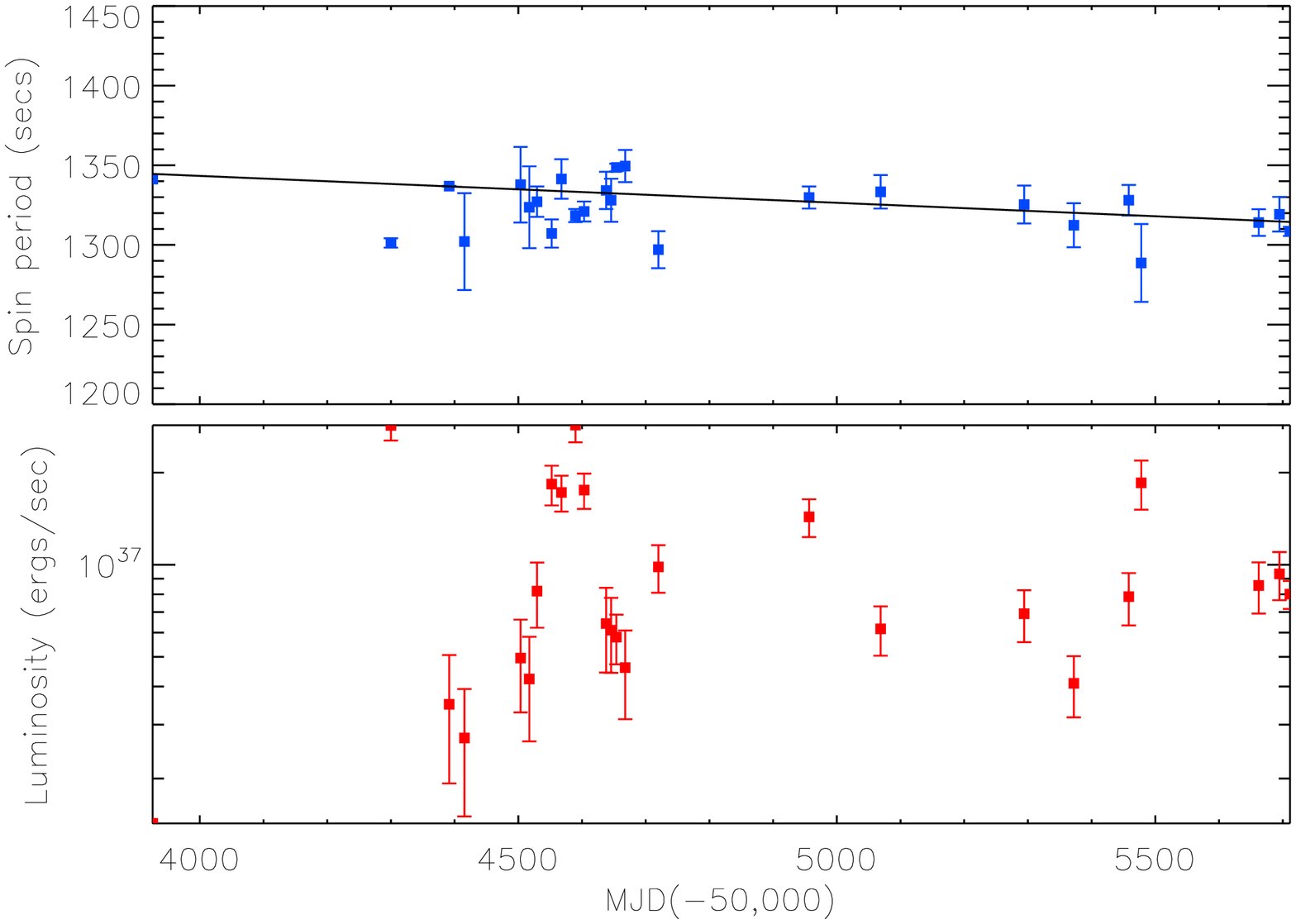}
\caption{The upper panel shows spin period as a function of MJD and the lower panel shows luminosity as a function of MJD for the source SXP1323. The line in the upper panel shows the best-fitting $\dot{P}$.}
\end{figure}
\label{lastpage}

\end{document}